\documentclass[aps,prd,twocolumn,showpacs,superscriptaddress,10pt]{revtex4-2}
\usepackage{xcolor}
\usepackage{hyperref}
\usepackage{amsmath}
\usepackage{amssymb}
\usepackage{graphicx}
\usepackage{graphicx}     
\usepackage{caption}      
\usepackage{subcaption}
\usepackage{bm}
\usepackage{appendix}
\usepackage{orcidlink}

\usepackage{mathtools}
\begin{document}

\title{Analog Gravity in Magneto-Viscous Fluids: Enhanced Analog Hawking Temperature in  Accretion Disks}
\author{Aliv Sahoo \orcidlink{0009-0008-4745-0902}}
\email{alivsahoo@iisc.ac.in}
\affiliation{Department of Physics, Indian Institute of Science}

\author{Mayank Pathak \orcidlink{0000-0002-4834-4211}}
\email{mayankpathak@iisc.ac.in}
\affiliation{Joint Astronomy Programme, Department of Physics, Indian Institute of Science}

\author{Banibrata Mukhopadhyay \orcidlink{0000-0002-3020-9513}}
\email{bm@iisc.ac.in}
\affiliation{Department of Physics, Indian Institute of Science}
\affiliation{Joint Astronomy Programme, Department of Physics, Indian Institute of Science}

\begin{abstract}
We present a three-dimensional visco-magnetoacoustic framework for analog gravity in magnetohydrodynamic (MHD) flows. While standard fluid dissipation typically breaks the Lorentzian signature of acoustic metrics, we demonstrate that evaluating wave perturbations in the eikonal limit alongside the Shakura-Sunyaev $\alpha$-viscosity prescription preserves a well-defined effective spacetime geometry for the fast magnetoacoustic mode. The slow-magnetoacoustic mode and Alfvén mode do not admit a non-degenerate metric. To investigate analog horizon thermodynamics, we utilize astrophysical accretion disks as background media, specifically modeling numerical magnetized advective accretion flows around rotating black holes and analytical advection-dominated inflow-outflow solutions (ADIOS). Standard self-similar ADIOS models strictly enforce a constant Mach number, precluding horizon formation. We therefore introduce a magnetic field perturbation that breaks self-similarity, generates a dynamic Mach number, and enables the formation of a visco-magnetoacoustic horizon. By evaluating the spontaneous phonon emission at these horizons,  we reveal that the analog Hawking temperature is highly sensitive to the magnetic field topology; the spatial orientation of the background magnetic gradients dictates whether the Hawking radiation is amplified or suppressed. Furthermore, we find that increasing the viscosity parameter  leads to a monotonic increase in the analog Hawking temperature.
\end{abstract}
\maketitle

\section{INTRODUCTION}

Black holes are among the most fascinating predictions of general relativity, providing a unique arena in which gravitation, quantum field theory, and thermodynamics intersect \cite{PhysRevD.7.2333, Bardeen:1973gs, Wald:1995yp}. One of the most remarkable theoretical predictions associated with black holes is Hawking radiation, according to which quantum fluctuations near the event horizon give rise to a thermal flux of particles emitted by the black hole \cite{1974Natur.248...30H,Hawking:1975vcx}. Although this phenomenon has profound implications for our understanding of gravity and quantum mechanics, its direct observation remains beyond the reach of current astrophysical experiments because the Hawking temperature of astrophysical black holes is extremely small. Any astrophysical signal of Hawking radiation is very likely swamped by the 2.7 K cosmic microwave background (CMB). 
This observational challenge 
has motivated the development of analog-gravity systems, in which suitably engineered laboratory media reproduce the kinematic properties of curved spacetime, allowing aspects of horizon physics and Hawking radiation to be investigated under controlled experimental conditions \cite{PhysRevLett.106.021302, Steinhauer2016}. Among the most prominent realizations are acoustic black holes, first proposed by Unruh, in which acoustic perturbations propagating through a moving fluid can experience an effective curved spacetime and form an analog horizon when the background flow becomes supersonic \cite{PhysRevLett.46.1351,2011LRR....14....3B,Pathak2022}.

Despite the  theoretical success of analog gravity, the vast majority of formulations strictly rely on ideal, inviscid fluids. Introducing fluid dissipation into the effective spacetime geometry 
is  challenging because  viscosity typically breaks the required Lorentzian signature of the acoustic metric. This breakdown occurs because standard Navier-Stokes viscosity introduces higher-order spatial derivatives \cite{1998CQGra..15.1767V}. Previous attempts to address this analog viscosity problem 
include treating  the dissipation as an explicit spacetime source term \cite{2012MPLA...2750185G}, or relying on  a double perturbation approach \cite{2022Univ....8..205P}. However, accretion disk systems around black holes provide an alternate framework of turbulent viscosity, namely, the $\alpha$-viscosity prescription \cite{1973A&A....24..337S}. This allows a natural setting where the $\alpha-$viscosity naturally removes the  extra derivatives in the fluid equations. This allows the possibility of a dissipative system to be written in a source-free form, successfully preserving Lorentzian effective geometry.


Analog-gravity descriptions of astrophysical accretion flows have been developed previously. Abraham et al. \cite{2006CQGra..23.2371A}, for example, studied the acoustic geometry associated with relativistic accretion in axially symmetric flows, while subsequent works extended this framework to black-hole accretion disks. These studies were, however, restricted to hydrodynamic and inviscid configurations \cite{das2007astrophysicalaccretionanaloguegravity,2007JCAP...06..009D,2006CQGra..23.2371A, Tarafdar:2013oqa, Pu:2012rv}. Noda et al. \cite{2017PhRvD..95j4055N} subsequently extended the analog-gravity framework to magnetohydrodynamic accretion, showing that magnetic fields modify the propagation of perturbations and give rise to a magnetoacoustic geometry. Their construction, however, was restricted to a two-dimensional, inviscid MHD flow, which  constrains the perturbations to a single plane and neglects the full three-dimensional wave propagation critical to many astrophysical systems, including accretion disks. Magnetic fields are nevertheless an essential ingredient of realistic astrophysical accretion systems, where magnetic stresses can efficiently transport angular momentum and significantly influence the structure and dynamics of the accreting plasma \cite{1991ApJ...376..214B,1991ApJ...376..223H,2011MNRAS.418L..79T, 2012MNRAS.423.3083M,2022ApJ...941...30C,2025ApJ...981..162P, 2015ApJ...807...43M}. Thus, while existing analog-gravity models have considered either purely hydrodynamic or ideal-MHD accretion flows, a three-dimensional analog geometry that simultaneously incorporates magnetic fields and viscous stresses remains unexplored.

Furthermore, a number of analytical models of  accretion flows,  some of which are well-known, have been developed with and without magnetic field using self-similar scalings \cite{narayan1994, 1995ApJ...452..710N,1999MNRAS.303L...1B, 2006PASJ...58..469A}. However, exact self-similarity poses a problem when such solutions are used as backgrounds for analog-gravity studies. In these solutions, the radial velocity and sound speed share the same radial dependence, forcing the Mach number to remain constant throughout the flow. As a result, the flow cannot be transonic; instead, it remains entirely subsonic or supersonic depending on the flow parameters. 
Therefore, even a magnetically consistent self-similar accretion solution cannot support horizon formation in an analog-gravity framework. A departure from exact self-similarity is therefore needed.

In this work, we develop a three-dimensional magneto-acoustic framework of analog gravity in fluids. We further use magnetized accretion disks around black holes as our background to introduce viscosity in our system using the $\alpha$-viscosity prescription. We consider numerical solutions in the lower magnetic field regime of the magnetically arrested advective accretion flow (MA-AAF) \cite{Mondal:2019lxg,Mondal:2018onm,Mondal:2018yjr,Mondal:2019xyg,2015ApJ...807...43M} and the magnetically perturbed ADIOS. We show that introducing a small perturbative magnetic field can break the self-similarity of ADIOS when considered as the background flow, generating a non-trivial radial variation in the Mach number that allows the flow to become transonic and form an analog horizon, even within the analytical framework. We investigate the analog Hawking temperature in both systems and study its variation with viscosity, magnetic-field strength, and magnetic-field topology.

The paper is organized as follows. In  Sec.~\ref{sec:eikonal}, we describe the theoretical reasoning used to obtain the effective acoustic metric. In Sec.~\ref{sec:magnetoviscous}, we derive the three-dimensional magnetoacoustic metric and then incorporate $\alpha$-viscosity to obtain the corresponding visco-magnetoacoustic metric, including the conditions for the horizon and ergoregion. In Secs.~\ref{sec:Numericaladaf} and \ref{ADIOS}, we detail the numerical and analytical background flow models used for our study, respectively. In Sec.~\ref{sec:hawking}, we calculate the analog Hawking temperature and discuss its dependence on the magnetic field and viscosity profiles. We conclude in  Sec.~\ref{sec:conclusion} with a brief discussion of future work needed to be done.

\section{Eikonal Approximation and Formalism for Effective Metric}
\label{sec:eikonal}
In general relativity, the propagation of a massless scalar field $\Phi$ in a curved spacetime is governed by the covariant Klein-Gordon equation:
\begin{equation}
    \label{eq:klein_gordon}
    \frac{1}{\sqrt{-g}} \partial_\mu \left( \sqrt{-g} g^{\mu\nu} \partial_\nu \Phi \right) = 0,
\end{equation}
where $g^{\mu\nu}$ is the contravariant metric tensor and $g$ is the determinant of the metric \cite{PhysRevLett.46.1351, 1998CQGra..15.1767V}.

To  analyze the analog geometry of our magnetohydrodynamic flow, we first introduce the effective metric tensor $M^{\mu\nu}$, which governs the high-frequency propagation of fast magnetoacoustic waves, analogously to how $g^{\mu\nu}$ determines the propagation of a massless scalar field  in curved spacetime.


To justify the interpretation of this matrix $M^{\mu\nu}$ as an effective metric tensor, we rely on the formal mathematical equivalence between the propagation of waves in the high-frequency limit in both general relativity and magnetohydrodynamics (MHD). This is the foundational principle of analog gravity.

To determine the trajectories of the wave rays, we apply the Wentzel-Kramers-Brillouin (WKB) or eikonal approximation \cite{2017PhRvD..95j4055N}. We assume the wave takes the form $\Phi(x) = A(x) e^{i S(x)}$, where $A(x)$ is a slowly varying amplitude and $S(x)$ is a rapidly varying phase. Substituting this ansatz into Eq. \eqref{eq:klein_gordon} and taking the high-frequency limit, the second derivatives of the phase dominate, and the equation simplifies to the eikonal equation:
\begin{equation}
    \label{eq:eikonal_gr}
    g^{\mu\nu} \partial_\mu S \partial_\nu S = 0.
\end{equation}
This demonstrates that the wave vector $k_\mu \equiv \partial_\mu S$ is a null vector, meaning the wave's phase propagates along null geodesics determined entirely by the spacetime geometry $g^{\mu\nu}$.

Applying the same WKB approximation to the wave perturbation  $\propto e^{i S(x)}$ in the background MHD flow, the highest-order derivative terms reduce the complex wave equation to a simple algebraic constraint \cite{2017PhRvD..95j4055N}:
\begin{equation}
    \label{eq:eikonal_mhd}
    M^{\mu\nu} \partial_\mu S \partial_\nu S = 0.
\end{equation}
Comparing Equations \eqref{eq:eikonal_gr} and  \eqref{eq:eikonal_mhd} reveals a direct mathematical isomorphism. Because the high-frequency fast magnetoacoustic wave obeys the exact same algebraic constraint as a massless scalar wave in a curved spacetime, the wave effectively ``perceives'' the moving background fluid as a curved geometry. 

Consequently, the coefficient matrix $M^{\mu\nu}$ plays a mathematical role identical to the contravariant metric tensor $g^{\mu\nu}$. Assuming $M^{\mu\nu}$ is invertible, its inverse $M_{\mu\nu}$ is taken to be proportional to  the covariant magnetoacoustic metric, allowing us to construct the effective line element $ds^2 = M_{\mu\nu} dx^\mu dx^\nu$, up to a conformal factor and identify analog horizons and ergoregions.

\section{Analog Gravity   for three-dimensional Magnetoviscous Flow}
\label{sec:magnetoviscous}

The governing equations are: 

\begin{align}
&\nabla \cdot \boldsymbol{E} = 4\pi\rho_e, \\
&\nabla \cdot \boldsymbol{B} = 0, \\
&\nabla \times \boldsymbol{E} = -\frac{1}{c}\frac{\partial \boldsymbol{B}}{\partial t}, \\
&\nabla \times \boldsymbol{B} = \frac{4\pi}{c}\boldsymbol{J} + \frac{1}{c}\frac{\partial \boldsymbol{E}}{\partial t},\label{maxwelleqn} \\
&\frac{\partial \rho}{\partial t} + \nabla \cdot (\rho \boldsymbol{v}) = 0,\label{massconserv}  \\
&\frac{\partial \boldsymbol{B}}{\partial t}= \nabla \times (\boldsymbol{v} \times \boldsymbol{B}),\label{fluxfreeze} \\
&\frac{\partial \boldsymbol{v}}{\partial t} + (\boldsymbol{v} \cdot \nabla)\boldsymbol{v} + \frac{\nabla p}{\rho} \nonumber \\
&\quad - \frac{1}{c} \boldsymbol{J} \times \boldsymbol{B} + \nabla \Phi_g  - \nabla \cdot \mathbf{W} = 0.
\end{align}
Here  $\Phi_g$ represents the gravitational potential  and $\mathbf{W}$ is the viscous stress tensor.

\textbf{Assumptions:}
\begin{itemize}
    \item The barotropic equation of state $P(\rho)$ is assumed.    
    \item We use WKB approximation assuming that the wavelength of perturbations is small
compared  to the scale of spatial variation of the background quantities.
    \item  We will consider systems  similar to accretion disks where the viscous stress tensor, when present, has only one nonzero component, which is  $W_{r\phi} = W_{\phi r} = -\alpha P $, when we choose $\alpha \neq 0$ in the paper.
   
\end{itemize}

\subsection{Magnetoacoustic metric for inviscid flow ($\alpha =0$)}
\subsubsection{Linear perturbation equations}
For a given background flow we use $\rho_0, \mathbf{V}_0, p_0, \mathbf{J}_0,~{\rm and}~\mathbf{B}_0$ to represent the unperturbed background density, velocity, pressure, current density, and magnetic field respectively.
Using the notation of \cite{335f2344-ccce-3243-bd16-64ce104c6ba0}, let $\bm{\xi}(\mathbf{r}_0,t)$ be the displacement of any fluid element from its initial position, $\mathbf{r}_0$, at initial time $t_0$ to its position at time $t$. Then let us introduce a perturbation in $\bm{\xi}(\mathbf{r}_0,t)$ by the amount $\delta\bm{\xi}$. This perturbation can be thought of due to the wave propagation.
\noindent That leads to the following perturbed variables,

\begin{align}
\rho &= \rho_0 + \delta\rho,\\
\mathbf{v} &= \mathbf{V}_0 + \delta\boldsymbol{v},\\
\mathbf{J} &= \mathbf{J}_0 + \delta\mathbf{J},\\
p &= p_0 + \delta p,\\
\mathbf{B} &= \mathbf{B}_0 + \delta \mathbf{B}.
\end{align}

\noindent Then the perturbed Navier-Stokes equation up to  first-order becomes:

\begin{align}
\delta\rho\,\frac{\partial \mathbf{V}_0}{\partial t}
+ \rho_0 \,\frac{\partial \delta\boldsymbol{v}}{\partial t}
+ \rho_0\,\mathbf{V}_0\!\cdot\!\nabla \,\delta\boldsymbol{v}+\rho_0\delta\boldsymbol{v}\cdot\nabla\mathbf{V_0}
+ \delta\rho\,\mathbf{V}_0\!\cdot\!\nabla \mathbf{V}_0\nonumber\\
= -\nabla\,\delta p
+ \frac{\delta\mathbf{J}\times \mathbf{B}_0}{c}
+\frac{\mathbf{J}_0\times \delta\mathbf{B}}{c} - \delta\rho\nabla\Phi_g . \label{eq:14}
\end{align}

The linearized perturbation relations used  throughout this work follow the  standard  MHD perturbation formalism as used in \cite{335f2344-ccce-3243-bd16-64ce104c6ba0}. The density perturbation is obtained from the continuity equation, Eq.~\eqref{massconserv}, while the pressure perturbation follows from the continuity equation 
together with the assumption of a barotropic equation of state. The velocity perturbation is derived by tracking the perturbed trajectory of a fluid element. Likewise, the ideal flux-freezing condition, Eq.~\eqref{fluxfreeze}, determines the magnetic field perturbation, and the perturbation in the current density follows directly from Maxwell's equation Eq.~\eqref{maxwelleqn}. The resulting first-order perturbation relations are \cite{335f2344-ccce-3243-bd16-64ce104c6ba0}:

\begin{align}
\delta P   &= -(c_{s}^2)_{0}\rho_0\nabla\!\cdot\!\delta\bm{\xi}
               -\;\delta\bm{\xi}\!\cdot\!\nabla P_0,
               \label{eq:15}\\[6pt]
\delta\boldsymbol{v}
            &= \frac{\partial\,\delta\bm{\xi}}{\partial t}
               + \mathbf{V}_0\!\cdot\!\nabla\,\delta\bm{\xi}
               - \delta\bm{\xi}\!\cdot\!\nabla\,\mathbf{V}_0,
               \label{eq:16}\\[6pt]
\delta\mathbf{B}
            &= \nabla\times\bigl(\delta\bm{\xi}\times\mathbf{B}_0\bigr),
               \label{eq:17}\\
\delta\mathbf{J}
            &= \frac{c}{4\pi}\,\nabla\times\delta\mathbf{B}
               = \frac{c}{4\pi}\,\nabla\times\Bigl[\nabla\times\bigl(\delta\bm{\xi}\times\mathbf{B}_0\bigr)\Bigr],
               \label{eq:18}\\[6pt]
\delta \rho 
            &= -\rho_0\,\nabla\!\cdot\!\delta\bm{\xi}
               -(\nabla \rho_0)\!\cdot\!\delta \bm{\xi}\,.
               \label{eq:19}
\end{align}
We adopt the eikonal ansatz
\begin{align}
\delta\boldsymbol{\xi}(\mathbf r,t)
&=\boldsymbol{\xi}_1(\mathbf r,t)e^{iS(\mathbf r,t)},
&
S(\mathbf r,t)&=\frac{\Phi(\mathbf r,t)}{\varepsilon},
\label{eq:21}
\end{align}
where $\varepsilon\ll1$   is of the order of the wavelength of the oscillations and is much smaller than the background variation length scale \cite{335f2344-ccce-3243-bd16-64ce104c6ba0,
Bender:1999box,Goedbloed_Keppens_Poedts_2010}. Defining
\[
\partial_t S=-\omega,\qquad \nabla S=\mathbf{k},
\]
and retaining only the leading-order $O(1/\varepsilon)$ terms, we obtain:
\begin{align}
\delta\mathbf{v} &\approx i(-\omega + \mathbf{V}_0 \cdot \mathbf{k})\delta\boldsymbol{\xi}, \label{eq:22} \\
\delta\mathbf{B} &\approx i\mathbf{k} \times (\delta\boldsymbol{\xi} \times \mathbf{B}_0), \label{eq:23} \\
\delta p &\approx -ic_s^2\rho_0(\mathbf{k} \cdot \delta\boldsymbol{\xi}), \label{eq:24} \\
\delta\mathbf{J} &\approx -\mathbf{k} \times \big(\mathbf{k} \times (\delta\boldsymbol{\xi} \times \mathbf{B}_0)\big), \label{eq:25} \\
\delta\rho &\approx -i\rho_0(\mathbf{k} \cdot \delta\boldsymbol{\xi}). \label{eq:26}
\end{align}
\subsubsection{Leading-order Terms}

Substituting Eqs.~\eqref{eq:22}--\eqref{eq:26} 
into Eq.~\eqref{eq:14},
we retain the leading $\mathcal{O}(\varepsilon^{-2})$ terms,
which include $\rho_0\partial_t\delta\mathbf v$,
$\rho_0(\mathbf V_0\cdot\nabla)\delta\mathbf v$, the pressure
gradient, and the Lorentz force. The remaining terms,
including $\rho_0(\delta\mathbf v\cdot\nabla)\mathbf V_0$,
$\delta\rho\,\mathbf V_0\cdot\nabla\mathbf V_0$,
$\delta\rho\nabla\Phi_g$, and
$\mathbf J_0\times\delta\mathbf B$, are of order
$\mathcal{O}(\varepsilon^{-1})$ and are therefore neglected. 
The resulting leading-order equation is
\begin{align}
-\rho_0(\omega-\mathbf{V}_0\cdot\mathbf{k})^2
\delta\boldsymbol{\xi}
={}&-(c_s)_0^2\rho_0\mathbf{k}
(\mathbf{k}\cdot\delta\boldsymbol{\xi}) \notag\\
&-\frac{1}{4\pi}
\left\{
\mathbf{k}\times
\left[\mathbf{k}\times
(\delta\boldsymbol{\xi}\times\mathbf{B}_0)\right]
\right\}\times\mathbf{B}_0 .
\label{eq:33}
\end{align}

This equation constitutes a homogeneous system of three linear
equations for the components of $\delta\boldsymbol{\xi}$. Non-trivial
solutions therefore require the determinant of the coefficient matrix
to vanish, yielding a cubic dispersion relation in
$(\omega-\mathbf{V}_0\cdot\mathbf{k})^2$ with three roots corresponding
to the fast, slow, and Alfv\'en modes. For notational simplicity, henceforth, we denote the background flow velocity $\mathbf{V}_0$ simply by $\boldsymbol{v}$.
\subsubsection{Metric calculation}
\paragraph{Alfvén mode:}
One root of the determinant corresponds to the Alfv\'en mode. We define the Alfvén speed
\begin{equation*}
    \mathbf{V}_A = \frac{\mathbf{B}_0}{\sqrt{4\pi\rho}}.
\end{equation*}
The dispersion relation becomes
\begin{equation*}
    \omega^2 - 2\omega(\mathbf{k} \cdot \boldsymbol{v}) + (\mathbf{k} \cdot \boldsymbol{v})^2 - (\mathbf{k} \cdot \mathbf{V}_A)^2 = 0.
\end{equation*}
Using Eq. \eqref{eq:21}, this can be written  as
\begin{equation}\label{eq:alfven_covariant}
\begin{split}
    (\partial_t S)^2 &+ 2 (\partial_t S)(\partial_i S) v_i + (\partial_i S)(\partial_j S) v_i v_j \\
    &- \frac{(B_i \partial_i S)(B_j \partial_j S)}{4\pi\rho} = 0.
\end{split}
\end{equation}
Thus, the symmetric effective metric tensor \(M^{\mu\nu}\) takes the block form
\begin{equation}
    M^{\mu\nu} =
    \begin{pmatrix}
        1 & v_j \\
        v_i & v_i v_j - \dfrac{B_i B_j}{4\pi\rho}
    \end{pmatrix}.
\end{equation}
Since \(\det M^{\mu\nu}=0\), the Alfv\'en effective metric is degenerate and admits no well-defined inverse.
\paragraph{Fast and slow modes:}
The fast mode gives: 
\begin{equation}
    \label{eq:34}
    (\omega - \vec{k} \cdot \boldsymbol{v})^2 = k^2 \left[ \frac{V_A^2 + c_s^2}{2} \pm \frac{1}{2} \sqrt{(V_A^2 - c_s^2)^2 + 4 c_s^2 V_A^2 \sin^2 \theta} \right],
\end{equation}
where $\theta$ represents the angle between the wave vector $\vec{k}$ and the background magnetic field $\vec{B}$.
Now using \eqref{eq:21}, the above can be written as 
\begin{equation}
\label{eq:35}
\begin{split}
&\left( \frac{\partial S}{\partial t} + \boldsymbol{v} \cdot \nabla S \right)^2\\
&\qquad \qquad= \frac{V_{\mathrm{M}}^2 |\nabla S|^2}{2} \left[
1 \pm \sqrt{
1 - 4
\left( \frac{c_s V_{\mathrm{A}}}{V_{\mathrm{M}}^2} \right)^2
\left( \frac{\boldsymbol{b} \cdot \nabla S}{|\nabla S|} \right)^2
}
\right],
\end{split}
\end{equation}
where $V_{\mathrm{M}} \equiv \sqrt{c_s^2 + V_{\mathrm{A}}^2}$, and $\boldsymbol{b} \equiv \boldsymbol{V}_{\mathrm{A}} / V_{\mathrm{A}}$ represents the direction of background magnetic field lines. 

Following \cite{2017PhRvD..95j4055N}, we define:
\begin{equation} \label{eq:35}
    \eta \equiv \left( \frac{c_s V_{\mathrm{A}}}{V_{\mathrm{M}}^2} \right)^2.
\end{equation}

By assuming this parameter to be sufficiently small ($\eta \ll 1$), we can apply a Taylor series expansion to simplify the expression involved with the square root in Eq.~\eqref{eq:35}, leading to
\begin{equation} \label{eq:36}
    \sqrt{1 - 4\eta \left( \frac{\boldsymbol{b} \cdot \nabla S}{|\nabla S|} \right)^2} = 1 - 2\eta \left( \frac{\boldsymbol{b} \cdot \nabla S}{|\nabla S|} \right)^2 + \mathcal{O}(\eta^2).
\end{equation}

This small-parameter assumption ($\eta \ll 1$) is physically justified in extreme pressure regimes. Specifically, it holds true when the system is either heavily magnetic pressure-dominated ($(c_s / V_{\mathrm{A}})^2 \ll 1$) or gas pressure-dominated ($(V_{\mathrm{A}} / c_s)^2 \ll 1$).
Hence,
\begin{equation}
    \eta \approx 
    \begin{cases} 
        (c_s/V_{\mathrm{A}})^2 \ll 1 & \text{(magnetic pressure-dominated case),} \\ 
        (V_{\mathrm{A}}/c_s)^2 \ll 1 & \text{(gas pressure-dominated case).} 
    \end{cases}
    \label{eq:38}
\end{equation}

Thus, Eq.~\eqref{eq:35} becomes

\begin{equation}
    \left( \frac{\partial S}{\partial t} + \boldsymbol{v} \cdot \nabla S \right)^2 \approx 
    \begin{cases} 
        V_{\mathrm{M}}^2 (|\nabla S|^2 - \eta(\boldsymbol{b} \cdot \nabla S)^2) & \text{(fast mode),} \\ 
        \eta V_{\mathrm{M}}^2 (\boldsymbol{b} \cdot \nabla S)^2 & \text{(slow mode).} 
    \end{cases}
    \label{eq:39}
\end{equation}

 By assigning the coefficients of these equations to the components of matrices $M_{\mathrm{fast}}^{\mu\nu}$ and $M_{\mathrm{slow}}^{\mu\nu}$, Eqs.~\eqref{eq:39} can be written as

\begin{align}
    M_{\mathrm{fast}}^{\mu\nu} \frac{\partial S}{\partial x^\mu} \frac{\partial S}{\partial x^\nu} &= 0 \qquad \text{(fast mode)}, \label{eq:40} \\
    M_{\mathrm{slow}}^{\mu\nu} \frac{\partial S}{\partial x^\mu} \frac{\partial S}{\partial x^\nu} &= 0 \qquad \text{(slow mode)}, \label{eq:41}
\end{align}

where $M_{\mathrm{fast}}^{\mu\nu}$ and $M_{\mathrm{slow}}^{\mu\nu}$ are:

\begin{equation}
    \begin{aligned}
        M_{\mathrm{fast}}^{\mu\nu} &= 
        \begin{pmatrix} 
            -1 & -v^i \\ 
            -v^i & V_{\mathrm{M}}^2 \delta^{ij} - (v^i v^j + \eta V_{\mathrm{M}}^2 b^i b^j) 
        \end{pmatrix}, \\
        M_{\mathrm{slow}}^{\mu\nu} &= 
        \begin{pmatrix} 
            -1 & -v^i \\ 
            -v^i & -(v^i v^j - \eta V_{\mathrm{M}}^2 b^i b^j) 
        \end{pmatrix}, \\
        i, j &= 1, 2,3,
    \end{aligned}
    \label{eq:42}
\end{equation}
which is exactly what is obtained in two-dimensional case in \cite{2017PhRvD..95j4055N}. The slow mode metric is not invertible but the fast mode metric is.

\subsection{Three-dimensional visco-magnetoacoustic metric with $\alpha$ viscosity}
Having established the effective acoustic metric for an ideal, inviscid magnetohydrodynamic flow, we now extend our framework to incorporate the effects of dissipative shear forces. In standard astrophysical accretion models, the transport of angular momentum is classically parameterized by the Shakura-Sunyaev $\alpha$-prescription, where the dominant off-diagonal components of the viscous stress tensor are assumed to be proportional to the local total pressure, such that $W_{r\phi} = W_{\phi r} = -\alpha P_{\text{total}}$ \cite{1973A&A....24..337S,Abramowicz}. Crucially, the applicability of this visco-magnetoacoustic geometry is not restricted to astrophysical accretion flows. The most direct laboratory analog of the $\alpha P$ closure is provided by dense granular flows  where the shear stress ($\tau$) scales with the confining pressure ($P$) according to $\tau = \mu(I) P$.  Here $\mu(I)$ is   the inertial-number-dependent friction coefficient \cite{2006Natur.441..727J}.
This system therefore offers an experimental setting in which the pressure-proportional stress assumption underlying our model can be tested. By evaluating the first-order wave perturbations of this viscous stress under the eikonal approximation, we will demonstrate how this dissipative mechanism introduces a non-trivial correction to the fast magnetoacoustic dispersion relation, ultimately yielding a modified effective metric tensor that reshapes the geometry of the analog event horizon.

To explicitly derive the modified metric, we now specialize our calculations to an accretion disk modeled in cylindrical coordinates.
The dynamics of the fluid are governed by the Navier-Stokes equation, extended to include  the magnetic field effects via Lorentz force:
\begin{equation} \label{eq:44}
\rho \frac{\partial \boldsymbol{v}}{\partial t} + \rho (\boldsymbol{v} \cdot \nabla) \boldsymbol{v} = -\nabla P +\rho \nabla\Phi_g + \nabla \cdot \mathbf{W} + \mathbf{J} \times \mathbf{B}.
\end{equation}

Introducing first-order perturbations, the linearized equation acquires an additional term $\nabla \cdot \delta \mathbf{W}$ compared to the inviscid case discussed above. We evaluate the viscous stress tensor $\mathbf{W}$ in cylindrical coordinates $(r, \phi, z)$. Following standard accretion disk theory, we assume the dominant viscous stress is the azimuthal shear. Because angular momentum is transported outward, this stress component is negative, setting $W_{r\phi} = W_{\phi r} = -\alpha P_{\text{total}}$ \cite{1973A&A....24..337S}, where $P_{\text{total}} = P_{\text{gas}} + P_{\text{rad}}$
\cite{Abramowicz}. All other components of $W_{ij}$ are assumed to be negligibly small. 

The divergence of the stress tensor in cylindrical coordinates simplifies to:
\begin{equation}
\nabla \cdot \mathbf{W} = -\frac{1}{r}\frac{\partial (\alpha P)}{\partial \phi}  \hat{r} -\left[ \frac{\partial (\alpha P)}{\partial r} + \frac{2(\alpha P)}{r} \right] \hat{\phi}.
\end{equation}

Applying a first-order perturbation, the divergence becomes:
\begin{equation}
\nabla \cdot \delta \mathbf{W} = -\alpha \left[ \frac{\partial \delta P}{\partial r} + \frac{2\delta P}{r} \right] \hat{\phi}   -\frac{\alpha}{r}\frac{\partial \delta P}{\partial \phi}\hat{r}.
\end{equation}

The total pressure perturbation is $\delta P = \delta P_{\text{gas}} + \delta P_{\text{rad}}$. As derived earlier under the  eikonal (WKB) approximation,  the density and gas pressure perturbations are obtained using Eqs. \eqref{eq:26} and \eqref{eq:24}: 
\begin{align}\label{eq:47}
\delta \rho &\simeq -i\rho (\mathbf{k} \cdot \boldsymbol{\xi}), \\
\delta P_{\text{gas}} &\simeq -i c_s^2 \rho (\nabla \cdot \boldsymbol{\xi}) \simeq -i c_s^2 \rho (\mathbf{k} \cdot \boldsymbol{\xi})\label{eq:48}.
\end{align}

Using the ideal gas law $T \propto P_{\text{gas}}/\rho$ and the radiation pressure definition, $P_{\text{rad}} \propto T^4$, 
the radiation pressure perturbation can be obtained as:
\begin{equation}\label{eq:49}
\delta P_{\text{rad}} = 4 P_{\text{rad}} \left( \frac{\delta P_{\text{gas}}}{P_{\text{gas}}} - \frac{\delta \rho}{\rho} \right).
\end{equation}

Using Equations \eqref{eq:47} -- \eqref{eq:49}, and eikonal approximation, we get:
\begin{align}
\nabla \cdot \delta \mathbf{W} =
-A (\mathbf{k} \cdot \boldsymbol{\xi}) (k_\phi \hat{r} + k_r \hat{\phi}),\nonumber
\end{align}
where we define 
\begin{equation}
A = \alpha c_s^2 \rho + 4 \alpha P_{\text{rad}} \left( \frac{c_s^2 \rho}{P_{\text{gas}}} - 1 \right).   
\end{equation}

Let $\omega' = \omega - \mathbf{k} \cdot \boldsymbol{v}_0$. Substituting the viscous force into the perturbed Navier-Stokes equation gives the full vector equation:
\begin{equation}
\begin{split}
-\rho(\omega')^2 \boldsymbol{\xi} &= -c_s^2 \rho \mathbf{k} (\mathbf{k} \cdot \boldsymbol{\xi}) \\
&\quad - \frac{1}{4\pi} [\mathbf{k} \times (\mathbf{k} \times (\boldsymbol{\xi} \times \mathbf{B}_0))] \times \mathbf{B}_0 \\
&\quad - A (\mathbf{k} \cdot \boldsymbol{\xi}) (k_\phi \hat{r} + k_r \hat{\phi}).
\end{split}
\label{eq:perturbed_NS}
\end{equation}
Letting $\Xi_k \equiv \mathbf{k}\cdot\boldsymbol{\xi}$ and $\Xi_B \equiv \mathbf{B}_0\cdot\boldsymbol{\xi}$, taking the dot product of the entire equation with $\mathbf{B}_0$ gives:
\begin{equation}
\begin{split}
\left(\omega'^2 - \frac{(\mathbf{k} \cdot \mathbf{B}_0)^2}{4\pi \rho}\right) \Xi_B &= \Xi_k \bigg[ c_s^2 (\mathbf{k} \cdot \mathbf{B}_0) \\
&\quad + \frac{A}{\rho} (\mathbf{B}_0 \cdot (k_\phi \hat{r} + k_r \hat{\phi})) \bigg].
\end{split}
\label{eq:sys_Y}
\end{equation}

Taking the dot product of Eq. \eqref{eq:perturbed_NS} with $\mathbf{k}$ gives:

\begin{equation}
\label{eq:sys_X}
\begin{split}
    \left( \omega'^2 - \frac{(\mathbf{k} \cdot \mathbf{B}_0)^2}{4\pi\rho} \right) \Xi_k
    &= \Xi_k \biggl[ \left( c_s^2 + \frac{B_0^2}{4\pi\rho} \right) k^2 \\
    &\qquad\quad + \frac{A}{\rho} \bigl( \mathbf{k} \cdot (k_\phi \hat{r} + k_r \hat{\phi}) \bigr) \biggr] \\
    &\quad - \Xi_B \frac{(\mathbf{k} \cdot \mathbf{B}_0) k^2}{4\pi\rho}.
\end{split}
\end{equation}

Equations \eqref{eq:sys_Y} and \eqref{eq:sys_X} form a homogeneous system of linear equations in terms of the variables $\Xi_k$ and $\Xi_B$. The solutions to this system dictate the physical wave modes.

\paragraph{Alfvén mode (trivial solution $\Xi_k= \Xi_B = 0$):}
Consider the case where the homogeneous system gives a trivial solution, $\Xi_k= 0$ and $\Xi_B = 0$. This implies $\mathbf{k} \cdot \boldsymbol{\xi} = 0$ (the wave is strictly incompressible) and $\mathbf{B}_0 \cdot \boldsymbol{\xi} = 0$. 

Plugging these conditions directly back into the Eq. \eqref{eq:perturbed_NS}, the acoustic and viscous terms vanish immediately.  Eq. \eqref{eq:perturbed_NS} thus naturally reduces to:
\begin{equation*}
-\rho(\omega')^2 \boldsymbol{\xi} = - \frac{(\mathbf{k} \cdot \mathbf{B}_0)^2}{4\pi}\boldsymbol{\xi},
\end{equation*}
from which  we naturally extract the Alfvén mode dispersion relation:
\begin{equation}\label{alfven}
(\omega - \mathbf{k} \cdot \boldsymbol{v}_0)^2 = \frac{(\mathbf{k} \cdot \mathbf{B}_0)^2}{4\pi \rho}.
\end{equation}
This essentially gives the same metric as in the case of $\alpha =0 $ and is degenerate.

\paragraph{Fast and slow modes (non-trivial solution):}
For non-trivial solutions ($\Xi_k, \Xi_B \neq 0$) representing compressible modes, the determinant of the coefficient matrix of the system of Equations \eqref{eq:sys_Y} and \eqref{eq:sys_X} must be zero.

Setting the determinant to zero gives the following highly structured quadratic equation in $\omega'^2$:
\begin{equation*}
\begin{split}
-4\pi \rho \omega'^4 &+ \omega'^2 \left[ k^2(B_0^2 + 4\pi \rho c_s^2) + 8\pi A k_r k_\phi \right] \\
&- k^2 \left[ c_s^2 (\mathbf{k} \cdot \mathbf{B}_0)^2 + \frac{A}{\rho} (\mathbf{k} \cdot \mathbf{B}_0) (B_r k_\phi + B_\phi k_r) \right] = 0.
\end{split}
\end{equation*}

Solving this polynomial for $(\omega - \mathbf{k} \cdot \boldsymbol{v})^2$ gives the exact dispersion relation for the fast and slow modes:
\begin{equation}\label{eq:54}
(\omega - \mathbf{k} \cdot \boldsymbol{v})^2 = \frac{[(B_0^2 + 4\pi \rho c_s^2)k^2 + 8\pi A k_r k_\phi]}{8\pi \rho} \left( 1 \pm \sqrt{1 - \mathcal{F}} \right),
\end{equation}
where the fraction $\mathcal{F}$ inside the radical is given by:
\begin{equation}\label{F}
\mathcal{F} = \frac{16\pi \rho k^2 \left[ c_s^2 (\mathbf{k} \cdot \mathbf{B}_0)^2 + \frac{A}{\rho} (\mathbf{k} \cdot \mathbf{B}_0) (B_r k_\phi + B_\phi k_r) \right]}{[(B_0^2 + 4\pi \rho c_s^2)k^2 + 8\pi A k_r k_\phi]^2}.
\end{equation}

\subsubsection{Metric calculation}

\textbf{Assumptions:} We utilize independent multi-parameter perturbation theory. This approach relies on two distinct dimensionless small parameters:

\begin{itemize}
    \item \textbf{The ideal extreme approximation \cite{2017PhRvD..95j4055N}:}  We assume the dimensionless  parameter $\eta$ is strictly small:
    \begin{equation}\label{eq:55}
        \eta \equiv \left( \frac{c_s V_A}{V_M^2} \right)^2 \ll 1.
    \end{equation}
    This assumption is physically justified in extreme regimes,  namely when the background flow is either heavily magnetic pressure-dominated ($V_A^2 \gg c_s^2$) or heavily gas pressure-dominated ($c_s^2 \gg V_A^2$).
    
    \item \textbf{The small viscous perturbation approximation:} We assume that the amplitude of the viscous shear stress is small compared to the fast mode restorative force. We define a new dimensionless viscous parameter $\sigma$:
    \begin{equation}{\label{eq:56}}
        \sigma \equiv \frac{A}{\rho V_M^2} \ll 1.
    \end{equation}
 
\end{itemize}

Assuming $\sigma \ll 1$ and $\eta \ll 1$ as independent first-order
small parameters, we neglect terms of order $\mathcal{O}(\sigma^2)$,
$\mathcal{O}(\eta^2)$, $\mathcal{O}(\sigma\eta)$, and higher. This
approximation reduces the dispersion relation to the quadratic form
$M^{\mu\nu}k_\mu k_\nu=0$, allowing us to define a valid Lorentzian
effective metric. To first-order in $\eta$ and $\sigma$, the quantity
$\mathcal{F}$, defined in Eq.~\eqref{F}, becomes
\begin{equation}
\mathcal{F}
\simeq
4\eta\frac{(\mathbf{k}\cdot\mathbf{b})^2}{k^2}
+
4\sigma\frac{V_A^2}{V_M^2}
\frac{(\mathbf{k}\cdot\mathbf{b})
(b_rk_\phi+b_\phi k_r)}{k^2}.
\end{equation}

Because $\eta, \sigma \ll 1$, it follows that $\mathcal{F} \ll 1$. Applying the approximation $\sqrt{1-\mathcal{F}} \simeq 1-\mathcal{F}/2$ yields the dispersion relation for the fast mode:
\begin{align}
(\omega-\mathbf{k}\cdot\boldsymbol{v})^2
&\simeq V_M^2k^2
+2\sigma V_M^2k_rk_\phi
-\eta V_M^2(\mathbf{k}\cdot\mathbf{b})^2 \notag\\
&\quad
-\sigma V_A^2(\mathbf{k}\cdot\mathbf{b})
(b_rk_\phi+b_\phi k_r),
\label{fastdispersion}
\end{align}
and for the slow mode:
\begin{align}
(\omega-\mathbf{k}\cdot\boldsymbol{v})^2
&\simeq
\eta V_M^2(\mathbf{k}\cdot\mathbf{b})^2
+\sigma V_A^2(\mathbf{k}\cdot\mathbf{b})
(b_rk_\phi+b_\phi k_r).
\label{slowdispersion}
\end{align}
To construct the effective  metric tensor $M^{\mu\nu}$, we map the cleanly isolated dispersion relations into the covariant eikonal equation:
\begin{equation}
    M^{\mu\nu} \partial_\mu S \partial_\nu S = 0 \quad \Rightarrow \quad M^{\mu\nu} k_\mu k_\nu = 0,
\end{equation}
where the covariant wavevector is defined as $k_\mu \equiv (-\omega, \mathbf{k})$. Expanding this explicit sum gives:
\begin{equation}
    M^{00}\omega^2 - 2M^{0i}\omega k_i + M^{ij}k_i k_j = 0.
\end{equation}

By expanding the $(\omega - \mathbf{k}\cdot\boldsymbol{v})^2$ terms on the left-hand side of our dispersion relations, Eqs.~\eqref{fastdispersion} and \eqref{slowdispersion}, and moving all terms to one side, we can directly read off the components of the symmetric contravariant metric tensor $M^{\mu\nu}$. 

To match the standard analog gravity signature, we multiply the resulting equation by $-1$, ensuring the temporal component is $M^{00} = -1$. The temporal and cross-components remain identical to the ideal, inviscid case:
\begin{align}
    M^{00} &= -1, \\
    M^{0i} &= M^{i0} = -v^i.
\end{align}

For the spatial components $M^{ij}$, we must  symmetrize the viscous $k_i k_j$ cross-terms to ensure the tensor remains symmetric ($M^{ij} = M^{ji}$). By doing so, the total  metric naturally splits into the ideal MHD background plus our novel anisotropic viscous corrections:
\begin{equation}
    M^{\mu\nu} = M^{\mu\nu}_{\text{ideal}} + \mathcal{V}^{\mu\nu},
\end{equation}
where $M^{\mu\nu}_{\text{ideal}}$ matches the standard ideal metric, and $\mathcal{V}^{\mu\nu}$ is a purely spatial correction tensor ($\mathcal{V}^{00} = \mathcal{V}^{0i} = 0$).

\paragraph{{Fast mode contravariant metric:}}
The spatial block for the fast mode is:
\begin{equation}
    M^{ij}_{\text{fast}} = V_M^2 \delta^{ij} - (v^i v^j + \eta V_M^2 b^i b^j) + \mathcal{V}^{ij}_{\text{fast}},
\end{equation}
where $i, j \in \{r, \phi, z\}$. By symmetrizing the previously derived viscous terms, the correction tensor is explicitly:
\begin{align}
    \mathcal{V}^{ij}_{\text{fast}} &= \frac{A}{\rho} \left( \delta^i_r \delta^j_\phi + \delta^i_\phi \delta^j_r \right) \nonumber \\
    &\quad - \frac{A}{8\pi\rho^2 V_M^2} \left[ B^i (B_r \delta^j_\phi + B_\phi \delta^j_r) \right. \nonumber \\
    &\qquad \qquad \qquad \left. + B^j (B_r \delta^i_\phi + B_\phi \delta^i_r) \right].
\end{align}

\paragraph{The slow mode contravariant metric:}
Applying the exact same mapping to the slow mode, the spatial block is:
\begin{equation}
    M^{ij}_{\text{slow}} = -(v^i v^j - \eta V_M^2 b^i b^j) + \mathcal{V}^{ij}_{\text{slow}}.
\end{equation}
Here, the viscous correction tensor is simply the second half of the fast mode correction, with a flipped sign:
\begin{align}
    \mathcal{V}^{ij}_{\text{slow}} &= \frac{A}{8\pi\rho^2 V_M^2} \left[ B^i (B_r \delta^j_\phi + B_\phi \delta^j_r) \right. \nonumber \\
    &\qquad \qquad \qquad \left. + B^j (B_r \delta^i_\phi + B_\phi \delta^i_r) \right].
\end{align}

\textbf{ Explicit components  in cylindrical coordinates:}
The components of the fast mode metric $M^{\mu\nu}_{\text{fast}}$ are explicitly:
\begin{widetext}
\begin{equation}\label{metricfast}
M^{\mu\nu}_{\text{fast}} = 
\begin{pmatrix}
-1 & -v_r & -v_\phi & -v_z \\[1ex]
-v_r & V_M^2 - v_r^2 - \eta V_M^2 b_r^2 - \sigma V_A^2 b_r b_\phi & -v_r v_\phi - \eta V_M^2 b_r b_\phi + \sigma V_M^2 - \frac{\sigma V_A^2(b_r^2 + b_\phi^2)}{2} & -v_r v_z - \eta V_M^2 b_r b_z - \frac{\sigma V_A^2 b_z b_\phi}{2} \\[2ex]
-v_\phi & -v_\phi v_r - \eta V_M^2 b_\phi b_r + \sigma V_M^2 - \frac{\sigma V_A^2(b_r^2 + b_\phi^2)}{2} & V_M^2 - v_\phi^2 - \eta V_M^2 b_\phi^2 - \sigma V_A^2 b_r b_\phi & -v_\phi v_z - \eta V_M^2 b_\phi b_z - \frac{\sigma V_A^2 b_r b_z}{2} \\[2ex]
-v_z & -v_z v_r - \eta V_M^2 b_z b_r - \frac{\sigma V_A^2 b_z b_\phi}{2} & -v_z v_\phi - \eta V_M^2 b_z b_\phi - \frac{\sigma V_A^2 b_r b_z}{2} & V_M^2 - v_z^2 - \eta V_M^2 b_z^2
\end{pmatrix}.
\end{equation}

The components of the slow mode metric $M^{\mu\nu}_{\text{slow}}$ are:
\begin{equation}\label{metricslow}
M^{\mu\nu}_{\text{slow}} = 
\begin{pmatrix}
-1 & -v_r & -v_\phi & -v_z \\[1ex]
-v_r & - v_r^2 + \eta V_M^2 b_r^2 + \sigma V_A^2 b_r b_\phi & -v_r v_\phi + \eta V_M^2 b_r b_\phi + \frac{\sigma V_A^2(b_r^2 + b_\phi^2)}{2} & -v_r v_z + \eta V_M^2 b_r b_z + \frac{\sigma V_A^2 b_z b_\phi}{2} \\[2ex]
-v_\phi & -v_\phi v_r + \eta V_M^2 b_\phi b_r + \frac{\sigma V_A^2(b_r^2 + b_\phi^2)}{2} & - v_\phi^2 + \eta V_M^2 b_\phi^2 + \sigma V_A^2 b_r b_\phi & -v_\phi v_z + \eta V_M^2 b_\phi b_z + \frac{\sigma V_A^2 b_r b_z}{2} \\[2ex]
-v_z & -v_z v_r + \eta V_M^2 b_z b_r + \frac{\sigma V_A^2 b_z b_\phi}{2} & -v_z v_\phi + \eta V_M^2 b_z b_\phi + \frac{\sigma V_A^2 b_r b_z}{2} & - v_z^2 + \eta V_M^2 b_z^2
\end{pmatrix}.
\end{equation}
\end{widetext}

From Eqs.~\eqref{metricfast} and \eqref{metricslow}, it is straightforward to decompose $M^{\mu \nu}$ as
$M^{\mu\nu} = M_{inv}^{\mu\nu} + \mathcal{V}^{\mu\nu}$,
where $\mathcal{V}^{\mu\nu}$ is the purely spatial viscous correction tensor and $M_{inv}^{\mu\nu}$ is the inviscid part of $M^{\mu\nu}$ without the $\sigma$ terms. 

Since the term $\mathcal{V}^{\mu\nu}$ scales as $\mathcal{O}\left(\sigma\right)$, retaining terms up to linear order in $\sigma$, we obtain:
\begin{equation}
M_{\mu\nu} = (M_{inv}^{-1})_{\mu\nu}
- (M_{inv}^{-1})_{\mu\alpha}
\mathcal{V}^{\alpha\beta}
(M_{inv}^{-1})_{\beta\nu}.
\end{equation}

\paragraph{Fast mode covariant metric:}
We define $\mathbf{C}$  as:

\begin{equation}
    \mathbf{C} = B_\phi \hat{\mathbf{r}} + B_r \hat{\boldsymbol{\phi}}.
\end{equation}

Using these auxiliary variables, the $3 \times 3$ viscous perturbation tensor $\mathcal{V}_{ij}$ simplifies to:
\begin{equation}
    \mathcal{V}_{ij} = \frac{A}{\rho} (\delta_{ir} \delta_{j\phi} + \delta_{i\phi} \delta_{jr}) - \frac{A}{8 \pi \rho^2 V_M^2} (B_i C_j + B_j C_i),
\end{equation}
where \textbf{B}  is the magnetic field.
Thus the components of $\delta M_{\mu\nu} = - (M_{\text{inv}}^{-1})_{\mu\alpha} \mathcal{V}^{\alpha\beta} (M_{\text{inv}}^{-1})_{\beta\nu}$ up to linear order in $\sigma$ and $\eta$ can now be expressed as: 

\begin{equation}
    \delta M_{tt} = \frac{1}{V_M^4} \left[ - \frac{2A}{\rho} ({v}_r {v}_\phi) + \frac{A}{4 \pi \rho^2 V_M^2} (\mathbf{B} \cdot {\boldsymbol{v}})(\mathbf{C} \cdot {\boldsymbol{v}}) \right],
\end{equation}
\begin{equation}
\begin{split}
    \delta M_{ti} &= \frac{1}{V_M^4} \Bigg[ \frac{A}{\rho} (v_\phi \delta_{ir} + v_r \delta_{i\phi}) \\
    &\quad - \frac{A}{8 \pi \rho^2 V_M^2} \Big( (\mathbf{B} \cdot \boldsymbol{v}) C_i + (\mathbf{C} \cdot \boldsymbol{v}) B_i \Big) \Bigg],
\end{split}
\end{equation}

\begin{equation}
\begin{split}
    \delta M_{ij} &= -\frac{1}{V_M^4} \Bigg[ \frac{A}{\rho} ( \delta_{ir} \delta_{j\phi} + \delta_{i\phi} \delta_{jr} ) \\
    &\quad - \frac{A}{8 \pi \rho^2 V_M^2} \Big( B_i C_j + C_i B_j \Big) \Bigg].
\end{split}
\end{equation}
Thus,  we obtain $M_{\mu\nu} = (M_{\text{inv}}^{-1})_{\mu\nu} + \delta M_{\mu\nu}$. 

To present the effective geometry explicitly, we construct the magnetoacoustic line element $ds^2 \propto M_{\mu\nu} dx^\mu dx^\nu$. Following  analog gravity conventions similar to those used in \cite{2017PhRvD..95j4055N}, we multiply the contravariant metric tensor by a conformal factor of $V_M^2$.
Combining the ideal MHD background with our linear-order viscous corrections $\delta M_{\mu\nu}$, the complete three-dimensional  visco-magnetoacoustic line element for the fast mode is:

\begin{widetext}
\begin{equation}
\begin{aligned}\label{eq:92}
ds_{fast}^2 \propto &\left[ -(V_M^2 - v^2 - \eta(b \cdot v)^2) + \sigma \left( -2v_r v_\phi + \frac{(B \cdot v)(C \cdot v)}{4\pi\rho V_M^2} \right) \right] dt^2 \\
&- 2 \left[ v_i + \eta b_i (b \cdot v) - \sigma \left( v_\phi \delta_{ir} + v_r \delta_{i\phi} - \frac{(B \cdot v)C_i + (C \cdot v)B_i}{8\pi\rho V_M^2} \right) \right] dt dx^i \\
&+ \left[ \delta_{ij} + \eta b_i b_j - \sigma \left( \delta_{ir}\delta_{j\phi} + \delta_{i\phi}\delta_{jr} - \frac{B_i C_j + C_i B_j}{8\pi\rho V_M^2} \right) \right] dx^i dx^j,
\end{aligned}
\end{equation}
\end{widetext}
where the indices $i,j \in \{r, \phi, z\}$.

\paragraph{Slow mode covariant metric:}
The determinant of $M^{\mu\nu}_{\text{slow}}$
is of  $\mathcal{O}(\sigma^2 \eta)$. However, our perturbative framework is strictly valid only up to linear order in the independent small parameters $\sigma$ and $\eta$. To maintain mathematical consistency, any quantity scaling as  $\mathcal{O}(\sigma^2 \eta)$ or higher must be treated as identically zero within our linear approximation. Consequently, like the ideal case, the visco-magnetoacoustic slow mode remains degenerate and does not admit a valid inverse metric tensor.

\paragraph{Metric signature:}
To establish the Lorentzian signature of the effective fast-mode metric
$M^{\mu\nu}_{\rm fast}$, we partition it as
\begin{equation}
M^{\mu\nu}_{\rm fast}
=
\begin{pmatrix}
M_{11} & X\\
X^{T} & Y
\end{pmatrix},
\qquad M_{11}=-1.
\end{equation}
By the Haynsworth inertia additivity formula, the inertia
$\operatorname{In}(A)\equiv(\pi(A),\nu(A),\delta(A))$, where
$\pi$, $\nu$, and $\delta$ denote the numbers of positive, negative,
and zero eigenvalues of $A$, respectively, satisfying 
\begin{equation}
\operatorname{In}\!\left(M^{\mu\nu}_{\rm fast}\right)
=
\operatorname{In}(M_{11})
+
\operatorname{In}(S),
\end{equation}
where the Schur complement of $M_{11}$ is
\begin{equation}
S
=
Y-X^{T}M_{11}^{-1}X
=
V_M^2
\left(
I-\eta bb^{T}
+\frac{\mathcal{V}_{\rm fast}}{V_M^2}
\right).
\end{equation}
Since $M_{11}=-1$, we have
$\operatorname{In}(M_{11})=(0,1,0)$. Furthermore, within the perturbative
regime $\eta\ll1$ and $\|\mathcal{V}_{\rm fast}\|/V_M^2\ll1$, by using Weyl's inequality, we can easily see that  the Schur
complement $S$ is positive definite, implying
$\operatorname{In}(S)=(3,0,0)$. Hence,
\begin{equation}
\operatorname{In}\!\left(M^{\mu\nu}_{\rm fast}\right)
=
(3,1,0),
\end{equation}
establishing the Lorentzian signature $(-,+,+,+)$.

\subsubsection{Physical interpretation of the metric degeneracy}


As established in the preceding derivations, the fast magnetoacoustic mode yields an invertible effective metric tensor ($M_{fast}^{\mu\nu}$), whereas both the slow mode and the Alfv\'en mode yield degenerate tensors ($\det M^{\mu\nu} = 0$) that do not admit an inverse. This mathematical distinction reflects a  physical difference in the spatial dimensionality of their wave propagation. If we consider the group velocity, $\mathbf{v}_g = \nabla_{\mathbf{k}} \omega'$, then that determines  the physical trajectory of the wave's energy.

For the fast mode, the characteristic equation under the $\eta \ll1$ and $\sigma \ll 1 $ limit,  the comoving  dispersion relation is given by Eq.~\eqref{fastdispersion}. Thus, we obtain  the comoving fast mode group velocity to be:
\begin{equation}
\mathbf{v'}_{g,\text{fast}} = \pm \frac{V_M^2}{\omega'} \left[ \mathbf{k} +\text{additional terms} \right].
\end{equation}
Because the group velocity retains a  dependence on the  three-dimensional wave vector  $\mathbf{k}$, the fast mode propagates in all spatial directions rather than being confined to the magnetic field lines, producing a three-dimensional expanding wavefront. 

In contrast, in the limits of $\eta \ll 1$ and $\sigma \ll 1$, the slow-mode dispersion relation simplifies to Eq.~\eqref{slowdispersion},
from which the comoving group velocity follows as:
\begin{equation}
\begin{split}
\mathbf{v'}_{g,\text{slow}}
= \frac{1}{\omega'} \Bigg[
&\eta V_M^2 (\mathbf{k} \cdot \mathbf{b})\mathbf{b} \\
&+ \frac{A}{8\pi\rho^2 V_M^2}
\Big( (\mathbf{k}\cdot\mathbf{C})\mathbf{B}_0
+ (\mathbf{k}\cdot\mathbf{B}_0)\mathbf{C} \Big)
\Bigg].
\end{split}
\end{equation}
The above reveals a dimensional restriction.
 The  entire group velocity vector is  a linear combination of only two vectors: the background magnetic field $\mathbf{B}_0$ and the  vector $\mathbf{C}$. Geometrically, these two vectors span only a two-dimensional plane. Consequently, the viscous slow mode has  zero propagation in the  direction orthogonal to both (i.e., along $\mathbf{B}_0 \times \mathbf{C}$). Unlike the ideal inviscid limit where the wave is restricted to a one-dimensional trajectory, the viscous perturbation expands the propagation into a two-dimensional sheet, but it still  remains blind to the third spatial dimension.
 
 An even stricter one-dimensional restriction governs the  Alfvén mode. Using Eq.~\eqref{alfven}, we obtain $\mathbf{v}'_{g,\text{Alfvén}}=\pm V_A \mathbf{b}$.
Thus, the group velocity is parallel to the magnetic field unit vector $\mathbf{b}$, confining wave propagation to one-dimensional magnetic field lines.

The dimensionality of the group velocity and the rank of the comoving spatial block of the effective metric tensor are equivalent consequences of the dispersion relation. To see this, we define the comoving spatial block $\tilde{M}^{ij} = M^{ij} + v^i v^j$. Differentiating the comoving quadratic dispersion relation,
\begin{equation}
    (\omega')^2 = \tilde{M}^{ij}k_i k_j,
\end{equation}
with respect to the wave vector yields the intrinsic group velocity:
\begin{equation}\label{grpupmatrix}
    \mathbf{v}'_g=\frac{1}{\omega'}\tilde{M}\mathbf{k}.
\end{equation}

Equation~\eqref{grpupmatrix} shows that the comoving group velocity lies in the image of the spatial block $\tilde{M}^{ij}$. For the fast magnetoacoustic mode, $\tilde{M}^{ij}$ has full rank, allowing propagation in all three spatial directions. Applying the Schur complement identity for block matrices,
\begin{equation}
    \det(M^{\mu\nu})=-\det(\tilde{M}^{ij}),
\end{equation}
hence, $M^{\mu\nu}$ is invertible. This permits the construction of a nondegenerate four-dimensional spacetime metric.
In contrast, the  Alfvén and slow  modes do not propagate in all three spatial directions, hence
\begin{equation}
    \mathrm{rank}(\tilde{M}^{ij})<3.
\end{equation}
Therefore,
\begin{equation}
    \det(M^{\mu\nu})=0,
\end{equation}
and therefore these modes do not admit a non-degenerate four-dimensional pseudo-Riemannian visco-magnetoacoustic metric. Only the fast  mode admits an effective spacetime geometry.

\subsubsection{Visco-magnetoacoustic horizon}
The visco-magnetoacoustic horizon defines the boundary from which fast visco-magnetoacoustic waves can no longer propagate upstream against the inward background flow. In our covariant framework, this one-way surface is located where the contravariant radial metric component vanishes: 
\begin{align}
M^{rr}(r_H) = 0.
\end{align}
Substituting our derived component for the viscous fast mode, the horizon distance $r_H$ is found by solving the following root equation:
\begin{align}\label{eq:97}
V_M^2 - v_r^2 - \eta V_M^2 b_r^2 - \sigma V_A^2b_r b_\phi = 0.
\end{align}
Physically, this corresponds to the radius where the inward radial fluid velocity exactly matches the outward radial phase velocity of the fast visco-magnetoacoustic wave, trapping the wave at a constant radius. 

\subsubsection {Visco-magnetoacoustic ergosurface}
The ergosurface defines the stationary limit, inside of which waves are frame-dragged and forced to co-rotate with the background fluid. Geometrically, it is located where the norm of the timelike Killing vector $\xi_{(t)}^\mu = (1, 0, 0, 0)$ vanishes, which corresponds to setting the temporal metric component to zero:
\begin{equation}
M_{tt}(r_E) = 0.
\end{equation}
Substituting our derived component for the viscous fast mode, the ergosurface distance $r_E$ is found by solving the following root equation: 
\begin{equation}
V_M^2 - v^2 - \eta(b \cdot v)^2 - \sigma \left( -2v_r v_\phi + \frac{(B \cdot v)(C \cdot v)}{4\pi\rho V_M^2} \right) = 0.
\end{equation}

The annular region bounded between the horizon and the ergosurface ($r_H < r < r_E$) constitutes the visco-magnetoacoustic ergoregion. Any wave entering this region can scatter with amplified energy, providing the geometric mechanism for visco-magnetoacoustic superradiance.

\paragraph*{Background Flow for the Analog-Gravity Framework:}
\label{sec:background}
To construct our analog geometry, we require a background flow model. We consider, for the present purpose, advective accretion flows that accurately capture the required dynamics. We approach this through two complementary frameworks to demonstrate the robustness of our three-dimensional  visco-magnetoacoustic metric. First, we establish a foundational framework using a rigorous numerical magnetohydrodynamic advective accretion flow around a rotating black hole as our background flow \cite{2015ApJ...807...43M}. This inherently captures the transonic nature of the flow required to form a visco-magnetoacoustic horizon. Subsequently, to account for the physical reality of powerful winds that carry away mass and energy in accretion flows, we extend our analysis to the analytical Advection-Dominated Inflow-Outflow Solutions (ADIOS) model \cite{1999MNRAS.303L...1B}. However, as we will demonstrate, the strict self-similarity of standard ADIOS models mathematically enforces a constant Mach number, which pathologically prevents the dynamic formation of an analog horizon. Therefore, we present a {\it novel, modified transonic framework}---introducing a  magnetic field perturbation---to break this self-similarity and successfully recover the visco-magnetoacoustic horizon.
\section{Numerical framework: 1.5-dimensional  advective accretion flow  model around a black hole}\label{sec:Numericaladaf}
To establish a robust numerical baseline for the transonic advective flow, we utilize the numerical implementation of the 1.5-dimensional, vertically averaged hydromagnetic accretion-flow model presented in \cite{2015ApJ...807...43M}, which is a variant of MA-AAF. This framework models an optically thin, axisymmetric, and steady-state accretion flow in a pseudo-Newtonian potential \cite{2002ApJ...581..427M}. 
The fluid dynamics are governed by the vertically averaged magnetized Navier-Stokes equations incorporating large-scale magnetic stresses.

The integrated continuity equation governing mass conservation is given by:
\begin{equation}\label{dotM}
\dot{M} = 4\pi x \rho h \vartheta,
\end{equation}
where $\dot{M}$ is the constant mass accretion rate, $x$ is the radial coordinate, $\rho$ is the vertically averaged mass density, $h$ is the disk half-thickness, and $\vartheta$ is the radial inward velocity. All the length, time and velocity are expressed in the units of $GM/c^2$, $GM/c^3$ and light speed $c$ respectively, where $G$ is Newton's gravitation constant and $M$ is the mass of the black hole.

The radial and azimuthal momentum balance equations are respectively:
\begin{equation}\label{rad_num_ns}
\begin{split}
\vartheta\frac{d\vartheta}{dx} + \frac{1}{\rho}\frac{dP}{dx} - \frac{\lambda^2}{x^3} + F &= \\
\frac{1}{4\pi\rho}&\left(B_x\frac{dB_x}{dx} + s_1\frac{B_z B_x}{h} - \frac{B_\phi^2}{x}\right),
\end{split}
\end{equation}
\begin{equation}\label{azimuth_num_ns}
\begin{split}
\vartheta\frac{d\lambda}{dx} &= \frac{1}{x\rho}\frac{d}{dx}(x^2 W_{x\phi}) \\
&\quad + \frac{x}{4\pi\rho}\left(B_x\frac{dB_\phi}{dx} + s_2\frac{B_z B_\phi}{h} + \frac{B_x B_\phi}{x}\right),
\end{split}
\end{equation}
where $\vartheta$ is the radial velocity, $P$ is the total pressure, $\lambda$ is the specific angular momentum, $F$ represents the magnitude of pseudo-Newtonian gravitational force \cite{2002ApJ...581..427M}, and $W_{x\phi}$ is the viscous shear stress. The variables $B_x$, $B_\phi$, and $B_z$ are the radial, azimuthal and vertical components of the magnetic field. Because the variables are vertically averaged, the scaling parameters $s_1$, $s_2$, and $s_3$ are introduced to account for the vertical gradients of these respective magnetic field components, particularly  $d B_x/dz =  s_1 B_x/h $, 
 $dB_\phi/dz = s_2 B_\phi/h$, and $dB_z/dz = s_3 B_z/h $.

Assuming vertical magnetostatic balance, the equilibrium thickness of the disk satisfies:
\begin{equation}\label{vert_balnce}
\frac{P}{\rho h} = \frac{Fh}{x} - \frac{1}{4\pi\rho}\left[B_x\frac{dB_z}{dx} + s_3\frac{B_z^2}{h}\right].
\end{equation}

Crucially, the advective nature of the flow is supplemented by the vertically averaged energy equation, which balances the advected entropy with the net heating and cooling, given by:
\begin{equation}\label{Q+Q-}
\begin{split}
\vartheta T\frac{ds}{dx} &= \frac{\vartheta}{\Gamma_3 - 1}\left(\frac{dP}{dx} - \frac{\Gamma_1 P}{\rho}\frac{d\rho}{dx}\right) \\
&= Q^+ - Q^- \\
&= Q_{vis}^+ + Q_{mag}^+ - Q_{vis}^- - Q_{mag}^-,
\end{split}
\end{equation}
where $T$ is the temperature, $s$ is the entropy, and $\Gamma_1$ and $\Gamma_3$ are the polytropic indices of the flow. The net energy released and radiated out per unit volume per unit time is partitioned into viscous and magnetic contributions. Assuming the Shakura-Sunyaev viscosity prescription and defining fractions of radiated heat $f_{vis}$ and $f_m$, the viscous and magnetic heating/cooling terms are given by:
\begin{equation}
Q_{vis}^+ - Q_{vis}^- = \frac{\alpha f_{vis}(P + \vartheta^2\rho)\lambda}{x^2},
\end{equation}
\begin{equation}\label{Qmag}
Q_{mag}^+ - Q_{mag}^- = \frac{3f_m |B|^2 \vartheta}{16\pi x}.
\end{equation}

Finally,  assuming a very large magnetic Reynolds number, the magnetic field is frozen into the flow. Under steady-state and axisymmetric conditions, the induction equation and the divergence-free magnetic field condition  reduce to the following radial relations:
\begin{equation}
\frac{d}{dx}\left(\vartheta B_{\phi}-\frac{B_x\lambda}{x}\right)=0,
\end{equation}
\begin{equation}
\frac{d}{dx}\left(x\vartheta B_z\right)=0,
\end{equation}
\begin{equation}
\frac{d}{dx}\left(xB_x\right)=0.
\end{equation}

For our analog gravity framework, this system of coupled differential equations accurately captures the inherently transonic nature of advective accretion. The flow dynamically passes through a critical point where the numerator and denominator of the radial velocity gradient vanish, naturally forcing the flow to become supersonic and providing the requisite physical conditions for the formation of an analog magnetoacoustic horizon.

\subsection{Solution procedure}
To solve this system of coupled differential equations, we reduce the radial velocity gradient ($d\vartheta/dx$) into a single rational expression. Because the advective accretion flow around a black hole is inherently transonic, the flow must pass smoothly through a critical point. For a continuous physical solution, we apply the regularity condition at this critical point, requiring both the numerator and denominator of the radial velocity gradient to vanish simultaneously. This allows us to algebraically determine the exact flow variables and gradients at the critical radius. Following Mukhopadhyay and Chatterjee~\cite{2015ApJ...807...43M},
 the magnetic field strength is prescribed at the critical point through the dimensionless parameter $f_A$ by imposing

\begin{equation}\label{Bcrit}
B_{xc}=B_{\phi c}=B_{zc}
=
\sqrt{4\pi\rho_c}\,
\frac{c_{sc}}{f_A\sqrt{3}},
\end{equation}
where $\rho_c$ and $c_{sc}$ denote the density and sound speed at the critical point, respectively. We consider these critical point values as  one of the important boundary conditions and we then numerically integrate the equations to obtain the entire sub-Keplerian flow profile. This dynamical crossing of the critical point naturally establishes the supersonic inner region of the flow, providing the requisite physical background conditions for the formation of the analog magnetoacoustic horizon. For a detailed description of the parameter space constraints, we refer the reader to \cite{2015ApJ...807...43M}.
\subsection{Background flow profiles}
\begin{figure*}[t] 
    \centering
    
    \begin{subfigure}[b]{0.45\textwidth}
        \centering
        \includegraphics[width=\textwidth]{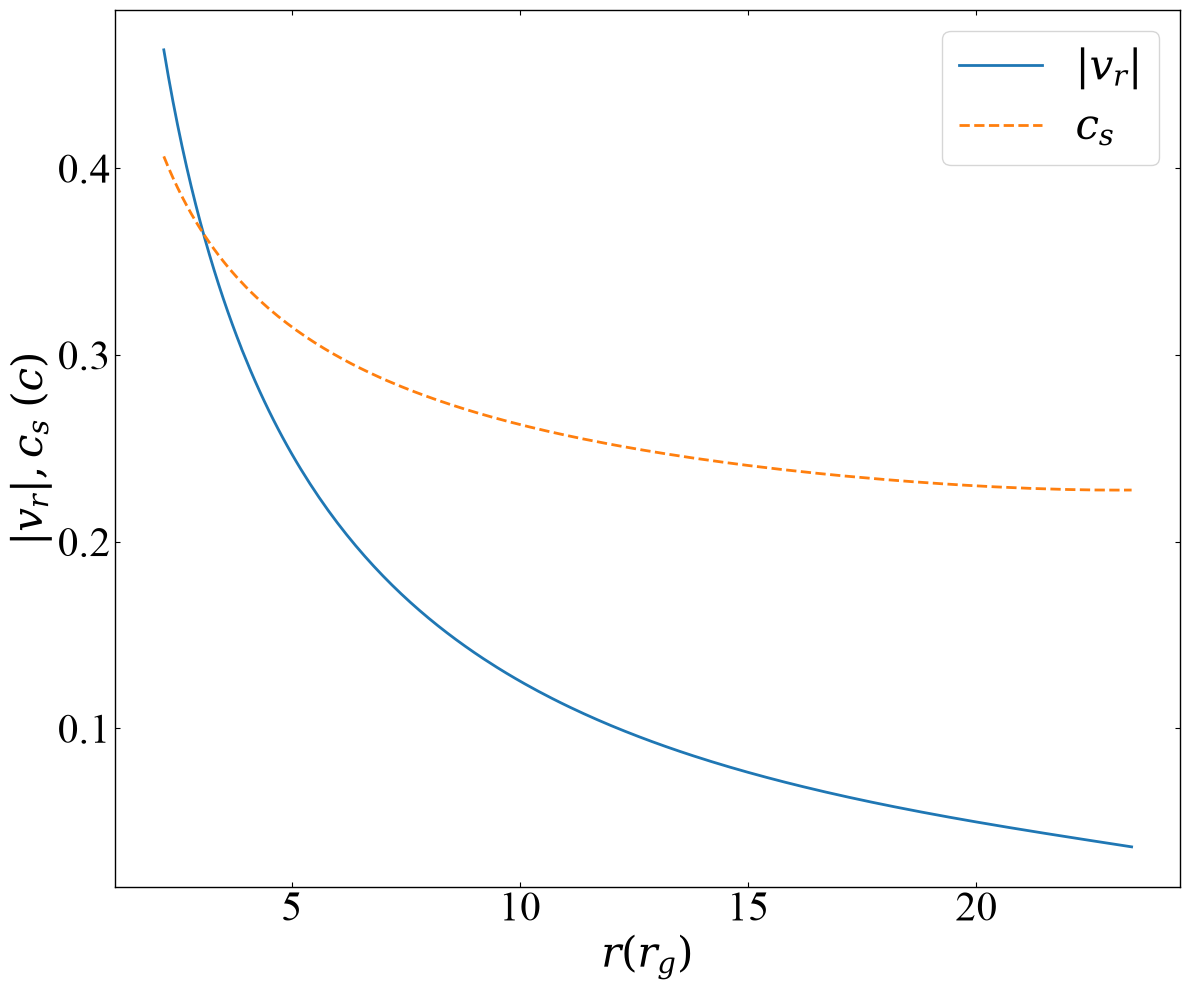}
        \caption{} 
        \label{fig:sub_v}
    \end{subfigure}
    \hfill
    \begin{subfigure}[b]{0.45\textwidth}
        \centering
        \includegraphics[width=\textwidth]{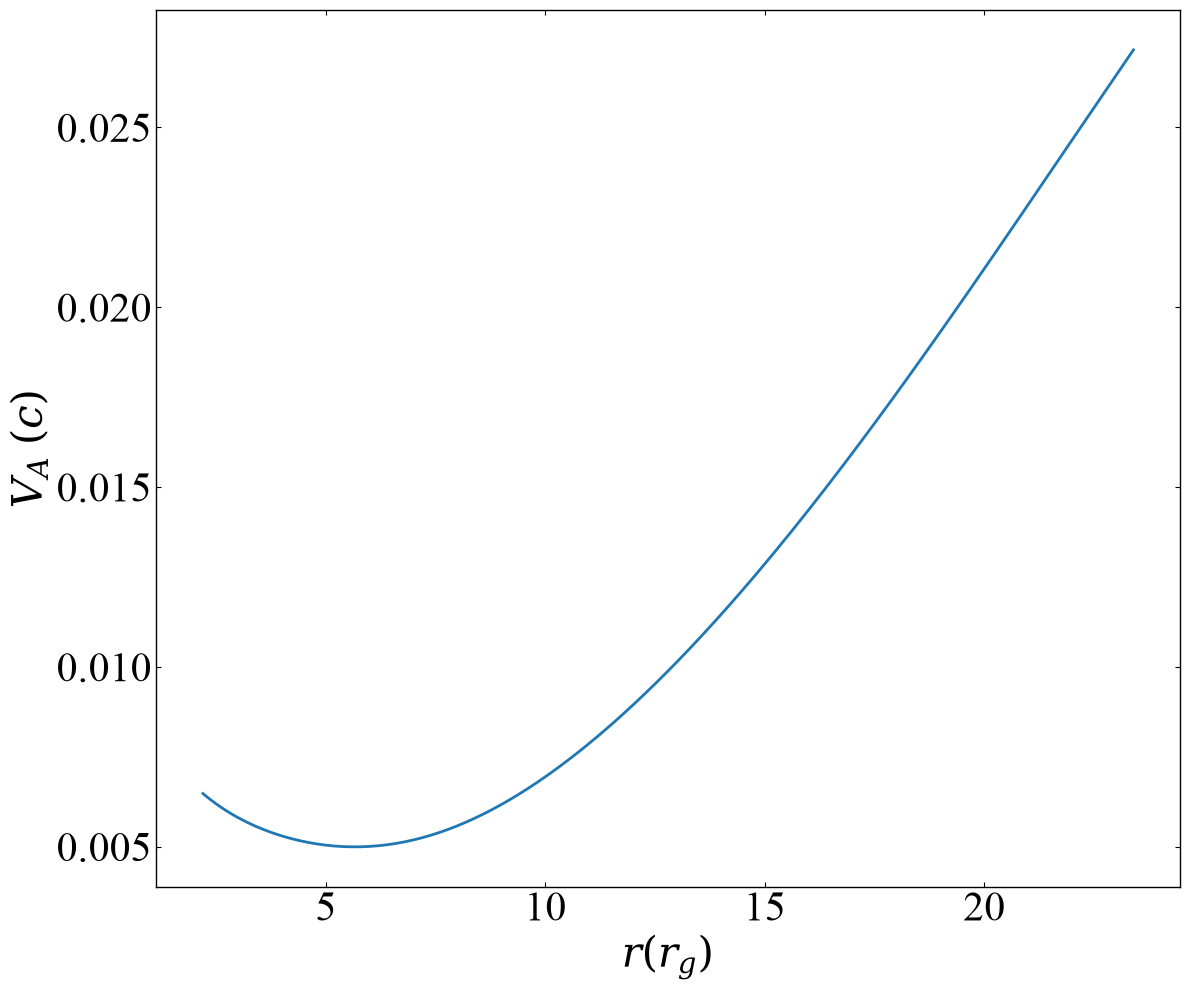}
        \caption{} 
        \label{fig:sub_vanum}
    \end{subfigure}
    
    \vspace{0.4cm} 
    
    \begin{subfigure}[b]{0.45\textwidth}
        \centering
        \includegraphics[width=\textwidth]{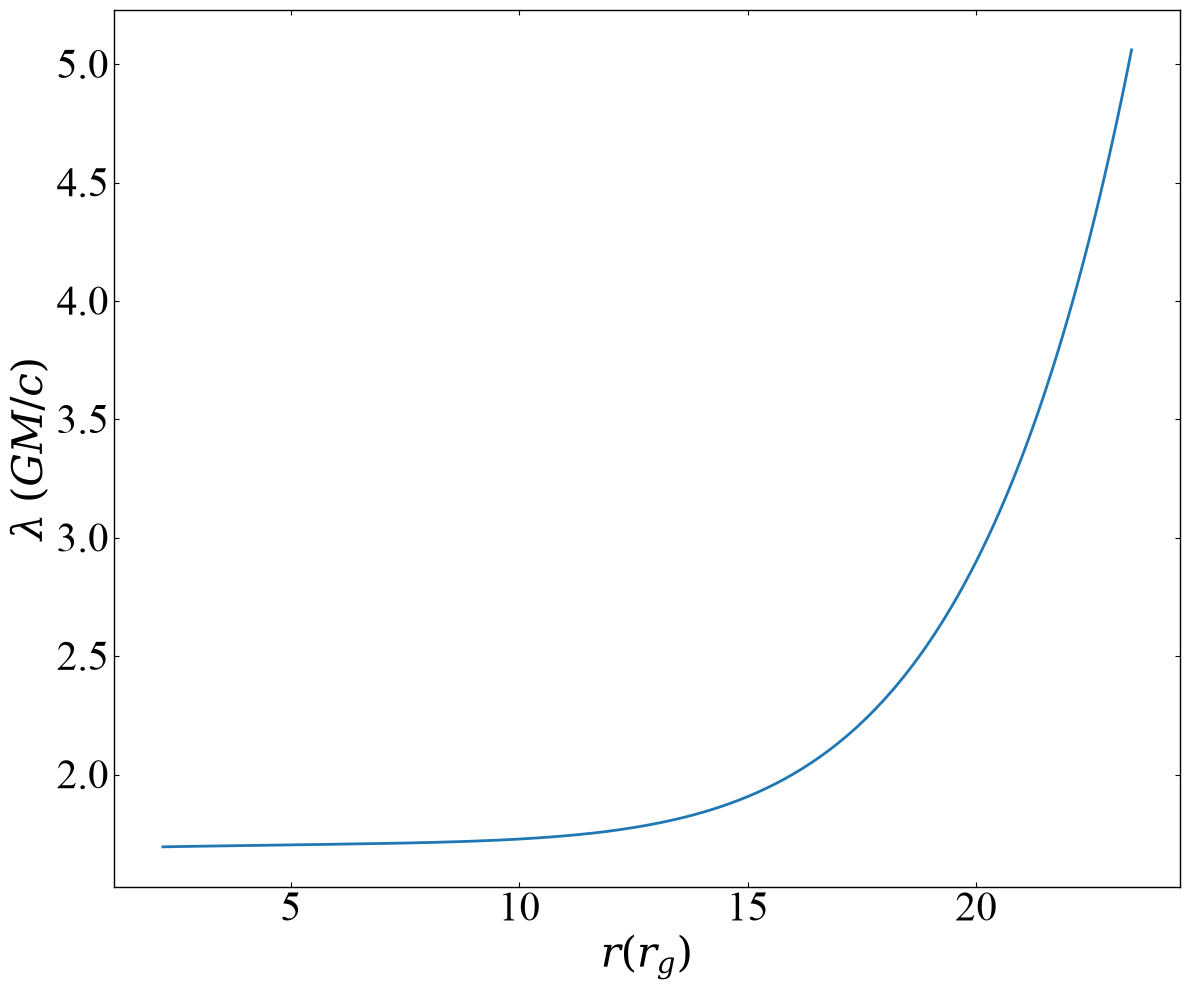}
        \caption{} 
        \label{fig:sub_lambda}
    \end{subfigure}
    \hfill
    \begin{subfigure}[b]{0.45\textwidth}
        \centering
        \includegraphics[width=\textwidth]{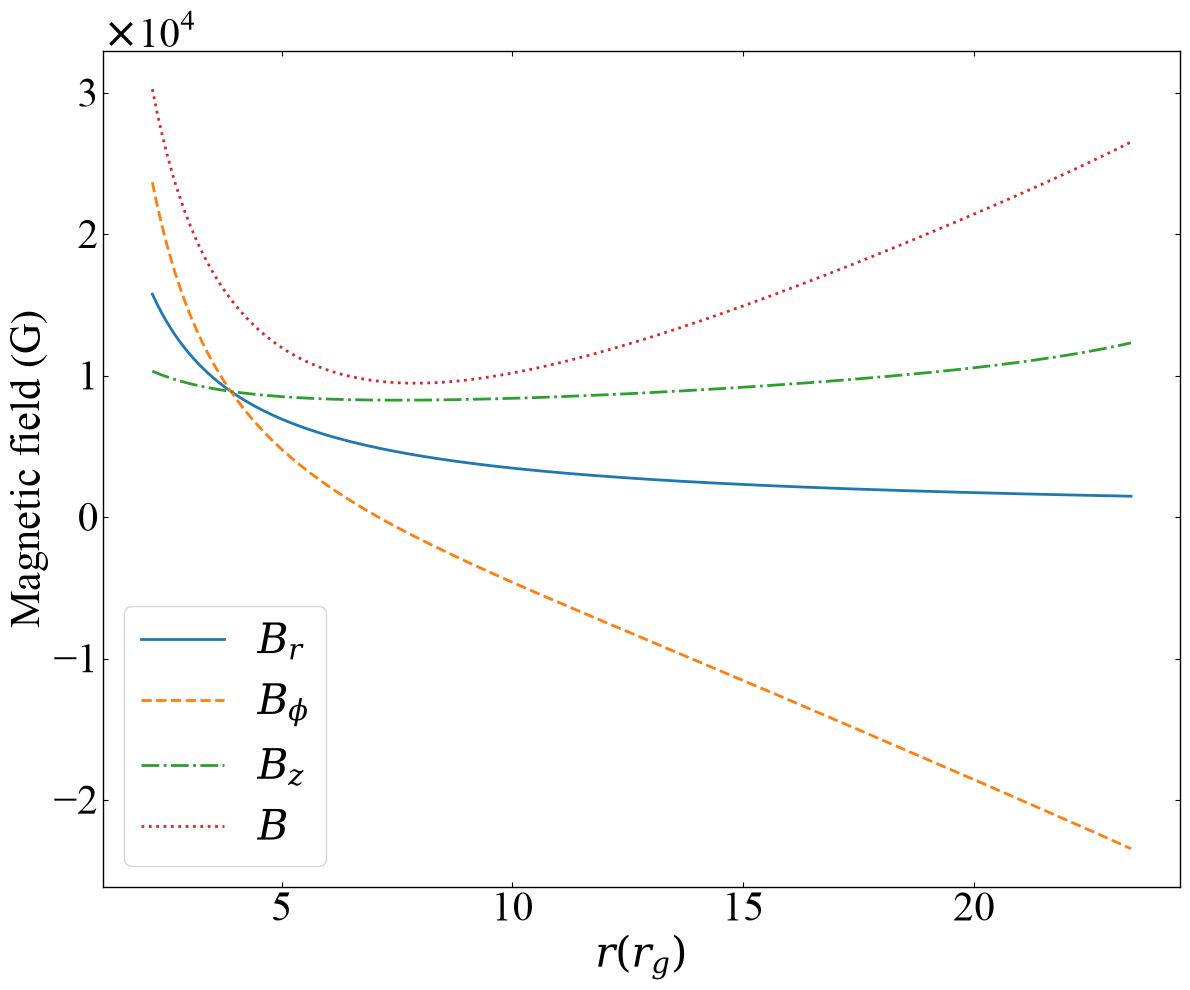}
        \caption{} 
        \label{fig:sub_Bnum}
    \end{subfigure}
    
    \caption{\raggedright Background flow profiles of (a) the radial inward velocity ($|v_r|$) and sound speed ($c_s$), (b) Alfvén speed ($V_A$), (c) specific angular momentum ($\lambda$), and (d) the net magnetic field ($|B|$) alongside its three components ($B_r$, $B_{\phi}$, $B_z$), for an accretion flow with $a = 0.998$, $\alpha = 0.01$,  $M = 10M_\odot$ and $\dot{M}=0.01\dot{M}_{\rm Edd}$.}
    \label{fig:numbackground_flow}
\end{figure*}

Figure~\ref{fig:numbackground_flow} presents the kinematic and magnetic profiles of the background flow. In panel (a), the flow smoothly
crosses the point at which the inward radial velocity, $|v_r|$,
exceeds the sound speed, $c_s$, indicating a transonic transition.
The radial profiles of the Alfvén speed, $v_A$, and the specific
angular momentum, $\lambda$, are given in panels (b) and (c),
respectively. The net magnetic field, $|B|$, and its individual
components, $B_r$, $B_\phi$, and $B_z$, are plotted as functions of
radial distance in panel (d). Importantly, comparing panels (a) and (b) reveals that $(c_s/ v_A)^2\ll1$ across the domain, ensuring the flow is gas pressure-dominated ($\eta \ll 1$). Furthermore, the chosen small viscosity parameter ($\alpha = 0.01$) ensures that $\sigma \ll 1$ is  satisfied.


\section{Analytical framework: generalization to the ADIOS model}\label{ADIOS}


While standard ADAF models conservatively assume a constant mass accretion rate, the ADIOS 
model physically accounts for a powerful wind that carries away mass, energy, and angular momentum \cite{1999MNRAS.303L...1B}. In this 1.5-dimensional self-similar framework, the mass accretion rate explicitly scales with radius as $\dot{M}(r) \propto r^p$, where $p$ ($0 \le p < 1$) is the mass-loss exponent. The physical kinematic quantities scale with the Keplerian velocity $v_K(r)$, giving the unperturbed profiles:
\begin{eqnarray}
    v_r^{(0)}(r) &=& -c_1 \alpha v_K(r), \label{eq:vr_adios} \\
    \Omega^{(0)}(r) &=& c_2 \Omega_K(r), \label{eq:omega_adios} \\
    c_s^{2(0)}(r) &=& \Gamma c_3 v_K^2(r), \label{eq:cs_adios}
\end{eqnarray}
where $c_1$, $c_2$, and $c_3$ are dimensionless constants that depend
on $p$, the specific heat ratio, and the advection fraction. Here $\Gamma$ denotes the polytropic index, defined through the polytropic equation of state $P\rho^{-\Gamma} = constant$. Because of
the outflow, the background density profile becomes shallower, scaling
as $\rho^{(0)}(r)\propto r^{-3/2+p}$. The corresponding pressure profile
scales as $P^{(0)}(r)=\rho^{(0)}c_s^{2(0)}/{\Gamma}\propto r^{p-5/2}$.

All variables (\(r,\,v,\,\rho,\,P,\,B,\) etc.) in the subsequent sections are strictly
dimensionless, having been scaled by their respective fundamental units.
\subsection{ ADIOS flow profile }
We follow the ADIOS formulation of \cite{1999MNRAS.303L...1B}, but relax the assumptions $v \ll r \Omega$ and $\gamma =5/3$ in deriving the flow profile without magnetic field, where $\gamma$ is the ratio of specific heats.
\subsubsection{Governing equations for the ADIOS}
\paragraph{ Continuity Equation:}
Assuming a steady-state ($\partial/\partial t = 0$) and axisymmetry ($\partial/\partial \phi = 0$) flow, the three-dimensional  mass conservation equation is:
\begin{equation}
 \frac{1}{r} \frac{\partial}{\partial r} (r \rho v_r) + \frac{\partial}{\partial z} (\rho v_z) = 0. 
\end{equation}
\paragraph{ Radial momentum balance:}
Using the radial Navier-Stokes equation and assuming that the radial
velocity is independent of height, we obtain:
\begin{equation}\label{vreqnns}
 \rho \left( v_r \frac{\partial v_r}{\partial r}  - \frac{v_\phi^2}{r} \right) = -\rho \frac{\partial \Phi_g}{\partial r} - \frac{\partial P}{\partial r}, 
 \end{equation}
where $\Phi_g = -GM/\sqrt{r^2 + z^2}$ denotes the gravitational potential.
\paragraph{ Angular Momentum Conservation:}
The transport of angular momentum is governed by the azimuthal Navier-Stokes equation given by
\begin{equation}
 \frac{1}{r} \frac{\partial}{\partial r} (\rho r^2 v_r v_\phi) + \frac{\partial}{\partial z} (\rho r v_z v_\phi) = \frac{1}{r} \frac{\partial}{\partial r} (r^2 W_{r\phi}) + \frac{\partial}{\partial z} (r W_{z\phi}). 
 \end{equation}
\paragraph{ Energy Conservation:}
From the general energy conservation law for a viscous fluid, we have
\begin{equation}
 \nabla \cdot \left[ \rho \boldsymbol{v} \left( \frac{1}{2}v^2 + \Phi_g + h \right) - \mathbf{W} \cdot \boldsymbol{v} \right] = 0, 
 \end{equation}
where $u$ is the specific internal energy, and $h = u + P/\rho$ is the specific enthalpy.



\subsubsection{Vertical integration and wind parameterization}
To reduce the three-dimensional  fluid equations to a one-dimensional macroscopic disk model, we integrate the equations vertically across the disk thickness, $z \in [-H, H]$ just as done in \cite{narayan1994}. We define the surface density as $\Sigma = \int_{-H}^{H} \rho \, dz =2\rho_{av} H$.


Crucially, in the ADIOS framework, the vertical velocity $v_z$ does not vanish at the disk boundaries. The integration of the $\partial/\partial z$ terms gives non-zero boundary evaluations, which mathematically represent the fluxes of mass, angular momentum, and energy siphoned away by the wind.

\paragraph{ Mass accretion and mass loss constant $p$:}
Integrating the continuity equation gives us
\begin{equation}\label{cont_vert}
\frac{\partial}{\partial r} (2\pi r \Sigma v_r) + 4\pi r (\rho v_z)_{H} = 0. 
\end{equation}
Defining the inward mass accretion rate as $\dot{M} = -2\pi r \Sigma v_r$, the boundary term represents the mass loss to the wind: $d\dot{M}/dr= 4\pi r (\rho v_z)_H$. Following Blandford and Begelman~\cite{1999MNRAS.303L...1B}, we adopt a power-law radial dependence of the mass accretion rate to parameterize mass loss through winds:
$$ \dot{M} \propto r^p \quad (0 \le p < 1). $$

\paragraph{Angular momentum flux and parameter $\Lambda$:}
Integrating the azimuthal equation and using the definitions of  the net inward flux of angular momentum to be  $F_l = \dot{M}r^2\Omega - G_{visc}$, where $G_{visc} = -\int_{-H}^H 2\pi r^2 W_{r\phi} \, dz$, is the viscous torque, we obtain
\begin{equation}\label{azcb}
 \frac{dF_l}{dr} = 4\pi r \left( \rho r^2 \Omega v_z - r W_{z\phi} \right)_H. 
 \end{equation}
The right-hand side is the angular momentum extracted by the wind (both advected by the gas and extracted by surface torques). Following Blandford and Begelman~\cite{1999MNRAS.303L...1B}, we parameterize this angular momentum loss using a dimensionless constant $\Lambda$:
\begin{equation}\label{FlGvisc}
F_l =\dot{M}r^2\Omega - G_{visc}= \Lambda \dot{M}v_Kr.
\end{equation}

\paragraph{ Energy flux and parameter $\epsilon_{\rm w}$:}
Similarly, integrating the energy equation under the assumptions of a
steady state, no radiative cooling, and the absence of extraneous energy
sources or sinks \cite{1999MNRAS.303L...1B}, and defining the net outward
energy flux as $F_E = G_{\rm visc}\Omega - \dot{M}Be$, where $Be$ is the
Bernoulli parameter, $Be = (1/2)v^2 + \Phi_g + h$, we adopt the
energy-flux parameterization introduced by Blandford and Begelman
\cite{1999MNRAS.303L...1B},
\begin{equation}\label{FeGvis}
   F_E = G_{visc}\Omega - \dot{M}Be = \epsilon_{\rm w} \dot{M} v_K^2,
\end{equation}
where $\epsilon_{\rm w}$ is the dimensionless energy-loss parameter.

By transforming the three-dimensional differential boundaries into these three self-similar algebraic closures ($p, \Lambda, \epsilon_{\rm w}$), we can now construct the exact, purely algebraic system for the unperturbed ADIOS flow.

\subsubsection{Reduction to an algebraic system:}
The vertically integrated equations admit the self-similar solution
introduced in Eqs.~\eqref{eq:vr_adios}, \eqref{eq:omega_adios}, and \eqref{eq:cs_adios}, together with
$\rho^{(0)}\propto r^{p-3/2}$ and $P^{(0)}\propto r^{p-5/2}$.
Substituting these scalings into the integrated conservation equations causes the common radial dependence to factor out, reducing the differential system to a set of algebraic equations for $c_1$, $c_2$, and $c_3$, which are given below.
\paragraph{ Radial momentum:}
Substituting the self-similar scalings into the radial Navier-Stokes equation, Eq.~\eqref{vreqnns}, gives us
\begin{equation} \label{eq:adios_c3}
c_3 = \frac{1 - c_2^2 - \frac{1}{2}c_1^2 \alpha^2}{5/2 - p}.
\end{equation}

\paragraph{ Angular momentum:}
Similarly, using the integrated angular-momentum equation, Eq.~\eqref{FlGvisc}, the definition of $\dot{M}$, and the
Shakura-Sunyaev prescription $W_{r\phi}=-\alpha P$
\cite{1973A&A....24..337S}, we obtain
\begin{equation}
    c_1=\frac{c_3}{c_2-\Lambda}.
    \label{eq:adios_c1}
\end{equation}
\paragraph{Energy conservation:}
Finally, substituting definition of $G_{visc}$
  and Eq.~\eqref{eq:adios_c3} into the integrated energy flux equation, Eq.~\eqref{FeGvis}, yields
\begin{equation} \label{eq:adios_eps}
\epsilon_{\rm w} = \frac{c_2 c_3}{c_1} - c_3 \left[ \frac{\gamma}{\gamma-1} - \frac{5}{2} + p \right] + \frac{1}{2}c_2^2.
\end{equation}

Eqs.~\eqref{eq:adios_c3}, \eqref{eq:adios_c1}, and ~\eqref{eq:adios_eps} form an exact, closed algebraic system. For a chosen set of wind boundary parameters ($p, \Lambda, \epsilon_{\rm w}$) and fluid properties ($\alpha, \gamma$), this system uniquely determines $c_1, c_2~{\rm and}~c_3$, thereby completely defining the unperturbed, steady-state, hydrodynamic ADIOS background upon which we will subsequently introduce magnetic perturbations.

Let us define two thermodynamic constants for mathematical brevity:
\begin{equation}
K = \frac{\gamma}{\gamma-1} - \frac{5}{2} + p, \quad M = \frac{5}{2} - p.
\end{equation}

Using Eqs.~\eqref{eq:adios_c3}, \eqref{eq:adios_c1}  and \eqref{eq:adios_eps}, we obtain

\begin{multline}
\frac{M}{K} \left( \frac{3}{2}c_2^2 - \Lambda c_2 - \epsilon_{\rm w} \right) = 1 - c_2^2 \\
-\frac{\alpha^2}{2} \Bigg[ \frac{ \frac{3}{2}c_2^2 - \Lambda c_2 - \epsilon_{\rm w} }{K(c_2-\Lambda)}  \Bigg]^2. 
\end{multline}
The above  algebraic equation can be solved numerically for $c_2$,
from which $c_1$ and $c_3$ follow directly using
Eqs.~\eqref{eq:adios_c1} and \eqref{eq:adios_c3}.

\subsubsection{Two special cases of our solution}
Under two special cases we can obtain the closed form analytical solution of the above equation.

\paragraph{ Narayan-Yi solution \cite{narayan1994}:}
If we put $p = 0$ (constant $\dot M$), $\Lambda =0$ (no wind loss) and $\epsilon_{\rm w} =0 $ (no mass loss) in Eqs.~\eqref{eq:adios_c3}, \eqref{eq:adios_c1}, and \eqref{eq:adios_eps}, then one can  recover the exact Narayan-Yi profile (for $f=1$ case) in a straightforward manner. 


\paragraph{ Blandford-Begelman solution \cite{1999MNRAS.303L...1B}:}
If we put $v\ll \Omega r$  which translates to  $ c_1^2 \alpha^2/2 \ll c_2^2 $ in Eq.~\eqref{eq:adios_c3}
and 
$\gamma =5/3$ in Eqs.~\eqref{eq:adios_c3}, \eqref{eq:adios_c1}, and \eqref{eq:adios_eps} as assumed in \cite{1999MNRAS.303L...1B}, then we recover the Blandford-Begelman  solution.

\subsubsection{Limitations of the Self-Similar Background:}
The background profile derived above is strictly self-similar, which introduces the following physical limitations:
\begin{itemize}
    \item \textbf{Constant Mach Number:} Accretion flows are inherently \textit{transonic}, requiring the radial Mach number, $\mathcal{M}\equiv |v_r|/c_s$, to vary along the flow and cross unity at a physical sonic point. However, self-similar scaling gives us:
    \begin{equation}
        \mathcal{M} = \frac{|-c_1 \alpha v_K(r)|}{\sqrt{\Gamma c_3} v_K(r)} = \frac{c_1 \alpha}{\sqrt{\Gamma c_3}} = \text{constant}.
    \end{equation}
    
    \item \textbf{Analog Horizon Pathology:} A globally constant-Mach number self-similar solution cannot capture the transonic topology of accretion, which requires a transition from subsonic to supersonic flow. This also prevents the horizon condition $M^{rr}=0$, Eq.~\eqref{eq:97}, from selecting a distinct radial location. Since all velocity-squared and Alfv\'en-speed-squared terms entering the condition share the same radial scaling, the radial dependence factors out, reducing the horizon condition to an algebraic relation among constants.
\end{itemize}
 


To break the exact self-similarity of the background flow and generate a dynamic Mach number, we introduce a  magnetic field $\mathbf{B} = (B_R, B_\phi, B_z)$ as a perturbation to the ADIOS profile without magnetic field, as derived below.

\subsection{Perturbation of the ADIOS due to  magnetic field}
We take the zeroth-order background state to be the unperturbed hydrodynamic ADIOS flow, whose profile is derived in the preceding subsection.
On adding a magnetic field of the form 
$$\mathbf{B} = \epsilon^{1/2}\mathbf{B}^{(0)} + \mathcal{O}(\epsilon^{3/2}),$$
where $\epsilon$ is a small dimensionless quantity.
The flow variables change as
\begin{align}
\boldsymbol{v} &= \boldsymbol{v}^{(0)} + \epsilon\boldsymbol{v}^{(1)} + \mathcal{O}(\epsilon^2),\\
\rho &= \rho^{(0)} + \epsilon\rho^{(1)} + \mathcal{O}(\epsilon^2),\\
P &= P^{(0)} + \epsilon P^{(1)} +\mathcal{O}(\epsilon^2),
\end{align}
and similarly for all the other flow variables.

Here
$\boldsymbol{v}^{(0)},\rho^{(0)}, P^{(0)}$, etc.,  are the ADIOS flow variables without magnetic field. 
$\mathbf{B}^{(0)}$ is a given  magnetic field (not assumed self-similar). The scaling of $\mathbf{B}$ with $\epsilon^{1/2}$ ensures magnetic pressure/forces enter the momentum equation at $\mathcal{O}(\epsilon)$.

We work in the steady limit for the hydrodynamic background and compute the \emph{steady} $\mathcal{O}(\epsilon)$ corrections to the flow variables by the Lorentz force $\mathbf{F}_{\rm L}^{(1)}=(1/4\pi)(\nabla\times\mathbf{B}^{(0)})\times\mathbf{B}^{(0)}$.

We define the vertically integrated surface density $\Sigma = \int_{-H}^{H} \rho dz = 2\rho_{av}H$, the  integrated pressure $P_{I} = \int_{-H}^{H} P dz = 2P_{av}H$, and the integrated Lorentz forces $(F_L)_i^{I} = \int_{-H}^{H} (F_L)_i dz = 2H(F_L)_i$. We introduce the first-order expansions: $\Sigma = \Sigma^{(0)} + \Sigma^{(1)}$, and $P_{I} = P_{I}^{(0)} + P_{I}^{(1)}$.

 We will  drop ``av" subscript from each variable to improve readability.

\subsubsection{Magnetic field topology: } 
Our objective is to construct a field that provides radial and azimuthal Lorentz forces ($(F_L)_r \neq 0, (F_L)_\phi \neq 0$) while strictly maintaining vertical magnetostatic balance ($(F_L)_z = 0$).

Under the assumption of axisymmetry, the condition $(F_L)_z = 0$ translates to
\begin{equation}
-\frac{\partial B_\phi}{\partial z}B_\phi
-\left(\frac{\partial B_r}{\partial z}
-\frac{\partial B_z}{\partial r}\right)B_r = 0.
\end{equation}
This condition is satisfied if we choose $B_\phi$ and $B_r$ to be independent of $z$ and adopt a constant vertical magnetic field, $B_z = W$. Thus, we obtain
\begin{equation}
B_r = l(r), \qquad
B_\phi = f(r), \qquad
B_z = W.
\end{equation}
Now, enforcing the standard divergence-free condition, $\nabla\cdot\mathbf{B}=  0$, 
gives $(1/r)\partial_r(r B_r)=0$, which 
immediately restricts the radial field to:
\begin{equation}
    B_r(r) = \frac{L}{r},
\end{equation}
where $L$ is a constant. 
Next, we require the magnetic field to satisfy the steady-state ideal induction equation to first-order,
\begin{equation}
    \nabla\times(\boldsymbol{v}^{(0)}\times\mathbf{B}) = 0,
\end{equation}
which gives
\begin{align*}
    (\nabla\times(\boldsymbol{v}^{(0)}\times\mathbf{B}))_\phi
    &= -\frac{d}{dr}\bigl(v_r^{(0)}f(r)-v_\phi^{(0)}l(r)\bigr)=0,\\
    (\nabla\times(\boldsymbol{v}^{(0)}\times\mathbf{B}))_z
    &= -\frac{1}{r}\frac{d}{dr}\bigl(r v_r^{(0)}(r)W\bigr)=0.
\end{align*}

Solving these equations yields the complete magnetic-field configuration up to $O(\epsilon^{1/2})$:
\begin{equation}\label{eq:132}
    B_r(r) = \epsilon^{1/2}\frac{L}{r}, \quad B_\phi(r) = \epsilon^{1/2}(F r^{1/2} + H r^{-1}), \quad B_z = 0.
\end{equation}
where  $H = -c_2 L/c_1 \alpha$ and $F$ is another constant.

The corresponding Lorentz force is therefore:
\begin{align}
    (F_L)_r(r) &= -\frac{3}{8\pi}\,F^2 - \frac{3}{8\pi}\,F H\,r^{-3/2}, \label{eq:FLR_explicit} \\[6pt]
    (F_L)_\phi(r) 
    &= \frac{3}{8\pi}\,F L\,r^{-3/2},\label{eq:FLphi_explicit}\\
    (F_L)_z(r) & = 0\label{eq:FLz_explicit}.
\end{align}
Consequently, the radial forcing contains both a constant effective pressure term (arising from $F^2$) and a decaying power-law piece ($\propto r^{-3/2}$), while the azimuthal Lorentz force follows a decaying power-law.

We assume the flow to follow a polytropic equation ($P \rho^{-\Gamma} =constant=K$).
The magnetic field is treated as a weak, externally imposed perturbation of the pre-existing ADIOS solution. As shown in Eq.~\eqref{eq:FLz_explicit}, this magnetic perturbation produces no Lorentz force in the vertical direction. Hence, we take the wind mass-loss rate through the disk surface ($\dot{\Sigma}_w=4\pi r(\rho v_z)_H$)  to remain unchanged by the addition of the magnetic field.
We assume that the external mass reservoir supplying mass to the outer edge of the accretion disk ($r=r_{out}$) remains completely unaffected by the small magnetic field introduced in the disk and thus,  the   perturbation in  velocity and surface density ($\Sigma$) due to the addition of magnetic field should  vanish at the outer boundary. Taking motivation from \cite{Blandford:1982xxl}, we adopt the  ansatz that the perturbation in the angular momentum loss parameter ($\Lambda^{(1)}$), induced by  the perturbative magnetic field scales as $\Lambda^{(1)} \propto R^{-2}$.

\subsubsection{Perturbations to the  vertically integrated equations}
\paragraph{Continuity equation: }
As explained and derived earlier in this paper,
the vertically integrated mass conservation equation, Eq.~ \eqref{cont_vert}, incorporates a wind mass-loss sink term $\dot{\Sigma}_w$:
\begin{equation}
    \frac{1}{r}\frac{d}{dr}(r \Sigma v_r) = \dot{\Sigma}_w.
\end{equation}

Since  $\dot{\Sigma}_w^{(1)}=0$, we obtain:
\begin{equation}
    \frac{1}{r}\frac{d}{dr} \left[ r (\Sigma^{(1)}v_r^{(0)} + \Sigma^{(0)}v_r^{(1)}) \right] = 0.
\end{equation}
Integrating this equation and applying the outer-boundary condition gives:
\begin{equation}
    \Sigma^{(1)}v_r^{(0)} + \Sigma^{(0)}v_r^{(1)} = 0 \implies \frac{\Sigma^{(1)}(r)}{\Sigma^{(0)}(r)} = -\frac{v_r^{(1)}(r)}{v_r^{(0)}(r)}.
    \label{eq:sigma_continuity}
\end{equation}

\paragraph{ Perturbations in height of the disk and volume density:}
Height of the disk is  given by \cite{narayan1994}:
\begin{equation}
H \approx \frac{\sqrt{P/\rho}}{\Omega_K}.
\end{equation}

Consequently, the fractional perturbation of the disk vertical scale height
is directly coupled to the fractional perturbation of the local sound speed:
\begin{equation}\label{H_perturb1}
\frac{H^{(1)}}{H^{(0)}} =
\frac{c_s^{(1)}}{c_s^{(0)}} .
\end{equation}

Using the assumption of polytropic equation of state, we obtain:

\begin{equation}\label{H_final}
\frac{H^{(1)}}{H^{(0)}}
=
\frac{\Gamma-1}{2}
\frac{\rho^{(1)}}{\rho^{(0)}} .
\end{equation}

Applying a first-order perturbative expansion to $\Sigma =2\rho H $  and using Eqs.~\eqref{eq:sigma_continuity} and \eqref{H_final}, we obtain:

\begin{equation}\label{sigma-v}
\frac{\rho^{(1)}}{\rho^{(0)}}
=
-\frac{2}{\Gamma +1}
\frac{v_r^{(1)}}{v_r^{(0)}} .
\end{equation}

\paragraph{ Perturbation in $P_{I}$:}

As explained before, $P_{I} =  2PH$. Using the  sound speed definition $P = \rho c_s^2/ \Gamma$ and the polytropic relation, we obtain:
\begin{align}\label{Pper}
    \frac{P^{(1)}}{P^{(0)}} &=  \Gamma  \frac{\rho^{(1)}}{\rho^{(0)}}=  \frac{2\Gamma}{\Gamma+1} \frac{ \Sigma^{(1)}}{\Sigma^{(0)}}.
\end{align}

\begin{equation}
    P_{I}^{(1)} = P_{I}^{(0)} \left( \frac{3\Gamma - 1}{\Gamma+1} \right) \frac{\Sigma^{(1)}}{\Sigma^{(0)}}=-\tilde{\Gamma}\,
c_3v_K^{2} \,
\Sigma^{(0)}
\frac{v_r^{(1)}}{v_r^{(0)}}.\label{P1exp}
\end{equation}
Where we defined  $\tilde{\Gamma} \equiv (3\Gamma - 1)/({\Gamma+1})$.The gradient of integrated pressure is evaluated as:
\begin{equation}\label{derivP1}
   -\frac{dP_{I}^{(1)}}{dr} = \tilde{\Gamma} \frac{\Sigma^{(0)} c_3v_K^{2}}{v_r^{(0)}} \left[ \frac{p-1}{r} v_r^{(1)} + \frac{d v_r^{(1)}}{dr} \right].
\end{equation}

\paragraph{Radial Navier-Stokes equation:}
The vertically integrated radial Navier-Stokes equation is:
\begin{equation}\label{prns}
    \Sigma v_r \frac{dv_r}{dr} - \Sigma \frac{v_\phi^2}{r} = -\frac{dP_I}{dr} - \frac{\Sigma }{r^2} + (F_L)_r^{I}.
\end{equation}
On perturbing the above equation and using Eqs.~\eqref{eq:sigma_continuity} and \eqref{derivP1}, we obtain:
\begin{equation}
\frac{dv_r^{(1)}}{dr} = \mathcal{C}_{rr} v_r^{(1)} + \mathcal{C}_{r\phi} v_\phi^{(1)} + \frac{(F_L)_r}{\Delta \rho^{(0)} v_r^{(0)}},
\end{equation}

where 
\begin{equation}\label{ODEr}
\mathcal{C}_{rr}
=
\frac{1+\tilde{\Gamma}c_3(p-1)-c_2^2}
{r(c_1\alpha)^2\Delta},
\end{equation}

\begin{equation}
\mathcal{C}_{r\phi} =-\frac{1}{r}
\frac{2c_2}{c_1\alpha\,\Delta},
\end{equation}
and $\Delta = 1 - \tilde{\Gamma} \frac{c_3}{(c_1\alpha)^2}$.

\paragraph{Perturbation in $\dot M$:}
We have $\dot M = -2\pi r \Sigma v_r$, thus,
\begin{equation} 
    \frac{\dot{M}^{(1)}}{\dot{M}^{(0)}} = \frac{\Sigma^{(1)}}{\Sigma^{(0)}} + \frac{v_r^{(1)}}{v_r^{(0)}}.
    \label{eq:Mdot_frac}
\end{equation}
Using Eqs.~\eqref{sigma-v} and ~\eqref{eq:Mdot_frac}, we have 
\begin{equation}\label{Mdotcha}
\frac{\dot{M}^{(1)}}{\dot{M}^{(0)}}=0.
\end{equation}

\paragraph{Angular Navier-Stokes equation:}
Upon introducing the Lorentz force, Eq.~\eqref{eq:FLphi_explicit}, the angular momentum balance, Eq.~\eqref{azcb}, becomes:
\begin{equation}\label{me}
\frac{dF_l}{dr} + 
\frac{d \mathcal{T}_{\mathrm{mag}}}{dr}= 4\pi r \left( \rho r^2 \Omega v_z - r W_{z\phi} \right)_H,
 \end{equation}
where 
\begin{align}\label{dT1maf}
\frac{d \mathcal{T}^{(1)}_{\mathrm{mag}}}{dr}
&=
4\pi r^2 H^{(0)} (F_L)_\phi \nonumber
\\
& = \frac{3}{2}
F L
c_3^{1/2}
r^{3/2}.
\end{align}

Hence,
\begin{equation}\label{T1mag}
\mathcal{T}^{(1)}_{\mathrm{mag}}=
\frac{3}{5}
FL
c_3^{1/2}
r^{5/2}.
\end{equation}

Thus, we obtain:
\begin{equation}\label{rdif}
F_l^{(1)}
+
\mathcal{T}^{(1)}_{\mathrm{mag}}
=
\Lambda^{(1)} \dot{M}^{(0)} v_K r
\end{equation}

We define  $X$, the total angular momentum flux modified by the magnetic torque as
\begin{equation}
X \equiv F_l + \mathcal{T}_{mag}
\end{equation}

Under  our  assumptions, the perturbation to the wind angular momentum parameter scales as $\Lambda^{(1)} = \Lambda_* r^{-2}$, where $\Lambda*$ is a constant amplitude. 

Differentiating Eq.~\eqref{rdif}, we obtain:
\begin{equation}\label{dX1defn}
\frac{dX^{(1)}}{dr} - \frac{(p - 1.5)}{r} X^{(1)} = 0.
\end{equation}
From the fundamental definition $F_l = \dot{M}r^2\Omega - G_{visc}$, the first-order perturbation $X^{(1)}$ can also be explicitly expanded as:
\begin{equation}\label{X1defn}
 X^{(1)} \equiv F_{l}^{(1)} + \mathcal{T}_{mag}^{(1)} = \dot{M}^{(0)} r v_{\phi}^{(1)} - 2\pi r^{2} \alpha_{SS} P_{I}^{(1)} + \mathcal{T}_{mag}^{(1)}. 
\end{equation}

Substituting Eq.~\eqref{X1defn} into Eq.~\eqref{dX1defn} and using
Eqs.~\eqref{P1exp}, \eqref{derivP1}, \eqref{T1mag}, and
\eqref{dT1maf}, we obtain
\begin{equation}
\begin{split}
\dot{M}^{(0)}r \frac{dv_{\phi}^{(1)}}{dr}
+ 2.5\dot{M}^{(0)}v_{\phi}^{(1)}
- \beta\dot{M}^{(0)}r \frac{dv_r^{(1)}}{dr} & \\
- 2.5\beta\dot{M}^{(0)}v_r^{(1)}
+ \frac{4-p}{r}\mathcal{T}_{\rm mag}^{(1)} & = 0,
\end{split}
\end{equation}
where we define the dimensionless constant$\beta =
\tilde{\Gamma}c_3/(c_1^2\alpha).$
Substituting
\begin{equation}
\frac{dv_r^{(1)}}{dr}
=
C_{rr}v_r^{(1)}
+
C_{r\phi}v_{\phi}^{(1)}
+
S_r
\end{equation}
then gives
\begin{align}\label{ODEphi}
\frac{dv_\phi^{(1)}}{dr}
&=
C_{\phi r}v_r^{(1)}
+
C_{\phi\phi}v_{\phi}^{(1)}
+
S_\phi,
\end{align}
where
\begin{align}\label{Cphir}
C_{\phi r}
&=
\beta C_{rr}
+
\frac{2.5\beta}{r},
\\
C_{\phi\phi}
&=
\beta C_{r\phi}
-
\frac{2.5}{r},
\\
S_{\phi}
&=
\beta S_r
-
\frac{4-p}{\dot{M}^{(0)}r^2}
\mathcal{T}_{\rm mag}^{(1)}.
\end{align}

\paragraph{Analytical solutions for velocity perturbations:}
The resulting perturbed flow equations are
\begin{equation}\label{matod}
\frac{d}{dr}\begin{pmatrix} v_r^{(1)} \\[4pt] v_\phi^{(1)} \end{pmatrix}
=
\frac{1}{r}\underbrace{\begin{pmatrix} A & B \\[4pt] C & D \end{pmatrix}}_{\equiv\,\mathbf{M}}
\begin{pmatrix} v_r^{(1)} \\[4pt] v_\phi^{(1)} \end{pmatrix}
+\begin{pmatrix} S_r(r) \\[6pt] S_\phi(r) \end{pmatrix},
\end{equation}
where 
\begin{align}
A &= \frac{1+\tilde{\Gamma}c_3(p-1)-c_2^2}{1 - \tilde{\Gamma}\frac{c_3}{(c_1\alpha)^2}}, \\[6pt]
B &= \frac{-2\frac{c_2}{c_1\alpha}}{1 - \tilde{\Gamma}\frac{c_3}{(c_1\alpha)^2}}, \\[6pt]
C &= \beta(A+2.5), \\[6pt]
D &= \beta B - 2.5.
\end{align}

The dimensional radial and azimuthal source terms are driven by the magnetic field perturbations and scale as linear combinations of power laws governed by $k_1 = 1.5 - p$ and $k_2 = 3 - p$:
\begin{align}
S_r(r) &= \Pi_1 \left( F^2 r^{k_2-1} - \frac{c_2}{c_1\alpha} F L r^{k_1-1} \right), \label{eq:Sr} \\
S_\phi(r) &= \beta S_r(r) - \Pi_2 F L r^{k_1-1}, \label{eq:Sphi}
\end{align}
where the dimensionless prefactors are $\Pi_1 \equiv 3\sqrt{c_3}/(2\dot{M}^{(0)}(1-\tilde{\Gamma}c_3/(c_1\alpha)^2))$ and $\Pi_2 \equiv 3(4-p)\sqrt{c_3}/(5\dot{M}^{(0)})$.

Because the system is linear and the source terms follow simple power laws, we bypass intermediate substitutions and directly construct the complete solution as the superposition of the particular and homogeneous solutions, \begin{equation}
\begin{pmatrix} v_r^{(1)}(r) \\[4pt] v_\phi^{(1)}(r) \end{pmatrix} 
= 
\begin{pmatrix} v_{r(part)}^{(1)}(r) \\[4pt] v_{\phi(part)}^{(1)}(r) \end{pmatrix} 
+ 
\begin{pmatrix} v_{r(hom)}^{(1)}(r) \\[4pt] v_{\phi(hom)}^{(1)}(r) \end{pmatrix}.
\end{equation}

The particular solutions take the exact algebraic form:
\begin{align}
v_{r(part)}^{(1)}(r) &= \frac{(k_1 - D)s_{1,r} + B s_{1,\phi}}{\Delta_{k_1}} \, r^{1.5-p} \nonumber \\
&\quad + \frac{(k_2 - D)s_{2,r} + B s_{2,\phi}}{\Delta_{k_2}} \, r^{3-p}, \label{eq:Vr_part} \\
v_{\phi(part)}^{(1)}(r) &= \frac{C s_{1,r} + (k_1 - A)s_{1,\phi}}{\Delta_{k_1}} \, r^{1.5-p} \nonumber \\
&\quad + \frac{C s_{2,r} + (k_2 - A)s_{2,\phi}}{\Delta_{k_2}} \, r^{3-p}. \label{eq:Vphi_part}
\end{align}
where $\Delta_{k_m} = (k_m - A)(k_m - D) - BC$, and the source amplitude components are defined inline as $\mathbf{s}_1 = (s_{1,r}, s_{1,\phi})^T = \left( -\Pi_1 \frac{c_2}{c_1\alpha} \tilde{F}_0 \tilde{L}, -\left(\beta \Pi_1 \frac{c_2}{c_1\alpha} + \Pi_2 \right) \tilde{F}_0 \tilde{L} \right)^T$ and $\mathbf{s}_2 = (s_{2,r}, s_{2,\phi})^T = \left( \Pi_1 \tilde{F}_0^2, \beta \Pi_1 \tilde{F}_0^2 \right)^T$.
The homogeneous solution is governed by the eigenvalues
$\lambda_1 = A + \beta B$ and $\lambda_2 = -2.5$ and the corresponding
eigenvectors of the coefficient matrix $\mathbf{M}$, appearing in Eq.~\eqref{matod}:
\begin{equation}
\begin{pmatrix} v_{r(hom)}^{(1)}(r) \\[4pt] v_{\phi(hom)}^{(1)}(r) \end{pmatrix} 
= 
D_1 \begin{pmatrix} 1 \\ \beta \end{pmatrix} r^{\lambda_1} 
+ 
D_2 \begin{pmatrix} B \\ -(A + 2.5) \end{pmatrix} r^{\lambda_2}.
\end{equation}

Crucially, the presence of the strictly negative eigenvalue ($\lambda_2 = -2.5$)  generates an $r^{-2.5}$ divergence profile at small radii. This mathematically captures the physically expected acceleration of the perturbed velocity as the flow approaches the central accretor.

Imposing the outer boundary condition $v_r^{(1)}(r_{\mathrm{out}}) = v_\phi^{(1)}(r_{\mathrm{out}}) = 0$ uniquely determines the integration constants:

\begin{equation}
\begin{split}
    D_1 &= \tilde{F}_0\tilde{L} \left[ \frac{\Pi_1\frac{c_2}{c_1\alpha}}{k_1 - A - \beta B} \right. \\
    &\quad \left. + \frac{B\Pi_2}{(A + \beta B + 2.5)(k_1 - A - \beta B)} \right] r_{\text{out}}^{k_1 - A - \beta B} \\
    &\quad - \frac{\Pi_1 \tilde{F}_0^2}{k_2 - A - \beta B} r_{\text{out}}^{k_2 - A - \beta B},
\end{split}
\end{equation}
\begin{equation}
D_2=
-\frac{\Pi_2\tilde{F}_0\tilde{L}}
     {(A+\beta B+2.5)(k_1+2.5)}
r_{\rm out}^{\,k_1+2.5}.
\end{equation}
Because the components of the particular solution ($\mathbf{V}_{\text{part}}^{(1)}$) are strictly proportional to the external magnetic field amplitudes ($F$ and $L$), the constants $D_1$ and $D_2$ automatically vanish when the magnetic perturbation is removed. This guarantees that the flow smoothly reverts to the unperturbed ideal ADIOS background without introducing unphysical artifacts.

\begin{figure*}
    \centering
    
    \begin{subfigure}{0.48\textwidth}
        \centering
        \includegraphics[width=\linewidth]{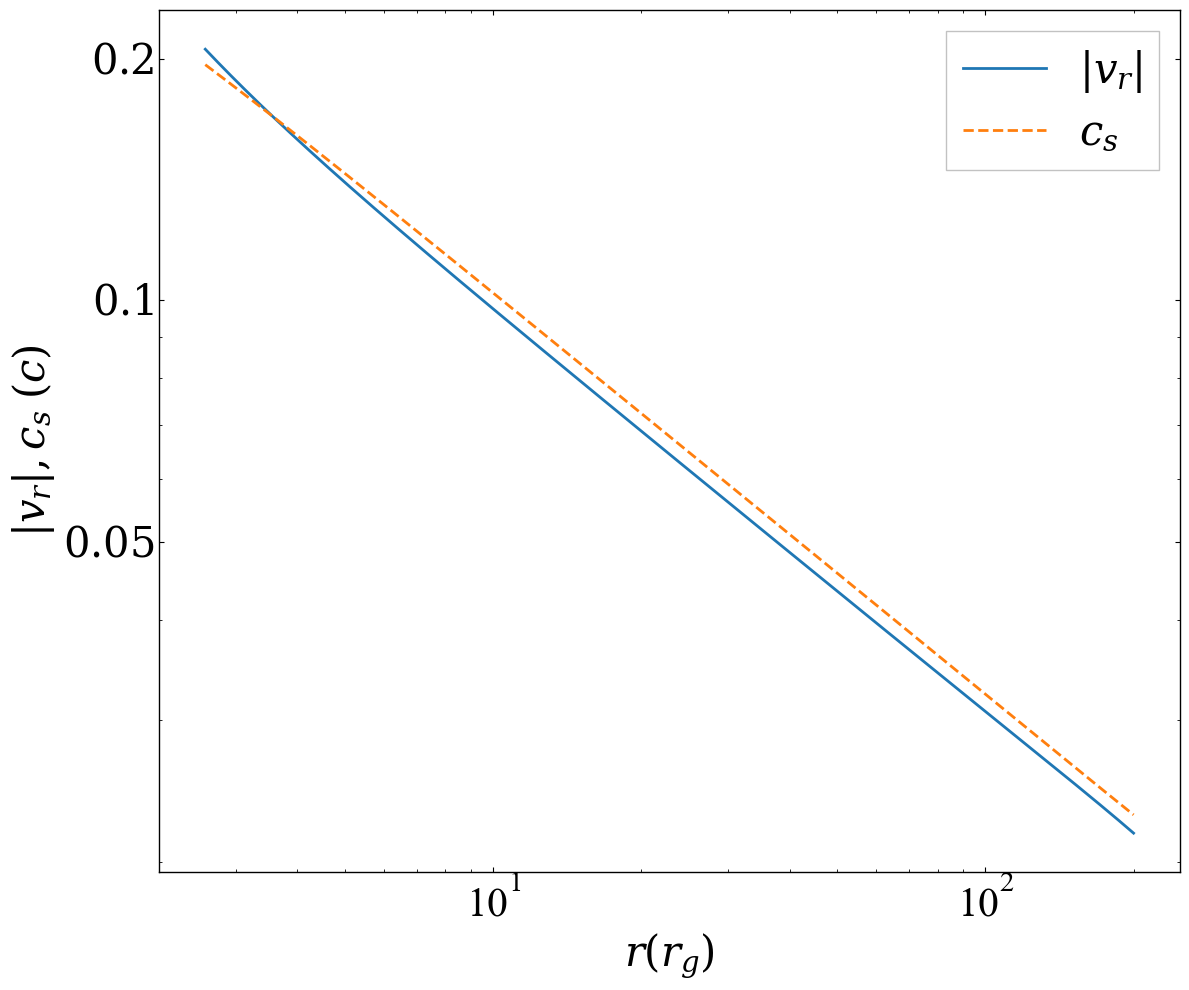}
        \caption{}
        \label{fig:panel_a}
    \end{subfigure}
    \hfill
    \begin{subfigure}{0.48\textwidth}
        \centering
        \includegraphics[width=\linewidth]{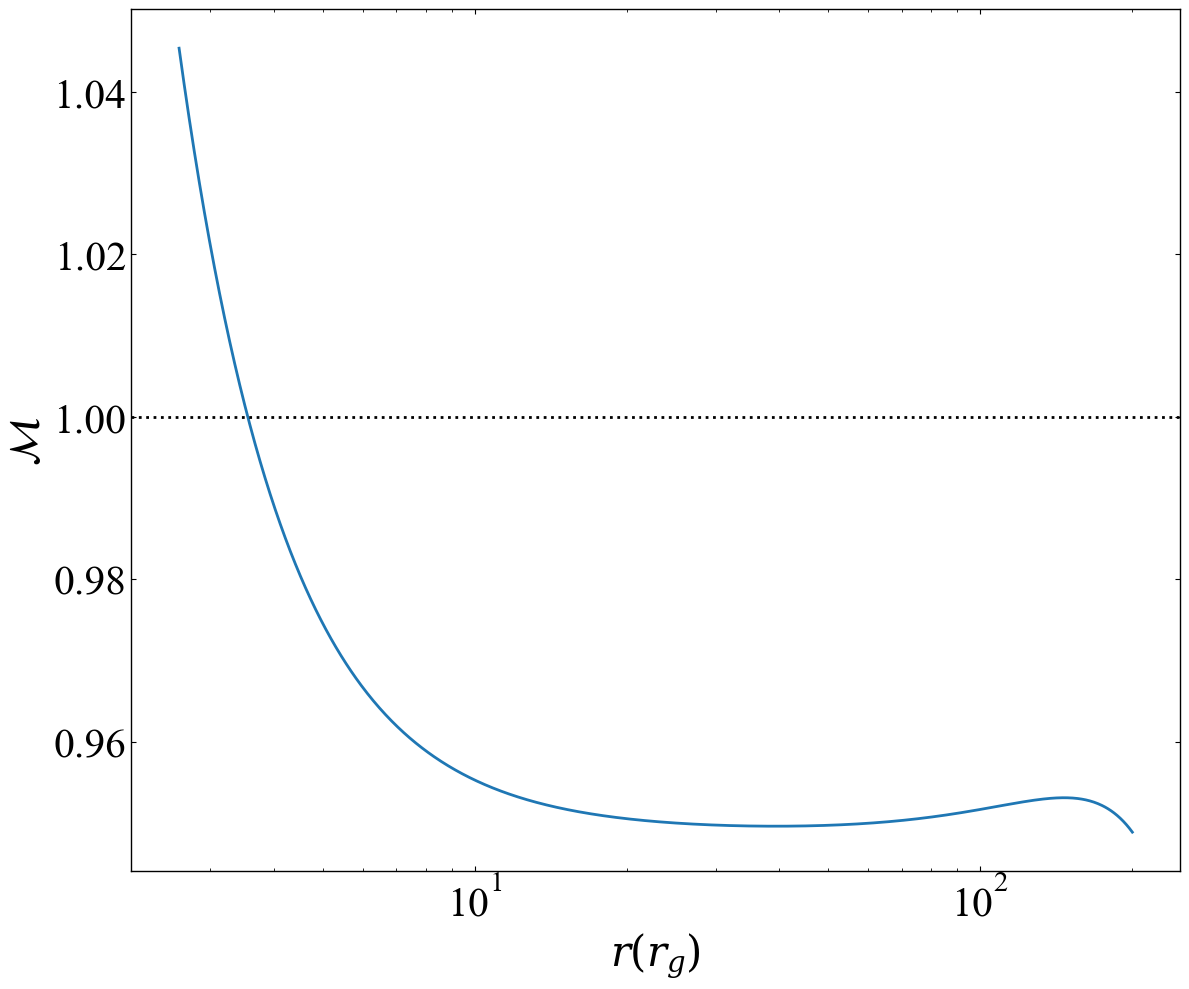}
        \caption{}
        \label{fig:panel_d}
    \end{subfigure}
    
    \vspace{0.4cm} 
    
    \begin{subfigure}{0.48\textwidth}
        \centering
        \includegraphics[width=\linewidth]{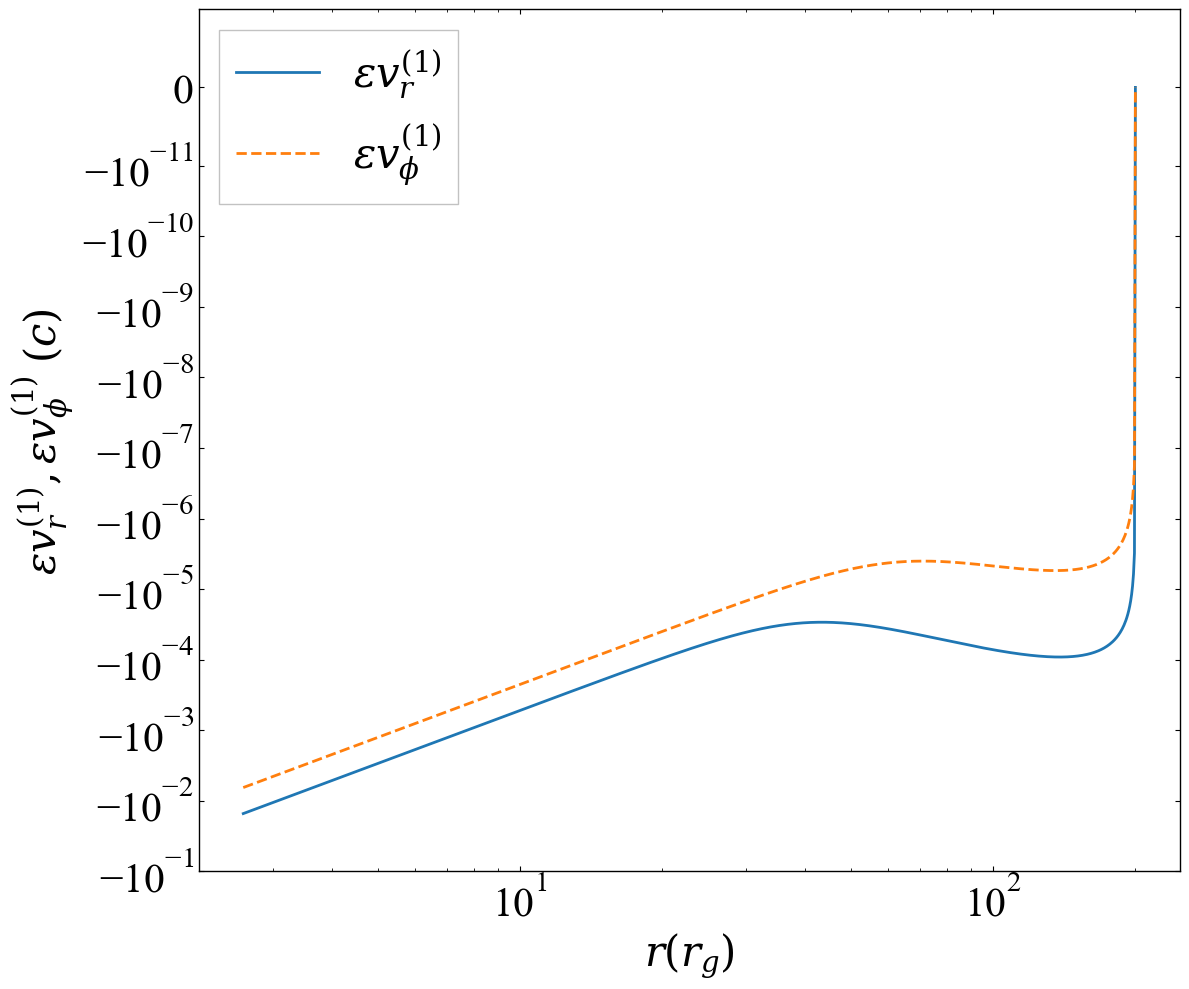}
        \caption{}
        \label{fig:panel_c}
    \end{subfigure}
    \hfill
    \begin{subfigure}{0.48\textwidth}
        \centering
        \includegraphics[width=\linewidth]{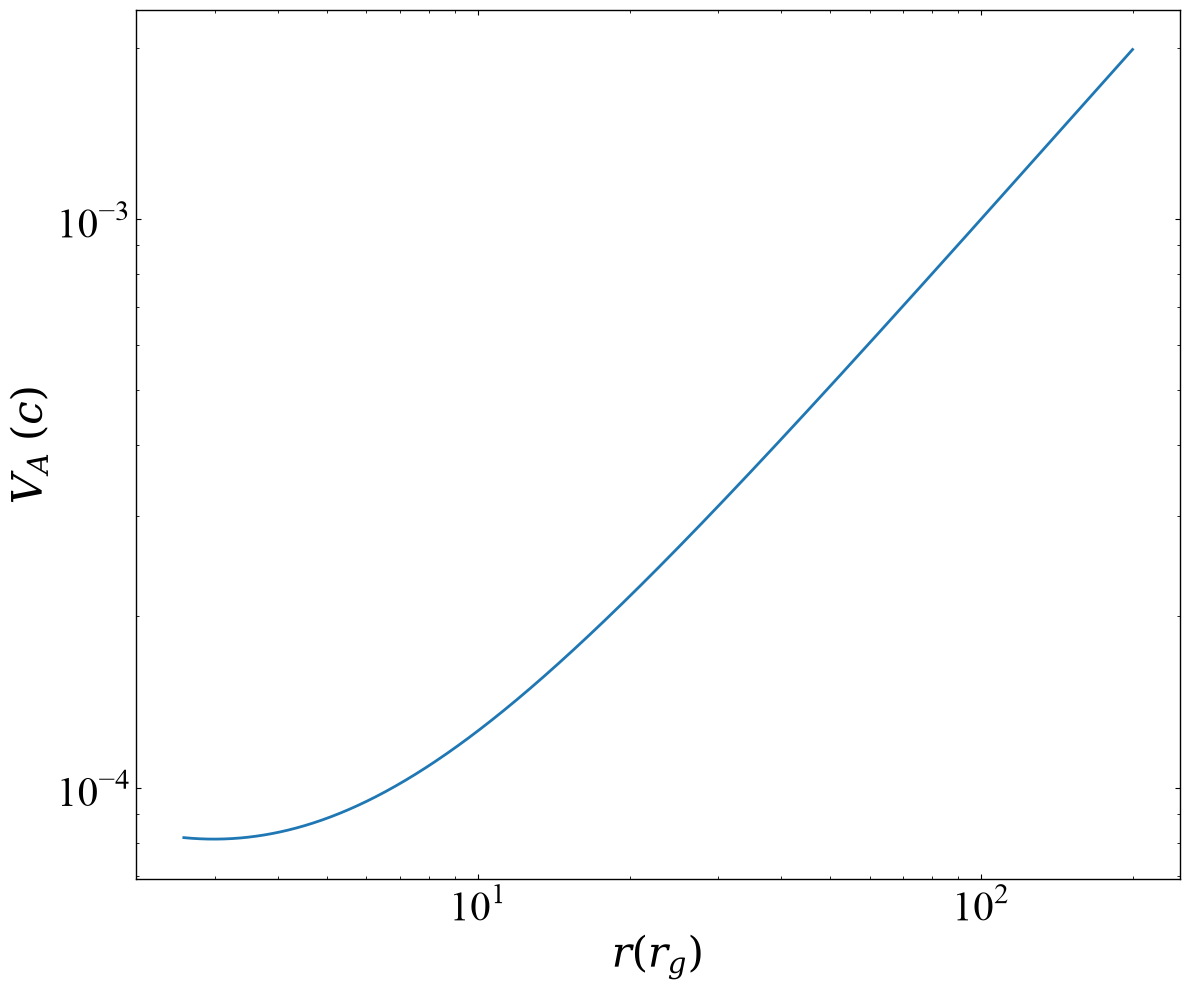}
        \caption{}
        \label{fig:panel_b}
    \end{subfigure}
    \caption{\raggedright
    Background flow profiles as functions of the radial coordinate  for ADIOS parameters: $p=0.5$, $\epsilon_{\rm w}=0.4$, $\Lambda = 0.890$, $\gamma=\Gamma=5/3$, $\alpha=0.070$, $M = 10M_\odot$ and  $\dot{M}=0.01\dot{M}_{\rm Edd}$, for a disk of size $200r_g$.
    \textbf{(a)} The radial inward velocity ($|v_r|$) and sound speed ($c_s$).
    \textbf{(b)} Mach number ($\mathcal{M}$) together with the
    $\mathcal{M}=1$ line.
    \textbf{(c)} Radial and azimuthal velocity perturbations.
    \textbf{(d)} Alfv\'en speed ($V_A$).}
    \label{fig:background_flow}
\end{figure*}

\begin{figure*}
    \centering
    
    \begin{subfigure}{0.48\textwidth}
        \centering
        \includegraphics[width=\linewidth]{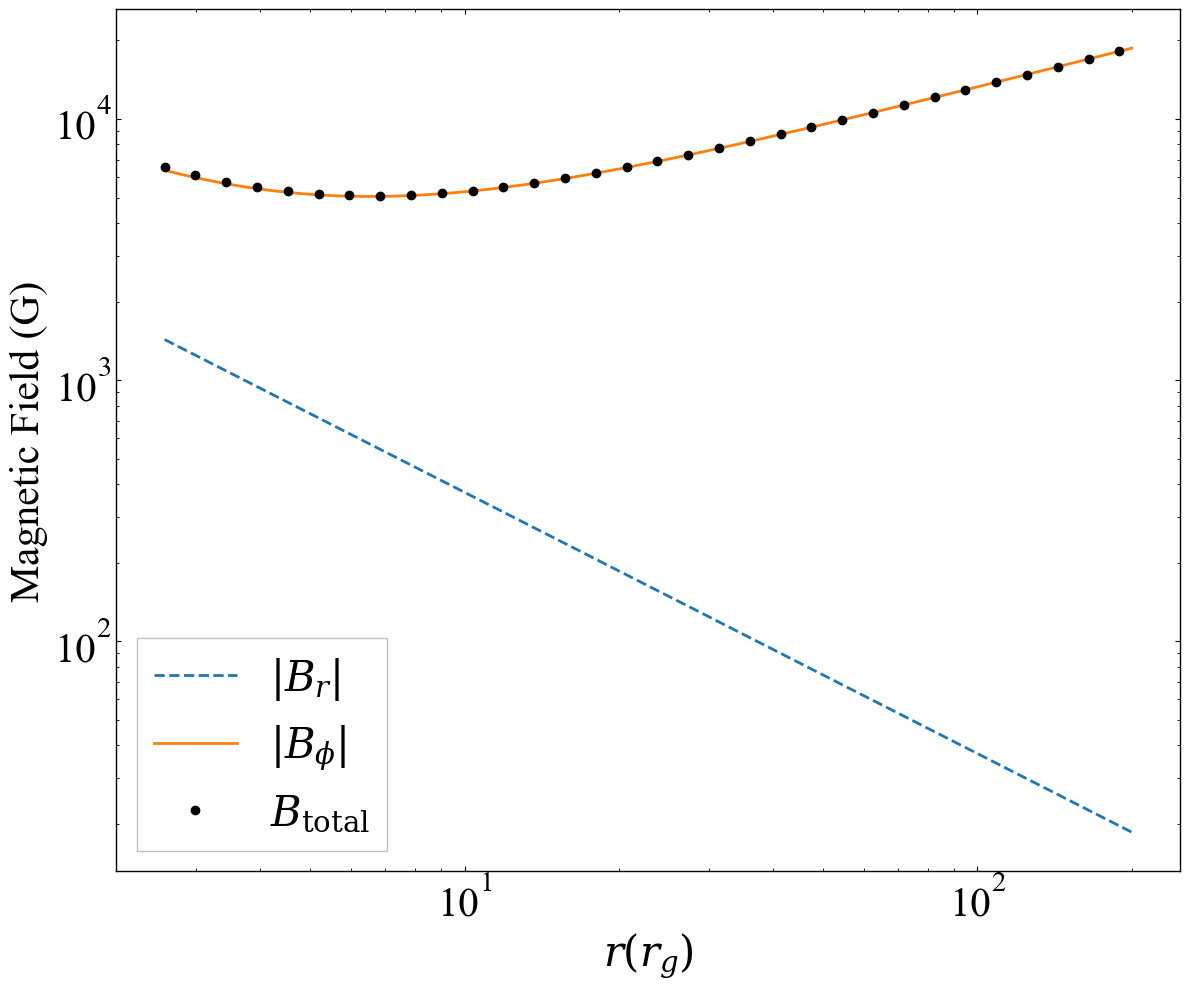}
        \caption{}
        \label{fig:panel_magnetic_field}
    \end{subfigure}
    \hfill
    \begin{subfigure}{0.48\textwidth}
        \centering
        \includegraphics[width=\linewidth]{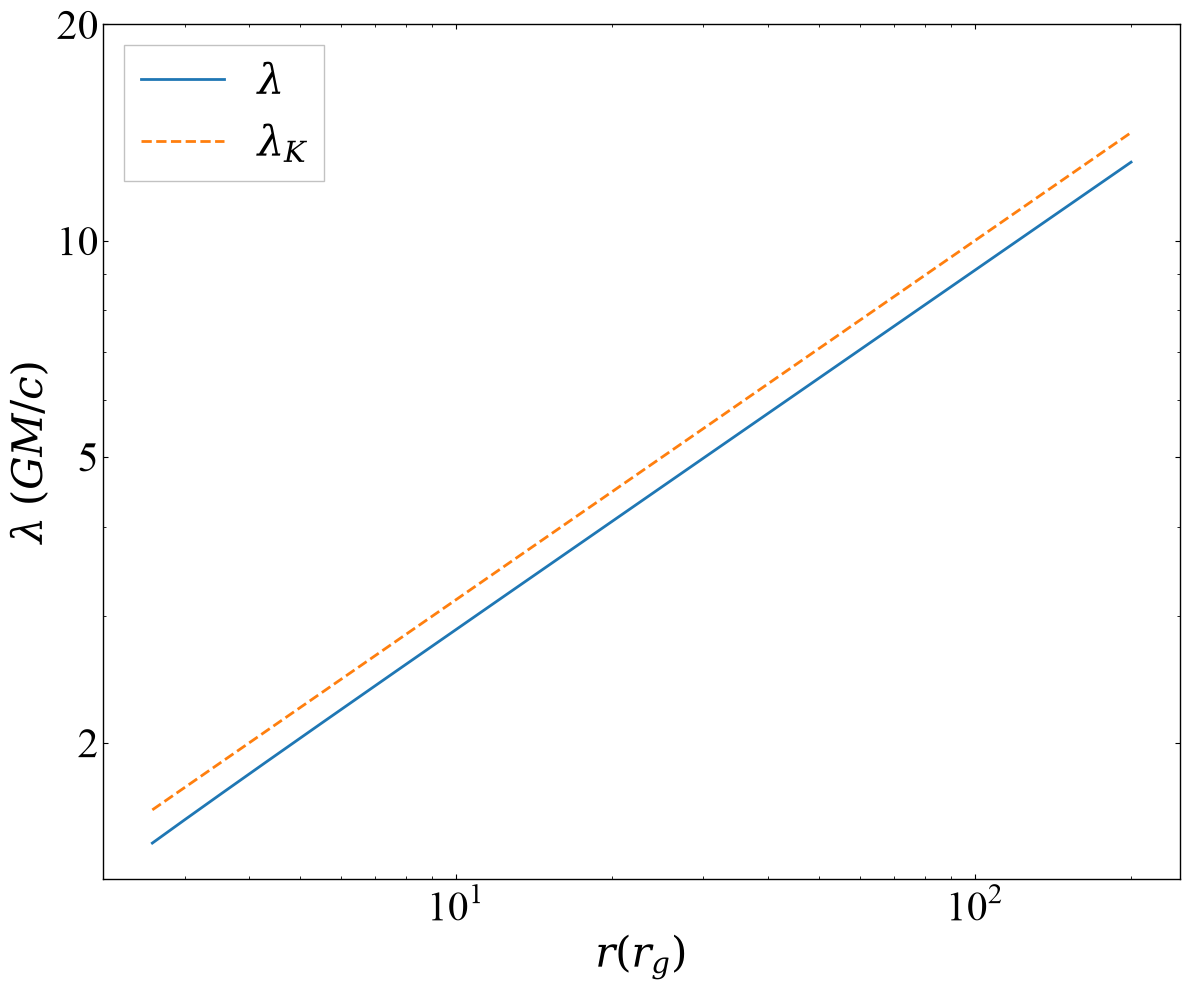}
        \caption{}
        \label{fig:panel_angular_momentum}
    \end{subfigure}
    
    \vspace{0.4cm}
    
    \begin{subfigure}{0.48\textwidth}
        \centering
        \includegraphics[width=\linewidth]{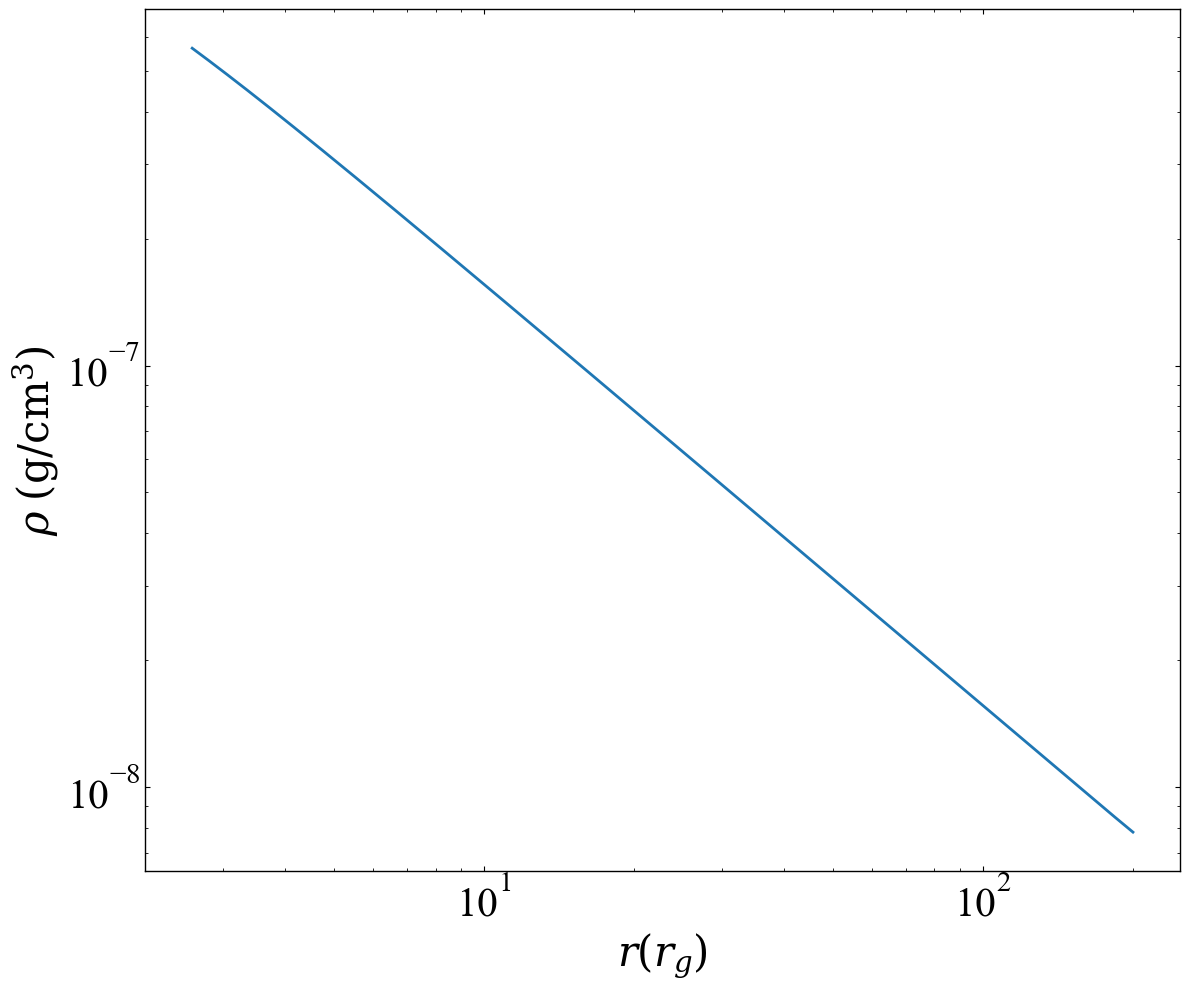}
        \caption{}
        \label{fig:panel_density}
    \end{subfigure}
    \hfill
    \begin{subfigure}{0.48\textwidth}
        \centering
        \includegraphics[width=\linewidth]{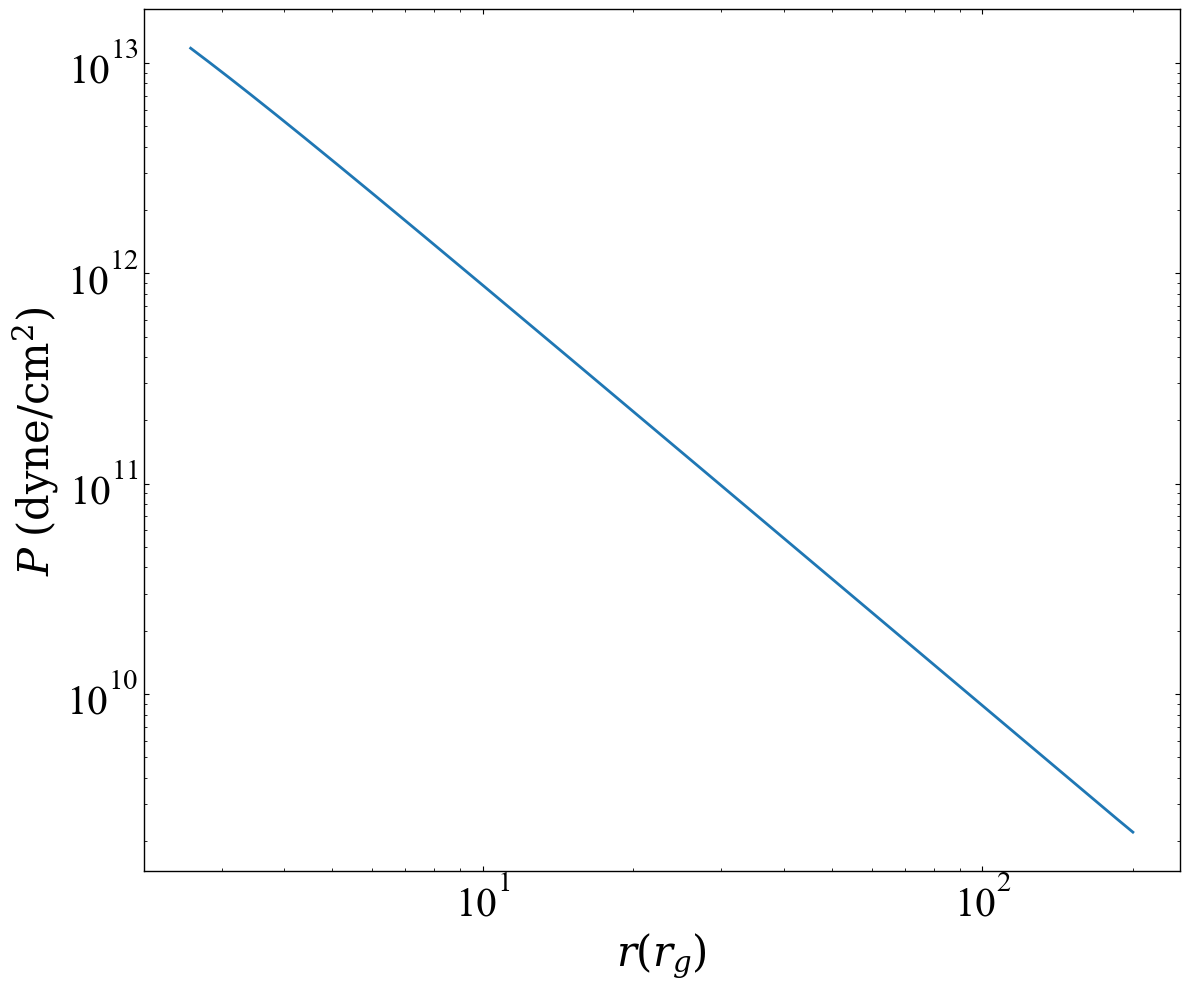}
        \caption{}
        \label{fig:panel_pressure}
    \end{subfigure}

    \caption{\raggedright
    Additional background flow profiles for the same parameters
    as in Fig.~\ref{fig:background_flow}.
    \textbf{(a)} Net magnetic field ($|B|$) alongside its components
    ($B_r$ and $B_\phi$; $B_z=0$ is assumed in our derivation).
    \textbf{(b)} Specific angular momentum ($\lambda$) compared to the
    Keplerian distribution ($\lambda_K$).
    \textbf{(c)} Total fluid density ($\rho$).
    \textbf{(d)} Total fluid pressure ($P$).}
    \label{fig:additional_profiles}
\end{figure*}

\subsubsection{Magnetically perturbed flow profiles}
The  fractional first-order perturbations for the volume density,  sound speed, and average pressure are completely determined by the radial velocity perturbation, which can be written concisely as:
\begin{equation}\label{perturbQ}
\frac{\mathcal{Q}^{(1)}}{\mathcal{Q}^{(0)}} =
C_{\mathcal{Q}}\frac{v_r^{(1)}}{v_r^{(0)}},
\qquad
\mathcal{Q}\in\{\rho,c_s,P\},
\end{equation}
where
\begin{equation}
C_{\mathcal{Q}} =
\left\{
-\frac{2}{\Gamma+1},
-\frac{\Gamma-1}{\Gamma+1},
-\frac{2\Gamma}{\Gamma+1}
\right\},
\end{equation}
for $\mathcal{Q}=\rho,c_s,P$, respectively.

Then we can say the profile up to first-order is 
\begin{equation}
\mathcal{X}=\mathcal{X}^{(0)}
+\epsilon\mathcal{X}^{(1)},
\qquad
\mathcal{X}\in\{v_r,v_\phi,\rho,P,c_s\}.
\end{equation}
As evident from the Eq.~\eqref{perturbQ}, the fractional perturbations of all thermodynamic quantities are directly proportional to the radial velocity perturbation ratio, $v_r^{(1)}/v_r^{(0)}$, scaled by coefficients of order unity. Consequently, if the radial velocity perturbation remains sufficiently small relative to the background flow, it mathematically guarantees that the fractional perturbations of the density, sound speed, and pressure will correspondingly remain small. Hence, the perturbative analysis is valid provided  the velocity perturbations  are small.

Figure~\ref{fig:background_flow} presents the background flow profiles resulting from our magnetically perturbed ADIOS framework. Panel \textbf{(a)} illustrates the radial inward velocity ($|v_r|$) and the  sound speed ($c_s$). Importantly, the curves intersect, showing transonic behavior. This behavior directly resolves the drawback of constant Mach number  associated with strictly self-similar flows which prevent them from being transonic. Panel \textbf{(b)} explicitly highlights our modified transonic behavior, where the Mach number ($\mathcal{M}$) smoothly crosses the $\mathcal{M}=1$ threshold. Because our magnetic field is introduced as a small perturbation over a self-similar profile of constant Mach number, the Mach number exhibits a small variation around the constant value that would be expected from a purely self-similar disk. 
Panel \textbf{(c)} maps the dimensionless radial and azimuthal velocity perturbations. As required by our analytical framework, the magnitudes of these first-order perturbations remain sufficiently smaller than the unperturbed background radial velocity ($|v_r|$). This relative smallness strictly validates our perturbative approach and ensures the linear truncation is accurate. Furthermore, panel \textbf{(d)} displays the Alfvén speed ($V_A$). $V_A$ is quite small in magnitude (compared to $c_s$), as the magnetic field is small.

Figure~\ref{fig:additional_profiles} presents additional background flow profiles, beginning with  panel \textbf{(a)} which plots the radial variation of the net magnetic field alongside its individual components. Panel \textbf{(b)} compares the flow's specific angular momentum ($\lambda$) against the Keplerian angular momentum ($\lambda_K$). Lastly, panels \textbf{(c)} and \textbf{(d)} show the total fluid density ($\rho$) and pressure ($P$), respectively, both of which exhibit the expected monotonic decrease as the radial distance from the central accretor increases.

\section{Analog Hawking temperature}
\label{sec:hawking}

A visco-magnetoacoustic black hole is characterized by the presence of a visco-magnetoacoustic  horizon.  Fast visco-magnetoacoustic perturbations propagating against the background flow are unable to escape once they cross this horizon, which causes the trapping of phonons in the region inside the horizon. In close analogy with gravitational black holes, the existence of the visco-magnetoacoustic horizon gives rise to Hawking radiation in the form of a thermal flux of phonons. In the analog-gravity framework, these wave perturbations are mapped to  a massless scalar field propagating in an effective curved spacetime. This continuum description is valid at wavelengths larger than the atomic  scale of the fluid, which naturally provides an ultraviolet (UV) cutoff. Within this  regime, these perturbations can be formally quantized using standard techniques for curved backgrounds \cite{Birrell:1982ix, 2007JCAP...06..009D}.  Consequently, the horizon is expected to emit a thermal flux of phonons, analogous to Hawking radiation. To quantify the thermal spectrum of this emission within our visco-magnetoacoustic model, we determine the analog Hawking temperature.

Previous works, like \cite{asenjo2011analogueblackholemagnetohydrodynamics} and \cite{2018EPJC...78..662G}, demonstrated that magnetic fields enhance the analog Hawking temperature. However, these models rely on uniform background fields in one-dimensional  configurations, entirely neglecting the magnetic topologies.
We study the effects of magnetic field magnitude as well as topology on the analog Hawking radiation. Specifically, we use the magnetically perturbed ADIOS flow (as described in $\S$ \ref{ADIOS}) and the numerical advective accretion flow around a black hole (as described in $\S$\ref{sec:Numericaladaf}) with varying magnetic field characteristics.
We additionally examine how the inclusion of $\alpha$-viscosity modifies the analog Hawking temperature.

We consider a stationary and axisymmetric flow, such as the two background models introduced in this paper. To calculate the analog surface gravity $\kappa$ and the subsequent Hawking temperature, we introduce the standard time and azimuthal Killing vectors $K^{\mu}=(1,0,0,0)$ and $L^{\mu}=(0,0,1,0)$. 

We construct a generalized Killing vector $X^{\mu} = K^{\mu} + \varpi L^{\mu}$ where $\varpi$ is a constant. We extract $\varpi$ by setting the generalized vector to become null at the horizon:
\begin{equation} \label{eq:null_condition}
    g_{\phi\phi}\varpi^2 + 2g_{t\phi}\varpi + g_{tt} = 0.
\end{equation}

To find the surface gravity, we apply the definition $-(1/{2})\partial_{\mu}(X^{\lambda}X_{\lambda}) = \kappa X_{\mu}$. Focusing on the radial component ($\mu=r$), we evaluate this at the horizon, given by
\begin{equation}
\begin{split}
-\frac{1}{2}\partial_{r}[g_{tt} + \varpi^{2}g_{\phi\phi} + 2\varpi &g_{t\phi}] \\
&= \kappa(g_{rt} + \varpi g_{r\phi}).
\end{split}
\end{equation}
Rearranging this gives the explicit expression for the surface gravity:
\begin{equation}
\kappa = \left. -\frac{\partial_{r}(g_{tt} + \varpi^{2}g_{\phi\phi} + 2\varpi g_{t\phi})}{2(g_{rt} + \varpi g_{r\phi})} \right|_{\text{horizon}}.
\end{equation}

Crucially, this surface gravity is strictly independent of any overall conformal factor $\Upsilon(r)$ of the metric. If we scale the metric components by $\Upsilon(r)$, the numerator of the surface gravity expression expands as:
\begin{equation}
\begin{split}
-\partial_{r}&\left[\Upsilon(g_{tt} + \varpi^{2}g_{\phi\phi} + 2\varpi g_{t\phi})\right] \\
&= -\Upsilon\partial_{r}\left[g_{tt} + \varpi^{2}g_{\phi\phi} + 2\varpi g_{t\phi}\right] \\
&\quad - (\partial_{r}\Upsilon)\left[g_{tt} + \varpi^{2}g_{\phi\phi} + 2\varpi g_{t\phi}\right].
\end{split}
\end{equation}
At the horizon, the term $g_{tt} + \varpi^{2}g_{\phi\phi} + 2\varpi g_{t\phi}$
identically vanishes by definition. The denominator similarly scales by $\Upsilon(r)$, resulting in the conformal factor canceling out entirely from the final expression for $\kappa$. Thus, we can directly utilize our  effective metric which is derived up to a conformal factor in this paper.  The analog Hawking temperature is then obtained via the standard relation $T_H = \hbar \kappa/({2\pi k_B})$. Because our metric is up to first-order in the viscous perturbation, we can extract the corrections to the Hawking temperature up to the first-order, as described below.

\subsection{Evaluation of the surface gravity}

\subsubsection{Perturbative expansion of $\varpi$}

We treat the non-magnetic inviscid  flow as the exact zeroth-order background and consider the viscous and magnetic corrections as first-order perturbations in the small parameters $\sigma$ and $\eta$, defined in Eqs.~\eqref{eq:55} and~\eqref{eq:56}. We  expand  both the Killing-vector coefficient and the metric components about their zeroth-order values:
\begin{align}
    \varpi &= \varpi_0 + \varpi_1 + \mathcal{O}(2), \\
    g_{\mu\nu} &= g_{\mu\nu}^{(0)} + g_{\mu\nu}^{(1)} + \mathcal{O}(2),
\end{align}
where $\mathcal{O}(2)$ denotes second order terms in $\sigma$ and $\eta$.
Substituting these expansions into Eq. \eqref{eq:null_condition} and grouping the  terms by their perturbative order, we obtain the following.\\

\paragraph{Zeroth-order equation:} 
Extracting the  unperturbed terms, we have:
\begin{equation}
    g_{\phi\phi}^{(0)}\varpi_0^2 + 2g_{t\phi}^{(0)}\varpi_0 + g_{tt}^{(0)} = 0.
\end{equation}
It is easy to see that the discriminant of the above quadratic equation vanishes identically, i.e.,
$D^{(0)} = (g_{t\phi}^{(0)})^2 - g_{tt}^{(0)}g_{\phi\phi}^{(0)} = 0$.
Hence, we obtain
\begin{equation}\label{eq:130}
    \varpi_0 = -\frac{g_{t\phi}^{(0)}}{g_{\phi\phi}^{(0)}}.
\end{equation}
\subsubsection{Independence of the first-order surface gravity from  $\varpi^{(1)}$ }

It may initially appear that calculating the first-order correction to the analog Hawking temperature requires strict knowledge of the first-order shift in the Killing-vector coefficient, $\varpi_1$.  However, a remarkable feature of the Killing horizon geometry strictly isolates the leading-order surface gravity from $\varpi_1$.

The analog surface gravity is extracted from the radial derivative of the generalized Killing vector norm evaluated at the horizon:
\begin{equation}
    \kappa = -\frac{\partial_r (X^\mu X_\mu)}{2(g_{rt} + \varpi g_{r\phi})} \Bigg|_{r_H},
\end{equation}
where $X^\mu X_\mu = g_{tt} + 2\varpi g_{t\phi} + \varpi^2 g_{\phi\phi}$. Let us perturbatively expand the numerator, defining the functional $F(r, \varpi) \equiv X^\mu X_\mu$. Introducing our linear expansions $\varpi = \varpi_0 + \varpi_1$ and $g_{\mu\nu} = g_{\mu\nu}^{(0)} + g_{\mu\nu}^{(1)}$, the variational shift in the norm at the horizon is given by the chain rule:
\begin{equation}\label{orderHawk}
    F(r, \varpi) \approx F^{(0)}(r, \varpi_0) + F^{(1)}(r, \varpi_0) + \left. \frac{\partial F^{(0)}}{\partial \varpi} \right|_{\varpi_0} \varpi_1.
\end{equation}

Here $F^{(0)}(r,\varpi_0)$ denotes the zeroth-order Killing-vector norm evaluated using the non-magnetic, inviscid metric and the zeroth-order coefficient $\varpi_0$. The quantity $F^{(1)}(r,\varpi_0)$ collects the first-order changes in the norm arising from the magnetic and viscous corrections to the metric, evaluated at fixed $\varpi=\varpi_0$.
Using $\varpi_0=-g_{t\phi}^{(0)}/g_{\phi\phi}^{(0)}$ from Eq.~\eqref{eq:130}, we find
\begin{equation}
\left.\frac{\partial F^{(0)}}{\partial\varpi}\right|_{\varpi_0}
=
2\left(g_{t\phi}^{(0)}+\varpi_0 g_{\phi\phi}^{(0)}\right)
=0.
\end{equation}
Consequently, the first-order shift in the Killing-vector coefficient, $\varpi_1$, drops completely out of the numerator of the surface gravity expression. We can therefore rigorously compute the first-order shift in the analog Hawking temperature using exclusively the zeroth-order Killing-vector coefficient $\varpi_0$ coupled with the first-order metric perturbations $g_{\mu\nu}^{(1)}$:
\begin{equation}
    X^\mu X_\mu \approx (g_{tt}^{(0)} + g_{tt}^{(1)}) + 2\varpi_0 (g_{t\phi}^{(0)} + g_{t\phi}^{(1)}) + \varpi_0^2 (g_{\phi\phi}^{(0)} + g_{\phi\phi}^{(1)}). 
\end{equation}

This mathematically guarantees that our linear effective metric framework is fully self-contained and sufficient to extract the thermodynamic properties of the visco-magnetoacoustic horizon without requiring higher-order metric expansions.

\subsection{ Hawking temperature in  ADIOS }

We begin by explaining the analytical ADIOS profile where the magnetic field is given by \begin{equation}\label{mag}
    B_R = \sqrt{\epsilon}\frac{L}{R}, \quad
    B_\phi =\sqrt{\epsilon}\left[  F R^{1/2} -\frac{c_2L}{c_1\alpha} R^{-1}\right], \quad
    B_z = 0,
\end{equation} 
where we can control the strength of the magnetic field using the $\epsilon$ parameter. We observed that if we change the magnetic field topology by changing the  relative sign of
$F$ and $L$, then the Hawking temperature may increase or decrease with magnetic field.

\begin{figure*}[t]
\centering

\begin{subfigure}{0.48\textwidth}
    \centering
    \includegraphics[width=\linewidth]{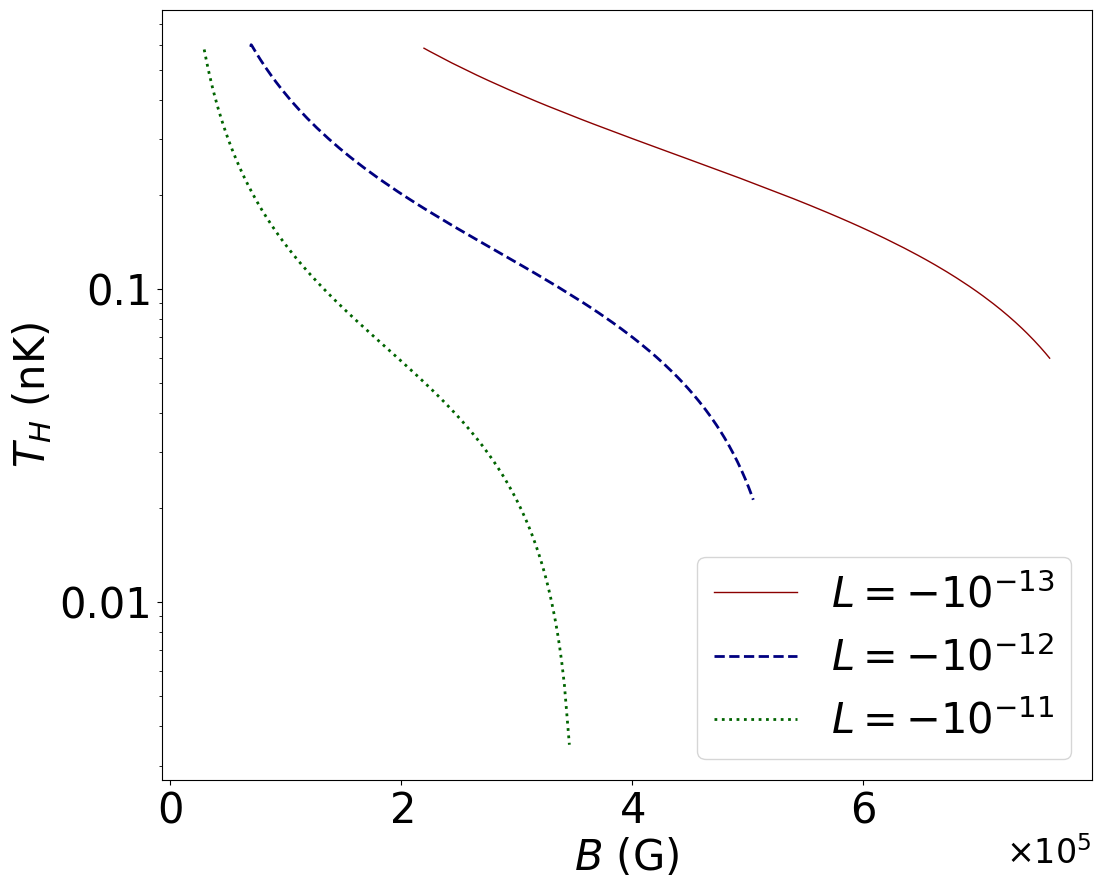}
    \caption{}
    \label{fig:th_a}
\end{subfigure}
\hfill
\begin{subfigure}{0.48\textwidth}
    \centering
    \includegraphics[width=\linewidth]{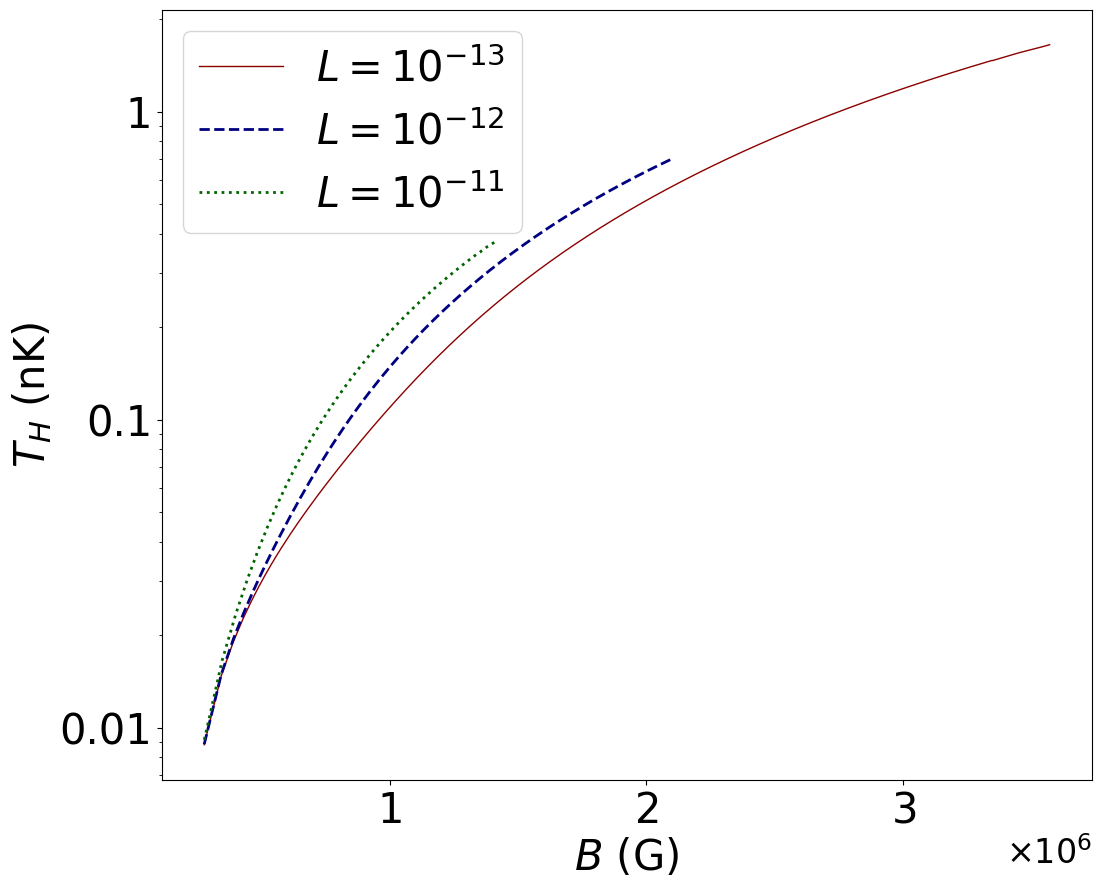}
    \caption{}
    \label{fig:th_b}
\end{subfigure}

\caption{\raggedright Hawking temperature $T_H$ as a function of the horizon magnetic field $B(r_H)$ for different $L$. The calculations are performed for $M=10M_\odot$,  $\dot{M}=0.01\dot{M}_{\rm Edd}$, $\alpha=0.046$, $\Gamma=1.596$, and ADIOS wind parameters $p=0.56$, $\Lambda=0.85$, and $\epsilon_{\rm w}=0.31$. The outer disk radius is fixed at $r_{\rm out}=80r_g$ and $F=10^{-11}$. (a) Decreasing behavior of $T_H$ with increasing $B(r_H)$ for negative values of $L$. (b) Increasing behavior of $T_H$ with increasing $B(r_H)$ for positive values of $L$.
}
\label{fig:analytical_topology}
\end{figure*}

The analytical ADIOS results shown in Fig.~\ref{fig:th_a} and Fig.~\ref{fig:th_b} exhibit a clear dependence of the Hawking temperature on the magnetic field at the horizon. In Fig.~\ref{fig:th_a}, corresponding to negative values of $L$, the temperature decreases monotonically with increasing $B(r_H)$. Furthermore, for a fixed value of $B(r_H)$, the Hawking temperature decreases as $L$ becomes more negative. In contrast, Fig.~\ref{fig:th_b}, corresponding to positive values of $L$, shows a monotonic increase of the Hawking temperature with increasing $B(r_H)$. For a fixed value of $B(r_H)$, the Hawking temperature also increases with increasing positive values of $L$. 

The curves begin only above a finite value of $B(r_H)$ because the unperturbed self-similar ADIOS solution possesses a constant subsonic Mach number and therefore does not admit a visco-magnetoacoustic horizon. The introduction of the magnetic perturbation modifies the velocity, density, and pressure profiles,  thereby breaking the exact self-similarity of the background solution. In addition, the magnetic field contributes directly to the horizon condition through the magnetic correction terms appearing in Eq.~\eqref{eq:97}. As the magnetic field strength increases, these combined effects eventually allow Eq.~\eqref{eq:97} to develop a physical root corresponding to a visco-magnetoacoustic horizon. The plotted curves therefore begin at the minimum magnetic field strength for which the perturbed flow first admits a horizon.

The curves terminate at finite values of $B(r_H)$ because our analytical treatment is based on a first-order perturbative expansion about the self-similar ADIOS background. As the magnetic field strength increases, the amplitudes of the magnetic corrections grow. We therefore restrict our analysis to the regime in which the magnetic perturbations remain sub-dominant to the background flow variables and our perturbative solution is valid at the horizon radius. The termination of the curves should thus be interpreted as the boundary of validity of the perturbative approximation rather than the absence of physical solutions beyond those magnetic field strengths.

\subsubsection*{Conditions for the enhancement or suppression of the Hawking temperature}

The surface gravity at the analog horizon can be expressed as
\begin{equation}
    \kappa
    =-\left.
    \frac{\partial_r\left(X^\mu X_\mu\right)}
    {2\left(g_{rt}+\varpi g_{r\phi}\right)}
    \right|_{r_H}
    \equiv
    \frac{\mathcal{N}}{\mathcal{D}}.
\end{equation}

Throughout this analysis, we work within the perturbative regime
\begin{equation}
    \eta \ll 1, \qquad
    \sigma \ll 1,
\end{equation}
and consider the gas-pressure-dominated limit, $c_s \gg V_A$. Under these assumptions, the magnetic and viscous contributions to $\mathcal{N}$ appear only as higher-order corrections, while the leading-order ($\mathcal{O}(1)$) contribution is given by
\begin{equation}
    \left|\partial_r\left(v_r^2-c_s^2\right)\right|.
\end{equation}
Consequently, the Hawking temperature is expected to be governed primarily by the behavior of this quantity, with the remaining terms producing only small perturbative corrections.

Therefore, determining whether the magnetic field enhances or suppresses the Hawking temperature reduces to understanding how it modifies the quantity
\begin{equation}
    \left|\partial_r\left(v_r^2-c_s^2\right)\right|.
\end{equation}

Under the limit of $\eta,\sigma\ll1$ and low magnetic field owing to $c_s\gg V_A $, the condition of horizon, Eq.~\eqref{eq:97}, becomes approximately equal to 
$$v_r^2 \approx c_s^2.$$
Now, looking at the radial Lorentz force, \begin{equation}
    (F_L)_r = - \frac{3}{8\pi} F^2 + \frac{3}{8\pi} F \left( \frac{c_2 L}{c_1 \alpha} \right) r^{-3/2}.
\end{equation}

Without loss of generality, we take \(F>0\), leaving two cases: \(L<0\) and \(L>0\).

\begin{itemize}
\item  \textbf{Case I: $L < 0$:} The second term in the radial Lorentz force becomes negative, thereby reinforcing the inward-directed magnetic pressure term,
($-{3F^2}/{8\pi}$). Consequently, the total radial Lorentz force becomes more negative, producing an additional inward pull that assists gravity. As a result, the magnitude of the radial velocity, $v_r$, increases.

From our analytical calculations, the first-order perturbation of the sound speed is given by
\begin{equation}
c_s^{(1)}
=
-c_s^{(0)}
\left(
\frac{\Gamma+1}{\Gamma-1}
\right)
\frac{v_r^{(1)}}{v_r^{(0)}}.
\end{equation}

Since both the background radial velocity, $v_r^{(0)}$, and its first-order perturbation, $v_r^{(1)}$, are negative, the perturbation $c_s^{(1)}$ is negative. Consequently, the sound speed $c_s$ decreases. This reduction in $c_s$ and increment in $|v_r|$ increases the Mach number across the flow. Because the Mach number profile is shifted upward, the flow reaches the sonic condition ($v_r = c_s$) at a larger radius. Therefore, increasing the magnetic field shifts the visco-magnetoacoustic horizon outward.

Because the velocity gradients become weaker farther from the black hole, this outward displacement of the  horizon results in a lower Hawking temperature. As $L$ becomes more negative, the inward Lorentz force is further enhanced, pushing the horizon to even larger radii. Consequently, the Hawking temperature decreases monotonically with increasing magnetic field strength, as illustrated in Fig.~\ref{fig:th_a}.

\item \textbf{Case II: $L>0$:} Conversely, in this scenario, the second term is positive and counteracts the inward-directed  magnetic force term proportional to $F^2$.  This opposition reduces the magnitude of the radial velocity, $v_r$, which in turn increases the magnitude of the sound speed, $c_s$. As a result, the sonic condition ($v_r = c_s$) is satisfied at a smaller radius, shifting the horizon inward. Because the horizon now forms in a region with steeper velocity gradients, the Hawking temperature rises with the addition of the magnetic field as higher magnetic field pushes the  horizon inwards. Furthermore, for a fixed magnetic field strength, increasing the positive value of $L$ shifts the horizon even further inward, leading to a correspondingly higher Hawking temperature, as illustrated in Fig.~\ref{fig:th_b}.

\end{itemize}

\subsection{Hawking temperature in  advective accretion  flow }
Having established the effective magnetoacoustic metric, constructed both analytical and numerical background flow models, and demonstrated  the influence of magnetic fields on the analog Hawking temperature within the analytical framework, we now turn to a more realistic investigation based on the numerical advective accretion flow solutions. Although the modified analytical ADIOS framework successfully breaks the exact self-similarity of the standard self-similar solution through a perturbative magnetic field, the underlying unperturbed background flow remains fundamentally self-similar. Consequently, the perturbation produces only a limited departure from the constant-Mach-number behavior characteristic of self-similar solutions.

 The numerical advective accretion flow  solutions obtained by integrating through the critical point naturally provide  realistic transonic radial profiles of the velocity, density, sound speed, and magnetic field.  In this section, we therefore compute the analog Hawking temperature using the numerical advective accretion flow solutions and demonstrate how the magnetic field affects the analog Hawking temperature.

To investigate the influence of magnetic fields on the analog Hawking temperature, we vary the magnetization parameter ($f_A$), introduced in Eq.~\eqref{Bcrit}. Throughout this analysis, $f_A$ is varied over the range $5\leq f_A\leq1000$, corresponding to progressively weaker background magnetic fields. In addition, we investigate the dependence of the analog Hawking temperature on the viscosity parameter ($\alpha$), considering $0\leq\alpha\leq0.1$ while ensuring that the  condition
$\sigma \ll1$
remains satisfied in every calculation.

\begin{figure*}[t]
\centering

\begin{subfigure}{0.48\textwidth}
    \centering
    \includegraphics[width=\linewidth]{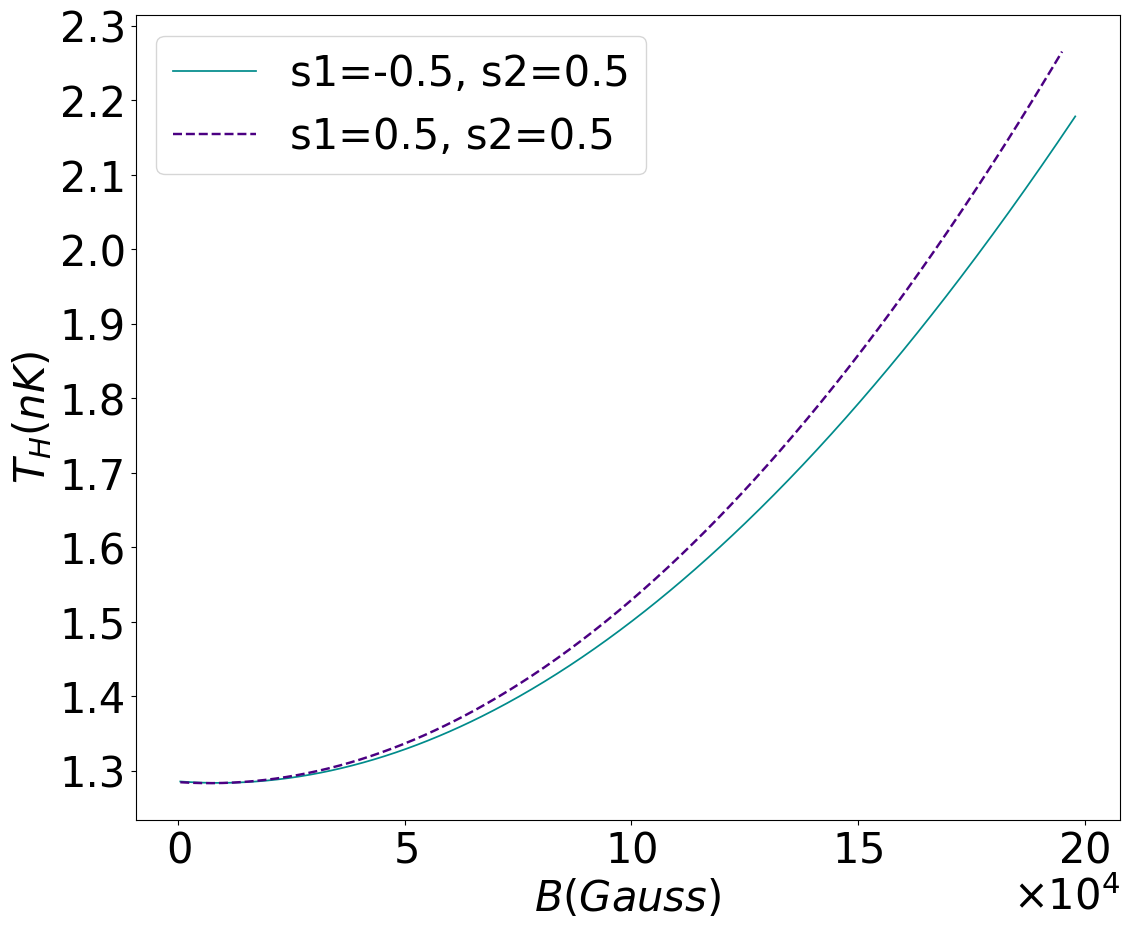}
    \caption{}
    \label{fig:numerical_inc}
\end{subfigure}
\hfill
\begin{subfigure}{0.48\textwidth}
    \centering
    \includegraphics[width=\linewidth]{{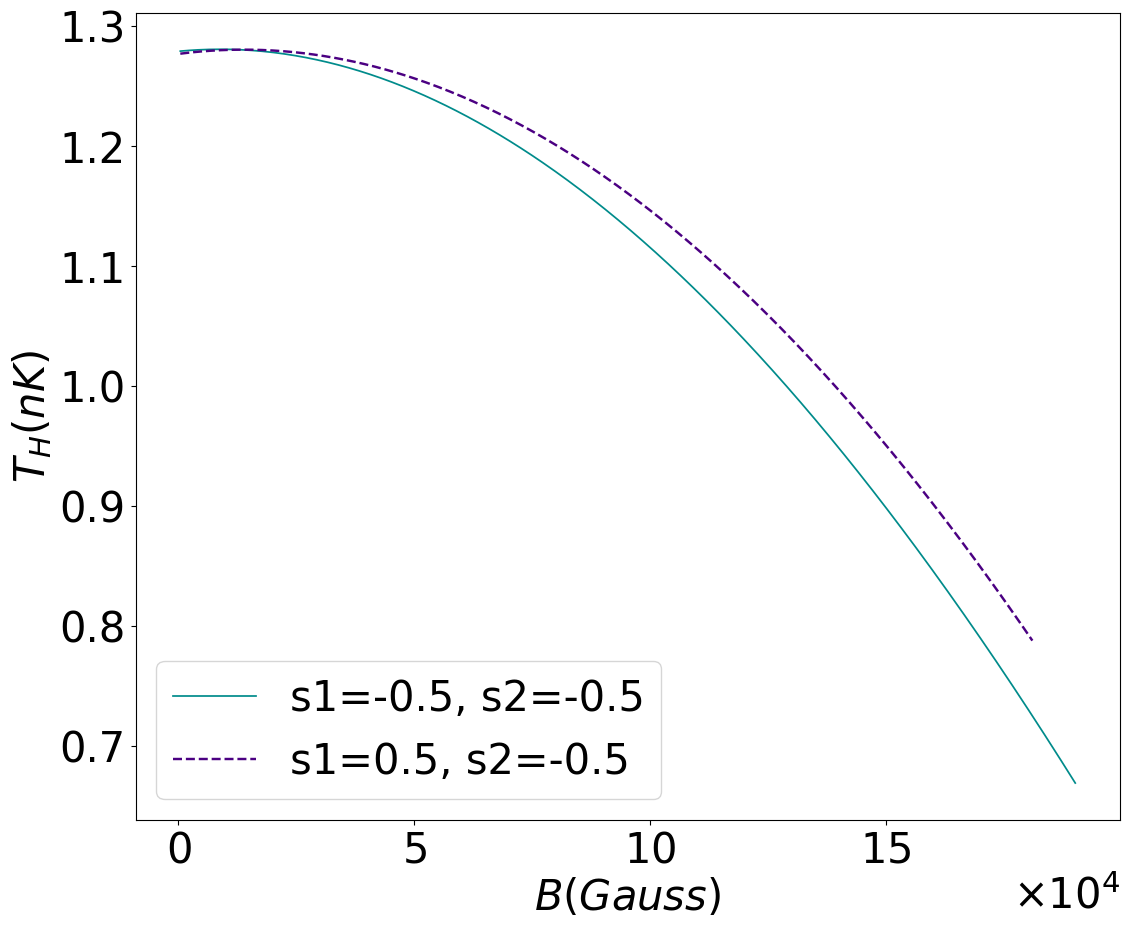}}
    \caption{}
    \label{fig:numerical_dec}
\end{subfigure}
\caption{\raggedright Variation of the analog Hawking temperature $T_H$ with the horizon magnetic field strength $B$ for different magnetic-field topologies. The calculations are performed for $a=0.998$, $\alpha = 0.01$, $\dot{M} = 0.01\dot{M}_{\rm Edd}$, and $M = 10M_\odot$, with the magnetic field components at the critical point fixed as $B_{xc} = B_{\phi c} = B_{zc}$. (a) For a positive azimuthal magnetic-field gradient ($s_2 = 0.5$), $T_H$ monotonically increases with $B$. (b) For a negative azimuthal gradient ($s_2 = -0.5$), $T_H$ monotonically decreases with $B$. In both panels, the curves correspond to $s_1 = \pm0.5$, demonstrating that the radial gradient parameter $s_1$ exerts a secondary influence on the temperature compared to $s_2$.}
\label{fig:numericalinc_dec}
\end{figure*}

Figure \ref{fig:numericalinc_dec} shows the variation of the analog Hawking temperature, $T_H$, with the magnetic field strength, $B$, evaluated at the horizon. 
The dependence of the Hawking temperature on the magnetic-field topology is characterized by the parameters $s_1$ and $s_2$, introduced in Eqs.~\eqref{rad_num_ns} and \eqref{azimuth_num_ns}. These parameters account for the vertical gradients of the radial and azimuthal magnetic-field components, respectively, within the numerical model. For Fig.~\ref{fig:numericalinc_dec}, the magnetic field components at the critical point are prescribed as
\begin{equation}\label{positiveBz}
B_{xc}=B_{\phi c}=B_{zc}=\sqrt{4\pi\rho_c}\frac{c_{sc}}{f_A\sqrt{3}}.
\end{equation}

Figure~\ref{fig:numerical_inc} corresponds to the case $s_2>0$, while Fig.~\ref{fig:numerical_dec} corresponds to $s_2<0$. For positive values of $s_2$, the analog Hawking temperature increases monotonically with increasing magnetic field strength for both choices of $s_1$. In contrast, when $s_2<0$, the Hawking temperature decreases monotonically as the magnetic field strength increases. In both panels, the curves corresponding to $s_1>0$ lie slightly above those for $s_1<0$. These results show  that the topology of the background magnetic field is an important factor which determines whether the analog Hawking temperature increases or decreases with magnetic field strength.

\begin{figure*}[t]
\centering

\begin{subfigure}{0.48\textwidth}
    \centering
    \includegraphics[width=\linewidth]{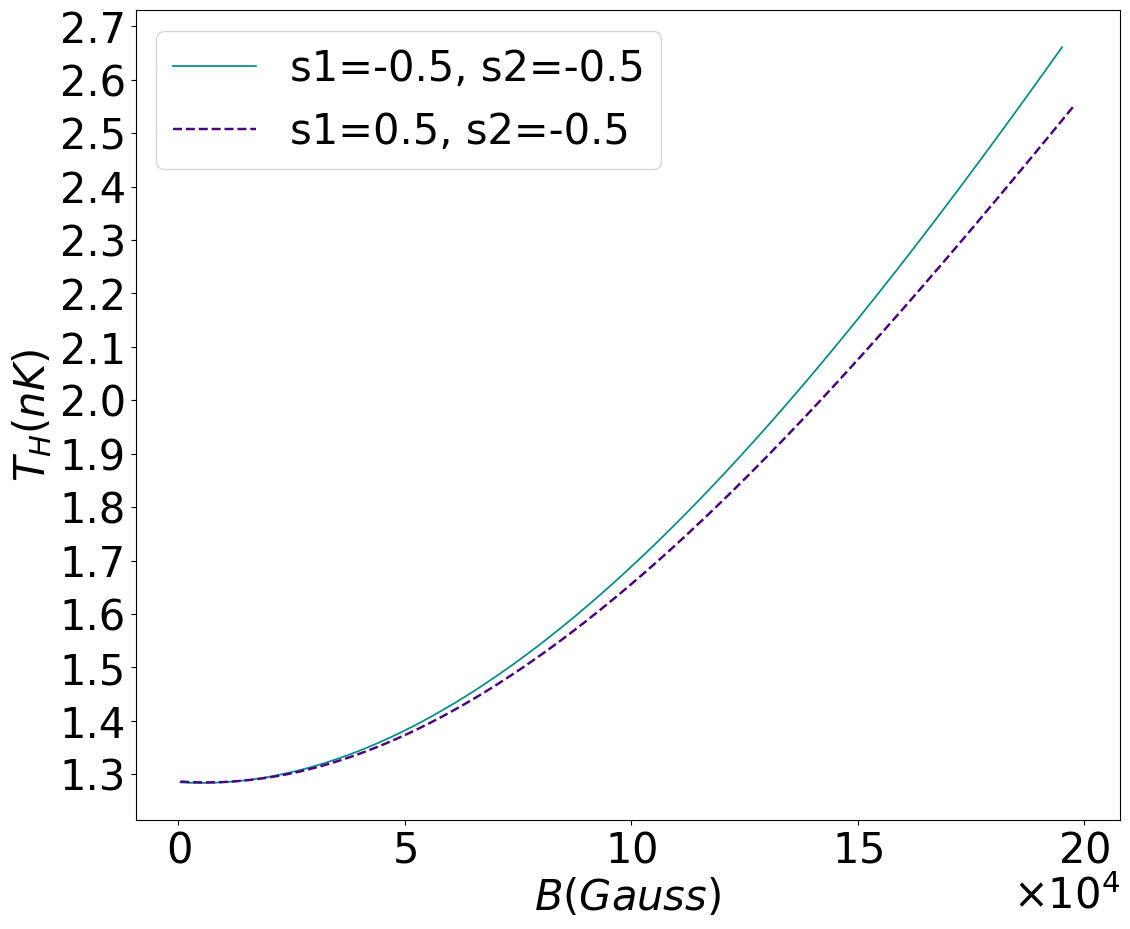}
    \caption{}
    \label{negBzinc}
\end{subfigure}
\hfill
\begin{subfigure}{0.48\textwidth}
    \centering
    \includegraphics[width=\linewidth]{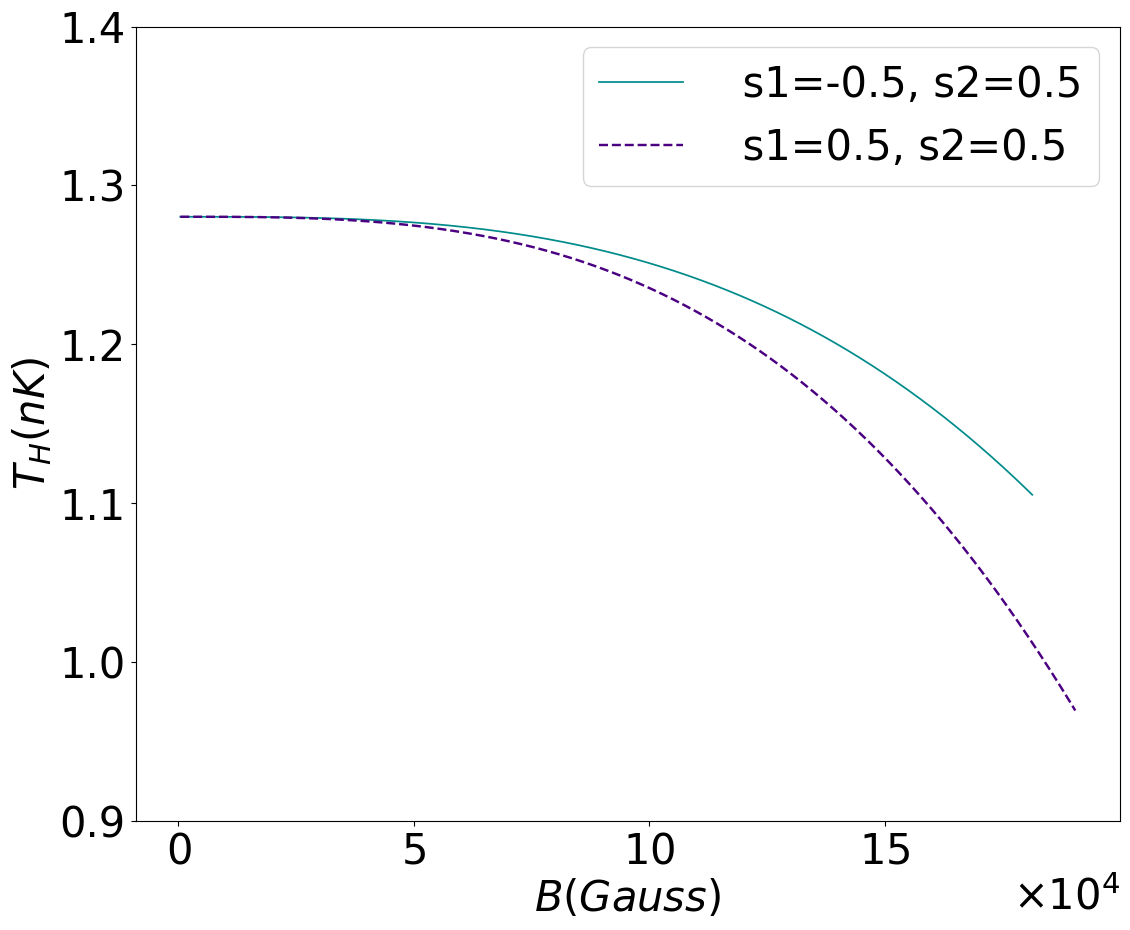}
    \caption{}
    \label{negBzdec}
\end{subfigure}

\caption{\raggedright   Variation of the analog Hawking temperature $T_H$ with the horizon magnetic field $B$ for the reversed vertical magnetic-field configuration, $B_{xc}=B_{\phi c}=-B_{zc}$. The calculations are performed using the same parameters as Fig.~\ref{fig:numericalinc_dec}. (a) For a negative azimuthal gradient ($s_2 = -0.5$), $T_H$ increases monotonically with $B$. (b) For a positive azimuthal gradient ($s_2 = 0.5$), $T_H$ decreases monotonically. Reversing the vertical magnetic field inverts the temperature's dependence on the  gradient topologies ($s_1$ and $s_2$) relative to Fig.~\ref{fig:numericalinc_dec}.}
\label{fig:negBzinc_dec}
\end{figure*}

Figure~\ref{fig:negBzinc_dec} shows the analog Hawking temperature for the alternative magnetic-field configuration in which the magnetic field components at the critical point are prescribed as
\begin{equation}
B_{xc}=B_{\phi c}=-B_{zc}=\sqrt{4\pi\rho_c},\frac{c_{sc}}{f_A\sqrt{3}}.
\end{equation}
Relative to the configuration shown in Fig.~\ref{fig:numericalinc_dec}, the dependence of the Hawking temperature on the magnetic field strength is reversed. For $s_2>0$, the analog Hawking temperature now decreases monotonically with increasing magnetic field strength, whereas for $s_2<0$ it increases monotonically. In addition, the relative ordering of the $s_1$ curves is also reversed: the curves corresponding to $s_1<0$ now lie  above those for $s_1>0$ in both panels. These results indicate that reversing the orientation of the vertical magnetic-field component at the critical point changes the dependence of the analog Hawking temperature on both the magnetic field strength and the magnetic-field topology. Moreover  the trends for positive and negative $s_1$ and $s_2$ are interchanged as compared to the case of Fig.~\ref{fig:numericalinc_dec}.

\begin{figure*}[t]
\centering

\begin{subfigure}{0.48\textwidth}
    \centering
    \includegraphics[width=\linewidth]{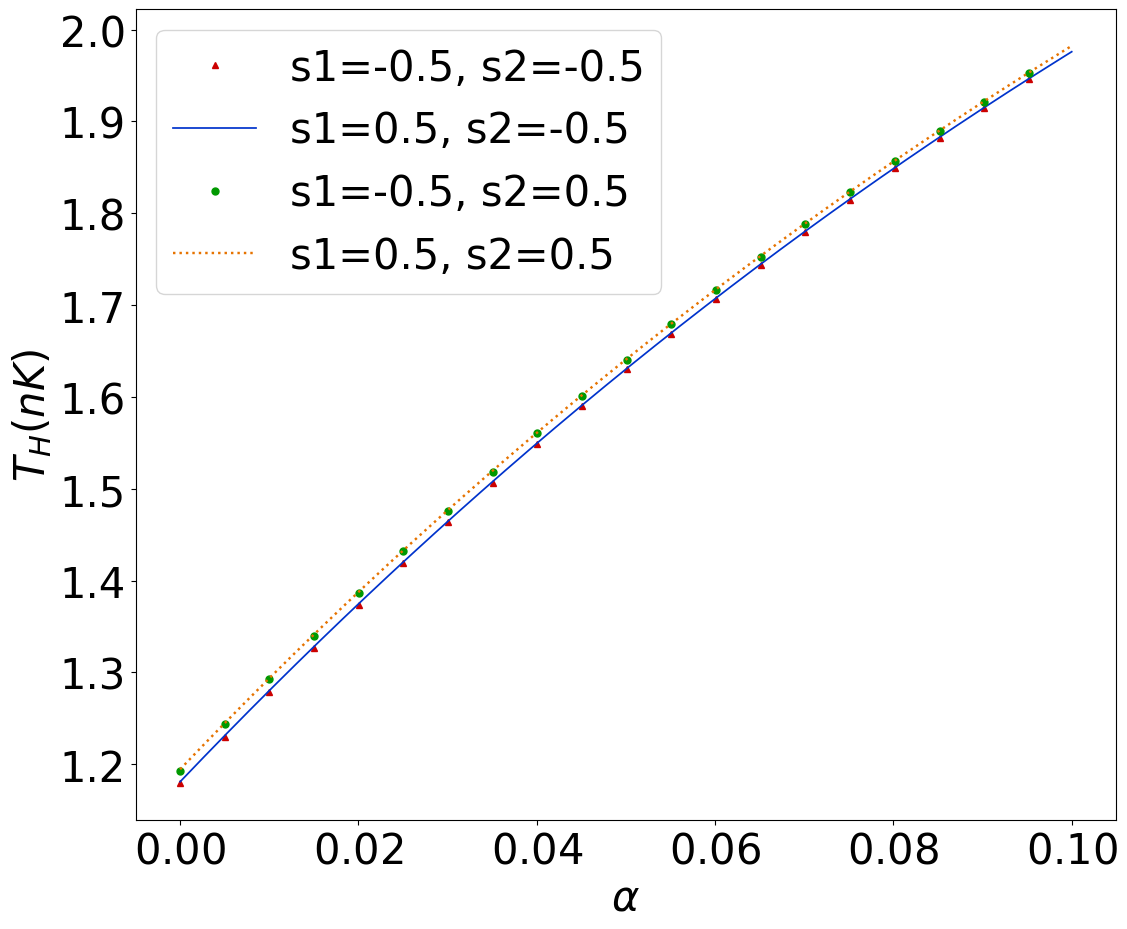}
    \caption{}
    \label{alphaTH}
\end{subfigure}
\hfill
\begin{subfigure}{0.48\textwidth}
    \centering
    \includegraphics[width=\linewidth]{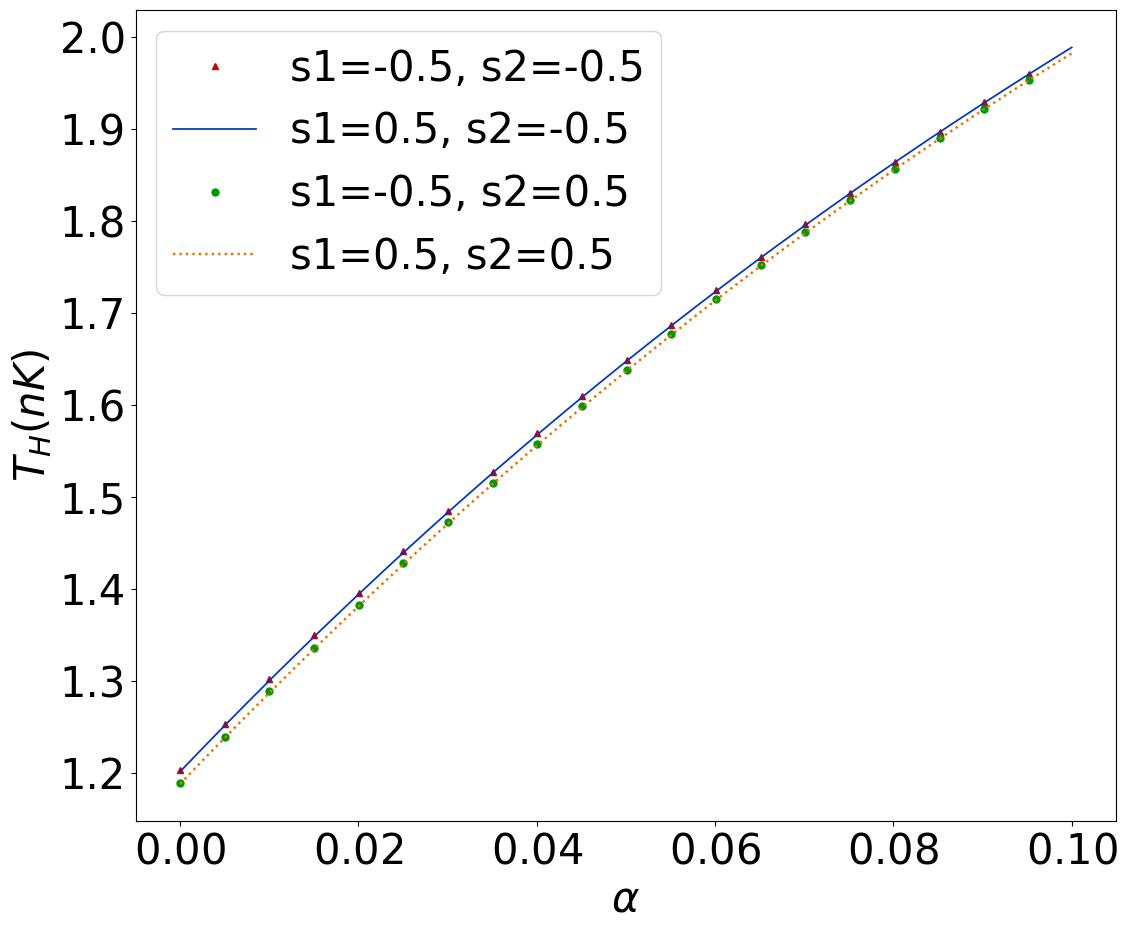}
    \caption{}
    \label{alphaBzneg}
\end{subfigure}

\caption{\raggedright Analog Hawking temperature, $T_H$, as a function of the viscosity parameter, $\alpha$, for different magnetic-field topologies. The magnetization parameter is fixed at $f_A = 20$ to keep the net magnetic field at the horizon approximately constant at $B(r_H) \approx 2 \times 10^4$~G across the sampled range of $\alpha$ for an accretion flow with $a = 0.998$, $M = 10M_\odot$, and $\dot{M} = 0.01\dot{M}_{\rm Edd}$. (a) Positive vertical magnetic-field configuration, $B_{xc} = B_{\phi c} = B_{zc}$. (b) Reversed vertical magnetic-field configuration, $B_{xc} = B_{\phi c} = -B_{zc}$. The curves correspond to the four combinations of the magnetic-field gradient parameters $s_1 = \pm0.5$ and $s_2 = \pm0.5$.}
\label{alphaplots}
\end{figure*}

Figure \ref{alphaplots} illustrates the variation of the analog Hawking temperature, $T_H$, with respect to the viscosity parameter, $\alpha$. The magnetization parameter is held fixed at $f_A = 20$, which ensures that the magnetic field strength at the analog horizon remains approximately constant across all sampled values of $\alpha$. Panel (a) presents the configuration where the magnetic field components at the critical point are $B_{xc} = B_{\phi c} = B_{zc}$, while panel (b) depicts the reversed vertical field scenario, $B_{xc} = B_{\phi c} = -B_{zc}$.

Several distinct trends emerge from these configurations. Most notably, the analog Hawking temperature exhibits a strong, monotonic increase as the viscosity parameter $\alpha$ grows. Furthermore, the curves cluster  based on the sign of the azimuthal gradient parameter, $s_2$, with variations in $s_1$ producing negligible splitting because we are studying the case of low magnetic field. 

Additionally, a clear topological symmetry is observed: in panel (a), configurations with $s_2 > 0$ produce  higher Hawking temperatures at low viscosity, whereas in panel (b), this behavior perfectly inverts, with $s_2 < 0$ producing the higher temperatures. This inversion numerically confirms the $B_z \rightarrow -B_z$ symmetry just as observed in Figs. \ref{fig:numericalinc_dec} and \ref{fig:negBzinc_dec}. Finally, as $\alpha$ approaches $0.1$, the temperature gap between different magnetic topologies visibly narrows, suggesting that viscous effects begin to dominate  differences produced by the vertical magnetic field gradients.

\subsubsection*{Explanation of the trends}
Just like the analytical case, we work in the limit of 
\begin{equation}
    \eta \ll 1, \quad
    \sigma \ll 1, \quad c_s \gg V_A.
\end{equation}
To  understand the physical origin of the trends we have to understand how  the dominant contribution to $T_H$  (i.e. $|\partial_r(v_r^2- c_s^2)|$) will
change with  the magnetic field. In the analytical ADIOS framework, the perturbative magnetic field is responsible for breaking self-similarity, thereby generating the sonic point and consequently largely determining the horizon location. In contrast, in the numerical framework, the non-magnetized advective flow is naturally transonic. Thus, introducing a small magnetic field does not significantly alter the locations of the sonic point and the analog horizon in the numerical framework.

We begin by  explaining Fig.~\ref{fig:numericalinc_dec} where the magnetic field components at the critical point are given by Eq.~\eqref{positiveBz}.
For $s_2>0$, the magnetic stress term $s_2B_zB_\phi/h$, 
appearing in the azimuthal momentum equation, Eq.~\eqref{azimuth_num_ns}, acts to enhance the outward transport of angular momentum. Consequently, the specific angular momentum of the accreting matter decreases more rapidly as it approaches the black hole. This weakens the centrifugal resistance provided by the term $\lambda^2/x^3$ in the radial momentum equation, Eq.~\eqref{rad_num_ns}, allowing the flow to accelerate more efficiently toward the horizon. As a result, the radial velocity profile becomes steeper at the  horizon, leading to an increase in the magnitude of  $-\partial_r(v_r^2)$ as can be seen from Eq.~\eqref{rad_num_ns}. 
Whereas for the case of $s_2<0$
, the $s_2 B_z B_\phi/h$ term becomes negative, which partially offsets the other magnetic torque components and hence the  angular momentum is removed less efficiently than the $s_2>0$ case, leaving a stronger residual centrifugal force to cushion the infall which  smooths out the radial acceleration, making the velocity gradient less steep at the  horizon. So the term $-\partial_rv_r^2$ does not increase much with the magnetic field unlike the $s_2>0$ case.

The influence of $s_1$ is qualitatively different. Rather
than modifying the angular momentum transport, $s_1$
enters directly through the magnetic tension term
$s_1B_zB_x/h$ in the radial momentum equation, Eq.~\eqref{rad_num_ns}. Flipping
the sign of $s_1$ therefore changes the radial force balance
locally. For $s_1<0$, the magnetic tension provides an
additional inward pull, leading to a slightly steeper
velocity gradient, i.e. it makes $\partial_r v_r^2$ more negative, whereas for $s_1>0$ it partially
counteracts the inward acceleration. The resulting
variation in $-\partial_r(v_r^2)$ due to the sign flip of $s_1$ is comparatively much  smaller than the case of sign flip of $s_2$.
This is because the $s_1$ term acts only as a linear
correction to the radial momentum balance, whereas $s_2$
modifies the global angular momentum distribution. Since
the centrifugal term depends quadratically on the
specific angular momentum through the $\lambda^2/x^3$
term, even a modest change in $\lambda$ produces a much
stronger  response than the  change due to  local magnetic tension.

 It is easy to see that $\partial_r c_s^2$ will be negative in sign and lesser in magnitude than $\partial_r v_r^2$ near the horizon and sonic point (because of the  transonic property).

The behavior of the acoustic contribution can be understood from the definition of the sound speed, and we obtain:
\begin{equation}\label{dcs2}
\frac{dc_s^2}{dr} = c_s^2 \left( \frac{d\ln P}{dr} - \frac{d\ln\rho}{dr} \right).
\end{equation}
Thus, the acoustic contribution is determined by the relative radial gradients of the pressure and density.

To analyze the behavior of the acoustic contribution, $\partial_r c_s^2$, we begin with the exact expression for the  sound speed gradient. By using Eqs.~\eqref{Q+Q-}, \eqref{dcs2} and the logarithmic derivative of Eq.~\eqref{dotM} and using $h = (\sqrt{P/\rho})/{\Omega_K}$, we obtain:
\begin{equation}
\begin{split}
\left(\frac{\Gamma_1+1}{2}\right)\frac{dc_s^2}{dr} 
&= -(\Gamma_1-1)c_s^2 \left( \frac{2}{5r} + \frac{d\ln v_r}{dr} \right) \\
&\quad + \frac{\Gamma_1(\Gamma_3-1)(Q^{+}-Q^{-})}{\rho\vartheta}.
\end{split}
\label{mastereq:dcs2}
\end{equation}

Now, the leading deciding factor is the term $ d\ln v_r/dr$ which decides whether, with increasing magnetic field $dc_s^2/dr $ increases or decreases.
Now, the Eq.~\eqref{rad_num_ns}, can be recast in the following form:
\begin{equation}
\frac{d\ln v_r}{dr} = \frac{
  \splitfrac{-F + \frac{\lambda^2}{x^3} - \frac{1}{\rho}\frac{dP}{dx}}
  {+ \frac{1}{4\pi\rho} \left( B_x\frac{dB_x}{dx} - \frac{B_\phi^2}{x} + s_1\frac{B_zB_x}{h} \right)}
}{v_r^2}.
\label{eq:logv}
\end{equation}

The influence of $s_2$ enters the numerator indirectly but powerfully through the angular momentum profile. Because the centrifugal support scales quadratically with specific angular momentum ($\lambda^2/x^3$), variations in $s_2$ produce a highly pronounced effect. For $s_2 > 0$, the more efficient outward transport of angular momentum reduces $\lambda$. This reduction quadratically decreases $\lambda^2$, thereby significantly increasing the negativity of the numerator. This quadratic reduction in the centrifugal barrier easily dominates the growth of the denominator, which is governed by \(v_r^2\) at the horizon. As a result, the entire fraction in the logarithmic velocity gradient attains more negative values.
 The opposite holds true for $s_2 < 0$, where angular momentum is retained, and the stronger centrifugal barrier prevents the gradient from becoming as negative and hence the growth in $v_r^2$ overpowers and  makes the fraction less negative. Consequently, because the acoustic contribution $\partial_r c_s^2$ depends directly on $-\partial_r \ln v_r$, it exhibits a pronounced increase for $s_2 > 0$ and a corresponding decrease for $s_2 < 0$.

The role of $s_1$, however, is qualitatively different because it only modifies the numerator linearly through the magnetic tension term ($s_1 B_z B_x / h$). This linear modification cannot overtake the  growth of the $v_r^2$ denominator at the horizon. Consider the $s_2 > 0$ configuration: when $s_1 < 0$, the magnetic tension provides an additional inward pull, making the numerator more negative than in the $s_1 > 0$ case. However, this extra inward acceleration causes $v_r$ (and thus the denominator $v_r^2$) to rise much more rapidly with the magnetic field. This aggressive growth in the denominator suppresses the overall magnitude of the fraction, causing the $-\partial_r \ln v_r$ profile to dip more slowly for $s_1 < 0$. Conversely, for $s_1 > 0$, the denominator does not grow as aggressively, which allows the numerator to dictate the fraction's behavior and causes $-\partial_r \ln v_r$ to rise  more strongly. Because the acoustic contribution, $\partial_r c_s^2$, is heavily dependent on this logarithmic velocity gradient, it consequently rises faster for the $s_1 > 0$ configuration. Similar competitive mechanics govern the $s_2 < 0$ case: while the acoustic contribution generally decreases with the magnetic field in this regime, it dips slower for $s_1 > 0$ than for $s_1 < 0$, as the interplay between the linear tension term and the denominator's growth continues to govern $\partial_rc_s^2$ term.

The term $Q^+ - Q^-$ is influenced by the magnetic field mostly through the 
$Q_{\rm mag}^{+} -Q_{\rm mag}^{-}$ part  which increases directly with magnetic field. Recall from Eq.~\eqref{Qmag} that:
\begin{equation}
Q_{\rm mag}^{+}
-
Q_{\rm mag}^{-}
=
\frac{f_m |B|^2 \vartheta}{16\pi r^3},
\end{equation}

where $f_m$ denotes the magnetic dissipation efficiency.
Thus, the heating term always contributes positively and increases monotonically with magnetic field strength. 
However this term turns out to vary by much lesser magnitude than $ d\ln v_r/dr$  in the case of low magnetic field, hence, the major deciding factor of the trend, whether $T_H$ will increase or decrease with magnetic field is decided by $ d\ln v_r/dr$.

Hence, for the $s_2>0$ configuration, the  contribution, $-\partial_rv_r^2$, increases rapidly with increasing magnetic field strength. Simultaneously, the acoustic contribution, $\partial_rc_s^2$, becomes progressively less negative, thereby reinforcing the increase in the total surface-gravity proxy.
\begin{equation}
-\partial_r\left(v_r^2-c_s^2\right)
=-\partial_rv_r^2+\partial_rc_s^2.
\end{equation}
Consequently, the analog Hawking temperature increases monotonically with magnetic field strength.

In contrast, for $s_2<0$, the increase in the $-\partial_rv_r^2$ contribution is comparatively weaker, while the acoustic contribution becomes increasingly negative with increasing magnetic field strength. The enhanced negative contribution from $\partial_rc_s^2$ outweighs the modest increase in $-\partial_rv_r^2$, resulting in a monotonic decrease of the analog Hawking temperature with magnetic field strength.

From Fig.~\ref{fig:numericalinc_dec}, we observed that changing the sign of $s_1$ produces a much weaker response than changing the sign of $s_2$. For $s_2>0$,  $-\partial_r v_r^2$, increases slightly more for $s_1<0$ than for $s_1>0$. While $\partial_r c_s^2$ becomes progressively less negative with increasing magnetic field for both cases,  the increase is considerably stronger for  $s_1>0$ than for $s_1<0$. Consequently, the analog Hawking temperature rises more rapidly for $s_1>0$.

A similar trend is observed for $s_2<0$. Although  $-\partial_r v_r^2$, increases slightly more for $s_1<0$, the dominant difference arises from the acoustic term. In this case, $\partial_r c_s^2$ becomes increasingly negative with increasing magnetic field, and this decrease is significantly stronger for $s_1<0$. Thus,  we obtain higher values of $T_H$ for the $s_1>0$ configuration.

The reversal of the observed trends in Fig.~\ref{fig:negBzinc_dec}, as compared to Fig.~\ref{fig:numericalinc_dec}, can be understood directly from the symmetry of the governing momentum equations. In the radial and azimuthal momentum balance equations, the vertical magnetic field gradients, $s_1$ and $s_2$, appear only through the magnetic tension terms, $s_1 B_z B_x/h$ and $s_2 B_z B_\phi/h$, respectively. Reversing the direction of the vertical magnetic field ($B_z \rightarrow -B_z$) simply changes the sign of these terms, making them $-s_1 B_z B_x/h$ and $-s_2 B_z B_\phi/h$. This is mathematically equivalent to keeping $B_z$ positive while reversing the signs of the gradient parameters ($s_1 \rightarrow -s_1$ and $s_2 \rightarrow -s_2$).

The same symmetry is preserved in the vertical magnetostatic balance equation, Eq.~\eqref{vert_balnce}. Here, the contribution involving $s_3$ depends on $B_z^2/h$, so it is unaffected by a reversal of the vertical magnetic field because it depends on the square of $B_z$. As a result, the full set of governing equations remains invariant under the transformation $B_z \rightarrow -B_z$ together with $s_1 \rightarrow -s_1$ and $s_2 \rightarrow -s_2$.

This symmetry explains why the flow dynamics in the negative-$B_z$ configuration are identical to those obtained with a positive $B_z$ and opposite signs of the magnetic field gradients.

It is important to note that although the trends are completely reversed, the curves for the negative-$B_z$ configuration do not perfectly overlap with the corresponding positive-$B_z$ curves that have opposite $s_1$ and $s_2$ signs. While the governing equations are invariant under the transformation $B_z \rightarrow -B_z$, $s_1 \rightarrow -s_1$, and $s_2 \rightarrow -s_2$, this discrepancy occurs because we can only enforce the reversed magnetic field component, $B_z$, at the critical point. As we numerically integrate the flow equations outward, we cannot guarantee that the field maintains exactly equal magnitude and opposite sign throughout the entire spatial extent of the disk.

The trends observed in Fig.~\ref{alphaplots} can be readily understood by examining the angular momentum transport, Eq.~\eqref{azimuth_num_ns}. As the viscosity parameter $\alpha$ increases, angular momentum is transported outward more efficiently. Similar to the mechanism described earlier for the $s_2 > 0$ configuration of Fig.~\ref{fig:numerical_inc}, this enhanced transport rapidly reduces the specific angular momentum of the inflow. Consequently, the centrifugal barrier in the radial momentum equation, Eq.~\eqref{rad_num_ns}, is significantly weakened. This weakened barrier allows the flow to accelerate more strongly toward the horizon, which steepens the radial velocity gradient and causes the analog Hawking temperature to rise monotonically with $\alpha$. 

Furthermore, the Hawking temperature gap between the $s_2 > 0$ and $s_2 < 0$ configurations noticeably reduces at higher values of $\alpha$. This convergence occurs because the substantial angular momentum removed by the elevated viscous shear ($\alpha$) heavily dominates the relatively minor torque variations introduced by flipping the sign of the magnetic azimuthal gradient parameter, $s_2$ at low magnetic fields.

\section{Conclusion}
\label{sec:conclusion}
We have discussed the magnetoacoustic geometry defined for wave propagation in three-dimensional MHD flows, extending the standard analog gravity framework to introduce the  effects of shear viscosity. By evaluating first-order wave perturbations under the eikonal approximation, we have found that in the  regimes $\eta \ll 1 $ and $\sigma \ll 1$, an effective Lorentzian metric can be successfully constructed exclusively for the fast magnetoacoustic mode. The slow mode as well as the Alfvén mode remain mathematically degenerate and do not admit a valid inverse metric tensor.  

To demonstrate the implications of the derived visco-magnetoacoustic geometry, we have considered two complementary background flow models: a numerical 1.5-dimensional magnetized advection-dominated accretion flow  solution (in a lower magnetic field regime of  MA-AAF) and an analytical ADIOS. The numerical MA-AAF 
 model provides a transonic background flow that admits the formation of an analog horizon. In contrast, the exact self-similarity of the standard ADIOS solution leads to a constant radial Mach number, forbidding the existence of a well-defined horizon. To overcome this limitation, we have introduced a steady large-scale magnetic field perturbation that breaks the  self-similarity while preserving the analytical framework, thereby recovering a dynamically formed analog horizon.

We have evaluated the analog Hawking temperature associated with spontaneous phonon emission at the visco-magnetoacoustic horizon. We have found that the Hawking temperature depends not only on the strength of the background magnetic field but also on its topology. In particular, the orientation of the magnetic-field components determines whether the magnetic field enhances or suppresses the temperature by modifying the location of the horizon and the corresponding velocity gradients. 
We have also examined the influence of the Shakura–Sunyaev viscosity parameter on the analog Hawking temperature. An increase in the viscosity parameter leads to a systematic increase in the Hawking temperature. This demonstrates that, in addition to magnetic-field topology, viscous angular momentum transport plays an important role in determining the thermodynamic properties of the effective spacetime.

As a natural extension of this work, future investigations will focus on superradiant wave scattering within the visco-magnetoacoustic ergoregion. While the  definitions of the horizon ($M^{rr} = 0$) and ergosurface ($M_{tt} = 0$) are sufficient for the thermodynamic analysis presented here, the study of superradiant scattering requires reducing the wave equation to a Schr\"odinger-like form. To facilitate such an analysis, future work will implement a generalized coordinate transformation that casts the effective metric into a Kerr-like form. 

The maximum Hawking temperature detected to date in classical fluid analogs  is of the order of $10^{-12}$ K \cite{PhysRevLett.106.021302}. Therefore, if a laboratory plasma system could be realized with an $\alpha$-viscosity and a magnetic field of order $10^4$ G, it could significantly enhance the Hawking temperature, making experimental detection more feasible.

\section{Acknowledgment}
AS acknowledges support from the Kishore Vaigyanik
Protsahan Yojana (KVPY), Department of Science and
Technology (DST), Government of India.
MP acknowledges the Prime
Minister’s Research Fellows (PMRF) scheme for providing  fellowship. BM acknowledges a partial support from the project funded by SERB/ANRF, India, with Ref. No. CRG/2022/003460.

\bibliography{references}

@ARTICLE{2017PhRvD..95j4055N,
       author = {{Noda}, Sousuke and {Nambu}, Yasusada and {Takahashi}, Masaaki},
        title = "{Analog rotating black holes in a magnetohydrodynamic inflow}",
      journal = {Physical Review D},
         year = 2017,
        month = may,
       volume = {95},
       number = {10},
          eid = {104055},
        pages = {104055},
          doi = {10.1103/PhysRevD.95.104055},
archivePrefix = {arXiv},
       eprint = {1610.06690},
 primaryClass = {gr-qc},
       adsurl = {https://ui.adsabs.harvard.edu/abs/2017PhRvD..95j4055N}
}

@ARTICLE{1999MNRAS.303L...1B,
       author = {{Blandford}, Roger D. and {Begelman}, Mitchell C.},
        title = "{On the fate of gas accreting at a low rate on to a black hole}",
     journal = {Monthly Notices of the Royal Astronomical Society},
         year = 1999,
        month = feb,
       volume = {303},
       number = {1},
        pages = {L1-L5},
          doi = {10.1046/j.1365-8711.1999.02358.x},
archivePrefix = {arXiv},
       eprint = {astro-ph/9809083},
 primaryClass = {astro-ph},
       adsurl = {https://ui.adsabs.harvard.edu/abs/1999MNRAS.303L...1B}
}

@ARTICLE{narayan1994,
       author = {{Narayan}, Ramesh and {Yi}, Insu},
        title = "{Advection-dominated Accretion: A Self-similar Solution}",
      journal = {The Astrophysical Journal Letters},
         year = 1994,
        month = jun,
       volume = {428},
        pages = {L13},
          doi = {10.1086/187381},
archivePrefix = {arXiv},
       eprint = {astro-ph/9403052},
 primaryClass = {astro-ph},
       adsurl = {https://ui.adsabs.harvard.edu/abs/1994ApJ...428L..13N}
}

@article{PhysRevLett.46.1351,
  title = {Experimental Black-Hole Evaporation?},
  author = {Unruh, W. G.},
  journal = {Phys. Rev. Lett.},
  volume = {46},
  issue = {21},
  pages = {1351--1353},
  numpages = {0},
  year = {1981},
  month = {May},
  publisher = {American Physical Society},
  doi = {10.1103/PhysRevLett.46.1351},
  url = {https://link.aps.org/doi/10.1103/PhysRevLett.46.1351}
}

@ARTICLE{1995ApJ...452..710N,
       author = {{Narayan}, Ramesh and {Yi}, Insu},
        title = "{Advection-dominated Accretion: Underfed Black Holes and Neutron Stars}",
      journal = {\apj},
         year = 1995,
        month = oct,
       volume = {452},
        pages = {710},
          doi = {10.1086/176343},
archivePrefix = {arXiv},
       eprint = {astro-ph/9411059},
 primaryClass = {astro-ph},
       adsurl = {https://ui.adsabs.harvard.edu/abs/1995ApJ...452..710N}
}

@ARTICLE{Abramowicz,
       author = {{Abramowicz}, Marek A. and {Fragile}, P. Chris},
        title = "{Foundations of Black Hole Accretion Disk Theory}",
      journal = {Living Reviews in Relativity},
         year = 2013,
        month = dec,
       volume = {16},
       number = {1},
          eid = {1},
        pages = {1},
          doi = {10.12942/lrr-2013-1},
archivePrefix = {arXiv},
       eprint = {1104.5499},
 primaryClass = {astro-ph.HE},
       adsurl = {https://ui.adsabs.harvard.edu/abs/2013LRR....16....1A}
}

@article{Blandford:1982xxl,
    author = "Blandford, R. D. and Payne, D. G.",
    title = "{Hydromagnetic flows from accretion discs and the production of radio jets}",
    doi = "10.1093/mnras/199.4.883",
    journal = "Mon. Not. Roy. Astron. Soc.",
    volume = "199",
    number = "4",
    pages = "883--903",
    year = "1982"
}

@ARTICLE{2015ApJ...807...43M,
       author = {{Mukhopadhyay}, Banibrata and {Chatterjee}, Koushik},
        title = "{Hydromagnetics of Advective Accretion Flows around Black Holes: Removal of Angular Momentum by Large-scale Magnetic Stresses}",
      journal = {\apj},
         year = 2015,
        month = jul,
       volume = {807},
       number = {1},
          eid = {43},
        pages = {43},
          doi = {10.1088/0004-637X/807/1/43},
archivePrefix = {arXiv},
       eprint = {1505.01281},
 primaryClass = {astro-ph.HE},
       adsurl = {https://ui.adsabs.harvard.edu/abs/2015ApJ...807...43M}
}

@misc{asenjo2011analogueblackholemagnetohydrodynamics,
      title={Analogue black hole in magnetohydrodynamics}, 
      author={Felipe A. Asenjo and Nelson Zamorano},
      year={2011},
      eprint={1102.2625},
      archivePrefix={arXiv},
      primaryClass={gr-qc},
      url={https://arxiv.org/abs/1102.2625}, 
}

@ARTICLE{2018EPJC...78..662G,
       author = {{Gheibi}, A. and {Safari}, H. and {Innes}, D.~E.},
        title = "{Magnetoacoustic and Alfv{\'e}nic black holes}",
      journal = {European Physical Journal C},
         year = 2018,
        month = aug,
       volume = {78},
       number = {8},
          eid = {662},
        pages = {662},
          doi = {10.1140/epjc/s10052-018-6109-1},
       adsurl = {https://ui.adsabs.harvard.edu/abs/2018EPJC...78..662G}
}

@book{335f2344-ccce-3243-bd16-64ce104c6ba0,
 ISBN = {9780691120737},
 URL = {http://www.jstor.org/stable/j.ctvzsmf0w},
  author = {{Kulsrud}, Russell M.},
 publisher = {Princeton University Press},
 title = {Plasma Physics for Astrophysics},
 urldate = {2026-07-10},
 year = {2005}
}

@ARTICLE{1973A&A....24..337S,
       author = {{Shakura}, N.~I. and {Sunyaev}, R.~A.},
        title = "{Black holes in binary systems. Observational appearance.}",
      journal = {Astronomy \& Astrophysics},
         year = 1973,
        month = jan,
       volume = {24},
        pages = {337--355},
       adsurl = {https://ui.adsabs.harvard.edu/abs/1973A&A....24..337S}
}

@book{Bender:1999box,
    author = "Bender, Carl M. and Orszag, Steven A.",
    title = "{Advanced Mathematical Methods for Scientists and Engineers I}",
    doi = "10.1007/978-1-4757-3069-2",
    publisher = "Springer",
    year = "1999"
}

@book{Goedbloed_Keppens_Poedts_2010, place={Cambridge}, title={Advanced Magnetohydrodynamics: With Applications to Laboratory and Astrophysical Plasmas}, publisher={Cambridge University Press}, author={Goedbloed, J. P. and Keppens, Rony and Poedts, Stefaan}, year={2010}}

@ARTICLE{2002ApJ...581..427M,
       author = {{Mukhopadhyay}, Banibrata},
        title = "{Description of Pseudo-Newtonian Potential for the Relativistic Accretion Disks around Kerr Black Holes}",
      journal = {\apj},
         year = 2002,
        month = dec,
       volume = {581},
       number = {1},
        pages = {427-430},
          doi = {10.1086/344227},
archivePrefix = {arXiv},
       eprint = {astro-ph/0205475},
 primaryClass = {astro-ph},
       adsurl = {https://ui.adsabs.harvard.edu/abs/2002ApJ...581..427M}
}

@ARTICLE{1998CQGra..15.1767V,
       author = {{Visser}, Matt},
        title = "{Acoustic black holes: horizons, ergospheres and Hawking radiation}",
      journal = {Classical and Quantum Gravity},
         year = 1998,
        month = jun,
       volume = {15},
       number = {6},
        pages = {1767-1791},
          doi = {10.1088/0264-9381/15/6/024},
archivePrefix = {arXiv},
       eprint = {gr-qc/9712010},
 primaryClass = {gr-qc},
       adsurl = {https://ui.adsabs.harvard.edu/abs/1998CQGra..15.1767V}
}

@ARTICLE{2006Natur.441..727J,
       author = {{Jop}, Pierre and {Forterre}, Yo{\"e}l and {Pouliquen}, Olivier},
        title = "{A constitutive law for dense granular flows}",
      journal = {Nature},
         year = 2006,
        month = jun,
       volume = {441},
       number = {7094},
        pages = {727-730},
          doi = {10.1038/nature04801},
archivePrefix = {arXiv},
       eprint = {cond-mat/0612110},
 primaryClass = {cond-mat.soft},
       adsurl = {https://ui.adsabs.harvard.edu/abs/2006Natur.441..727J}
}

@article{PhysRevD.7.2333,
  title = {Black Holes and Entropy},
  author = {Bekenstein, Jacob D.},
  journal = {Phys. Rev. D},
  volume = {7},
  issue = {8},
  pages = {2333--2346},
  numpages = {0},
  year = {1973},
  month = {Apr},
  publisher = {American Physical Society},
  doi = {10.1103/PhysRevD.7.2333},
  url = {https://link.aps.org/doi/10.1103/PhysRevD.7.2333}
}

@article{Bardeen:1973gs,
    author = "Bardeen, James M. and Carter, B. and Hawking, S. W.",
    title = "{The Four laws of black hole mechanics}",
    doi = "10.1007/BF01645742",
    journal = "Commun. Math. Phys.",
    volume = "31",
    pages = "161--170",
    year = "1973"
}

@book{Wald:1995yp,
    author = "Wald, Robert M.",
    title = "{Quantum Field Theory in Curved Space-Time and Black Hole Thermodynamics}",
    isbn = "978-0-226-87027-4",
    publisher = "University of Chicago Press",
    address = "Chicago, IL",
    series = "Chicago Lectures in Physics",
    year = "1995"
}

@ARTICLE{1974Natur.248...30H,
       author = {{Hawking}, S.~W.},
        title = "{Black hole explosions?}",
      journal = {Nature},
         year = 1974,
        month = mar,
       volume = {248},
       number = {5443},
        pages = {30-31},
          doi = {10.1038/248030a0},
       adsurl = {https://ui.adsabs.harvard.edu/abs/1974Natur.248...30H}
}

@article{Hawking:1975vcx,
    author = "Hawking, S. W.",
    editor = "Gibbons, G. W. and Hawking, S. W.",
    title = "{Particle Creation by Black Holes}",
    doi = "10.1007/BF02345020",
    journal = "Commun. Math. Phys.",
    volume = "43",
    pages = "199--220",
    year = "1975",
    note = "[Erratum: Commun.Math.Phys. 46, 206 (1976)]"
}

@ARTICLE{2011LRR....14....3B,
       author = {{Barcel{\'o}}, Carlos and {Liberati}, Stefano and {Visser}, Matt},
        title = "{Analogue Gravity}",
      journal = {Living Reviews in Relativity},
         year = 2011,
        month = may,
       volume = {14},
       number = {1},
          eid = {3},
        pages = {3},
          doi = {10.12942/lrr-2011-3},
       adsurl = {https://ui.adsabs.harvard.edu/abs/2011LRR....14....3B}
}

@article{PhysRevLett.106.021302,
  title = {Measurement of Stimulated Hawking Emission in an Analogue System},
  author = {Weinfurtner, Silke and Tedford, Edmund W. and Penrice, Matthew C. J. and Unruh, William G. and Lawrence, Gregory A.},
  journal = {Phys. Rev. Lett.},
  volume = {106},
  issue = {2},
  pages = {021302},
  numpages = {4},
  year = {2011},
  month = {Jan},
  publisher = {American Physical Society},
  doi = {10.1103/PhysRevLett.106.021302},
  url = {https://link.aps.org/doi/10.1103/PhysRevLett.106.021302}
}

@article{Steinhauer2016,
  title = {Observation of quantum Hawking radiation and its entanglement in an analogue black hole},
  author = {Steinhauer, Jeff},
  journal = {Nature Physics},
  volume = {12},
  number = {10},
  pages = {959--965},
  year = {2016},
  publisher = {Nature Publishing Group},
  doi = {10.1038/nphys3863},
  url = {https://doi.org/10.1038/nphys3863}
}

@ARTICLE{2012MPLA...2750185G,
       author = {{Gonz{\'a}lez-Fern{\'a}ndez}, B. and {Camacho}, A.},
        title = "{Fluid-Gravity Correspondence Under the Presence of Viscosity}",
      journal = {Modern Physics Letters A},
         year = 2012,
        month = oct,
       volume = {27},
       number = {32},
          eid = {1250185},
        pages = {1250185},
          doi = {10.1142/S0217732312501854},
archivePrefix = {arXiv},
       eprint = {1206.4023},
 primaryClass = {gr-qc},
       adsurl = {https://ui.adsabs.harvard.edu/abs/2012MPLA...2750185G}
}

@ARTICLE{2022Univ....8..205P,
       author = {{Pathak}, Mayank and {Majumdar}, Parthasarathi},
        title = "{Towards an Acoustic Geometry in Slightly Viscous Fluids}",
      journal = {Universe},
         year = 2022,
        month = mar,
       volume = {8},
       number = {4},
          eid = {205},
        pages = {205},
          doi = {10.3390/universe8040205},
archivePrefix = {arXiv},
       eprint = {2204.00605},
 primaryClass = {physics.flu-dyn},
       adsurl = {https://ui.adsabs.harvard.edu/abs/2022Univ....8..205P}
}

@ARTICLE{2006CQGra..23.2371A,
       author = {{Abraham}, Hrvoje and {Bili{\'c}}, Neven and {Das}, Tapas K.},
        title = "{Acoustic horizons in axially symmetric relativistic accretion}",
      journal = {Classical and Quantum Gravity},
         year = 2006,
        month = apr,
       volume = {23},
       number = {7},
        pages = {2371-2393},
          doi = {10.1088/0264-9381/23/7/010},
archivePrefix = {arXiv},
       eprint = {gr-qc/0509057},
 primaryClass = {gr-qc},
       adsurl = {https://ui.adsabs.harvard.edu/abs/2006CQGra..23.2371A}
}

@ARTICLE{2011MNRAS.418L..79T,
       author = {{Tchekhovskoy}, Alexander and {Narayan}, Ramesh and {McKinney}, Jonathan C.},
        title = "{Efficient generation of jets from magnetically arrested accretion on a rapidly spinning black hole}",
      journal = {Monthly Notices of the Royal Astronomical Society},
         year = 2011,
        month = nov,
       volume = {418},
       number = {1},
        pages = {L79-L83},
          doi = {10.1111/j.1745-3933.2011.01147.x},
archivePrefix = {arXiv},
       eprint = {1108.0412},
 primaryClass = {astro-ph.HE},
       adsurl = {https://ui.adsabs.harvard.edu/abs/2011MNRAS.418L..79T}
}

@ARTICLE{1991ApJ...376..214B,
       author = {{Balbus}, Steven A. and {Hawley}, John F.},
        title = "{A Powerful Local Shear Instability in Weakly Magnetized Disks. I. Linear Analysis}",
      journal = {The Astrophysical Journal},
         year = 1991,
        month = jul,
       volume = {376},
        pages = {214},
          doi = {10.1086/170270},
       adsurl = {https://ui.adsabs.harvard.edu/abs/1991ApJ...376..214B}
}

@ARTICLE{1991ApJ...376..223H,
       author = {{Hawley}, John F. and {Balbus}, Steven A.},
        title = "{A Powerful Local Shear Instability in Weakly Magnetized Disks. II. Nonlinear Evolution}",
      journal = {\apj},
         year = 1991,
        month = jul,
       volume = {376},
        pages = {223},
          doi = {10.1086/170271},
       adsurl = {https://ui.adsabs.harvard.edu/abs/1991ApJ...376..223H}
}

@ARTICLE{2012MNRAS.423.3083M,
       author = {{McKinney}, Jonathan C. and {Tchekhovskoy}, Alexander and {Blandford}, Roger D.},
        title = "{General relativistic magnetohydrodynamic simulations of magnetically choked accretion flows around black holes}",
    journal = {Monthly Notices of the Royal Astronomical Society},
         year = 2012,
        month = jul,
       volume = {423},
       number = {4},
        pages = {3083-3117},
          doi = {10.1111/j.1365-2966.2012.21074.x},
archivePrefix = {arXiv},
       eprint = {1201.4163},
 primaryClass = {astro-ph.HE},
       adsurl = {https://ui.adsabs.harvard.edu/abs/2012MNRAS.423.3083M}
}

@ARTICLE{2006PASJ...58..469A,
       author = {{Akizuki}, Chizuru and {Fukue}, Jun},
        title = "{Self-Similar Solutions for ADAF with Toroidal Magnetic Fields}",
      journal = {Publications of the Astronomical Society of Japan},
         year = 2006,
        month = apr,
       volume = {58},
        pages = {469-475},
          doi = {10.1093/pasj/58.2.469},
archivePrefix = {arXiv},
       eprint = {astro-ph/0602274},
 primaryClass = {astro-ph},
       adsurl = {https://ui.adsabs.harvard.edu/abs/2006PASJ...58..469A}
}

@ARTICLE{2025ApJ...981..162P,
       author = {{Pathak}, Mayank and {Mukhopadhyay}, Banibrata},
        title = "{Simulating ULXs and Blazars as GRMHD Accretion Flows Around a Black Hole}",
      journal = {\apj},
         year = 2025,
        month = mar,
       volume = {981},
       number = {2},
          eid = {162},
        pages = {162},
          doi = {10.3847/1538-4357/adb286},
archivePrefix = {arXiv},
       eprint = {2502.03538},
 primaryClass = {astro-ph.HE},
       adsurl = {https://ui.adsabs.harvard.edu/abs/2025ApJ...981..162P}
}

@ARTICLE{2022ApJ...941...30C,
       author = {{Chatterjee}, K. and {Narayan}, R.},
        title = "{Flux Eruption Events Drive Angular Momentum Transport in Magnetically Arrested Accretion Flows}",
      journal = {\apj},
         year = 2022,
        month = dec,
       volume = {941},
       number = {1},
          eid = {30},
        pages = {30},
          doi = {10.3847/1538-4357/ac9d97},
archivePrefix = {arXiv},
       eprint = {2210.08045},
 primaryClass = {astro-ph.HE},
       adsurl = {https://ui.adsabs.harvard.edu/abs/2022ApJ...941...30C}
}

@book{Birrell:1982ix,
    author = "Birrell, N. D. and Davies, P. C. W.",
    title = "{Quantum Fields in Curved Space}",
    doi = "10.1017/CBO9780511622632",
    isbn = "978-0-511-62263-2, 978-0-521-27858-4",
    publisher = "Cambridge University Press",
    address = "Cambridge, UK",
    series = "Cambridge Monographs on Mathematical Physics",
    year = "1982"
}

@ARTICLE{2007JCAP...06..009D,
       author = {{Das}, Tapas Kumar and {Bilic}, Neven and {Dasgupta}, Surajit},
        title = "{A black-hole accretion disc as an analogue gravity model}",
     journal = {Journal of Cosmology and Astroparticle Physics},
         year = 2007,
        month = jun,
       volume = {2007},
       number = {6},
          eid = {009},
        pages = {009},
          doi = {10.1088/1475-7516/2007/06/009},
archivePrefix = {arXiv},
       eprint = {astro-ph/0604477},
 primaryClass = {astro-ph},
       adsurl = {https://ui.adsabs.harvard.edu/abs/2007JCAP...06..009D}
}

@article{Pathak2022,
  author  = {Pathak, Mayank},
  title   = {Sonic Black Holes: A Perspective},
  journal = {Resonance},
  year    = {2022},
  volume  = {27},
  number  = {7},
  pages   = {1117--1125},
  issn    = {0973-712X},
  doi     = {10.1007/s12045-022-1408-0},
  url     = {https://doi.org/10.1007/s12045-022-1408-0},
  month   = jul
}

@misc{das2007astrophysicalaccretionanaloguegravity,
      title={Astrophysical Accretion as an Analogue Gravity Phenomena}, 
      author={Tapas Kumar Das},
      year={2007},
      eprint={0704.3618},
      archivePrefix={arXiv},
      primaryClass={astro-ph},
      url={https://arxiv.org/abs/0704.3618}, 
}

@article{Tarafdar:2013oqa,
    author = "Tarafdar, Pratik and Das, Tapas K. and Majumdar, Archan S.",
    title = "{Dependence of acoustic surface gravity on geometric configuration of matter for axially symmetric background flows in the Schwarzschild metric}",
    eprint = "1305.7134",
    archivePrefix = "arXiv",
    primaryClass = "gr-qc",
    doi = "10.1142/S0218271815500960",
    journal = "Int. J. Mod. Phys. D",
    volume = "24",
    number = "14",
    pages = "1550096",
    year = "2015"
}

@article{Pu:2012rv,
    author = "Pu, Hung-Yi and Maity, Ishita and Das, Tapas Kumar and Chang, Hsiang-Kuang",
    title = "{On Spin Dependence of Relativistic Acoustic Geometry}",
    eprint = "1204.1347",
    archivePrefix = "arXiv",
    primaryClass = "gr-qc",
    doi = "10.1088/0264-9381/29/24/245020",
    journal = "Class. Quant. Grav.",
    volume = "29",
    pages = "245020",
    year = "2012"
}

@article{Mondal:2019lxg,
    author = "Mondal, Tushar and Mukhopadhyay, Banibrata",
    title = "{Role of magnetically dominated disc-outflow symbiosis on bright hard-state black hole sources: ultra-luminous X-ray sources to quasars}",
    eprint = "1910.08564",
    archivePrefix = "arXiv",
    primaryClass = "astro-ph.HE",
    doi = "10.1093/mnras/staa1161",
    journal = "Mon. Not. Roy. Astron. Soc.",
    volume = "495",
    number = "1",
    pages = "350--364",
    year = "2020"
}

@article{Mondal:2018onm,
    author = "Mondal, Tushar and Mukhopadhyay, Banibrata",
    title = "{Ultraluminous X-ray sources as magnetically powered sub-Eddington advective accretion flows around stellar mass black holes}",
    eprint = "1808.10461",
    archivePrefix = "arXiv",
    primaryClass = "astro-ph.HE",
    doi = "10.1093/mnrasl/sly165",
    journal = "Mon. Not. Roy. Astron. Soc.",
    volume = "482",
    number = "1",
    pages = "L24--L28",
    year = "2019"
}

@article{Mondal:2018yjr,
    author = "Mondal, Tushar and Mukhopadhyay, Banibrata",
    title = "{Magnetized advective accretion flows: formation of magnetic barriers in Magnetically Arrested Discs}",
    eprint = "1802.01594",
    archivePrefix = "arXiv",
    primaryClass = "astro-ph.HE",
    doi = "10.1093/mnras/sty332",
    journal = "Mon. Not. Roy. Astron. Soc.",
    volume = "476",
    number = "2",
    pages = "2396--2409",
    year = "2018"
}

@article{Mondal:2019xyg,
    author = "Mondal, Tushar and Mukhopadhyay, Banibrata",
    title = "{FSRQ/BL Lac dichotomy as the magnetized advective accretion process around black holes: a unified classification of blazars}",
    eprint = "1904.05898",
    archivePrefix = "arXiv",
    primaryClass = "astro-ph.HE",
    doi = "10.1093/mnras/stz1062",
    journal = "Mon. Not. Roy. Astron. Soc.",
    volume = "486",
    number = "3",
    pages = "3465--3472",
    year = "2019"
}

\end{document}